\documentclass[twocolumn, tighten, times, twocolappendix]{aastex631}

\usepackage{epstopdf}
\usepackage{amsmath}
\usepackage{enumitem}

\newcommand{\HD}{HD 107146}
\newcommand{\HDN}{HD 92945}
\newcommand{\HDO}{HD 15115}
\newcommand{\HDnew}{HD 206893}
\def\inn{_{\rm in}}
\def\out{_{\rm out}}
\def\res{_{\rm res}}
\newcommand{\EQPP}[1]{Equation ({#1}, PI)}

\submitjournal{ApJ}

\shorttitle{Interactions between planets and massive debris disks}
\shortauthors{A. A. Sefilian, R. R. Rafikov, \& M. C. Wyatt}

\begin{document}

\title{Formation of Gaps in Self-gravitating Debris Disks by Secular Resonance in a Single-planet System. II. Towards a Self-consistent Model}

\correspondingauthor{Antranik A. Sefilian}
    \email{sefilian.antranik@gmail.com}

\author[0000-0003-4623-1165]{Antranik A. Sefilian}
    \altaffiliation{Alexander von Humboldt Postdoctoral Fellow.}
    \affil{Astrophysikalisches Institut und Universit{\"a}tssternwarte, Friedrich-Schiller-Universit\"at Jena, Schillerg\"a{\ss}chen~2--3, D-07745 Jena, Germany}
    \affil{Department of Applied Mathematics and Theoretical Physics, University of Cambridge, Wilberforce Road, Cambridge CB3 0WA, UK}

\author[0000-0002-0012-1609]{Roman R. Rafikov}
    \affil{Department of Applied Mathematics and Theoretical Physics, University of Cambridge, Wilberforce Road, Cambridge CB3 0WA, UK}
    \affiliation{Institute for Advanced Study, Einstein Drive, Princeton, NJ 08540, USA}

\author[0000-0001-9064-5598]{Mark C. Wyatt}
    \affil{Institute of Astronomy, University of Cambridge, Madingley Road, Cambridge CB3 0HA, UK}

\begin{abstract}
\noindent 
High-resolution observations of several debris disks  reveal  structures such as gaps and spirals, suggestive of gravitational perturbations induced by underlying planets.  Most existing studies of planet--debris disk interactions ignore the gravity of the disk, treating it as a reservoir of massless planetesimals. In this paper,  we continue our investigation into the long-term interaction between a single eccentric planet and an external, massive debris disk. Building upon our previous work, here we consider not only the axisymmetric component of the disk's gravitational potential, but also the non-axisymmetric torque that the disk exerts on the planet (ignoring for now only the non-axisymmetric component of the disk \textit{self}-gravity). To this goal, we develop and test a  semi-analytic `$N$-ring' framework that is based on a generalized (softened) version of the classical Laplace--Lagrange secular theory. Using this tool, we demonstrate that even when the disk is less massive than the planet,  not only can a secular resonance be established within the disk that leads to the formation of  a wide gap, but that the very same resonance also damps the planetary eccentricity $e_p$ via a process known as resonant friction. The resulting gap is initially non-axisymmetric (akin to those observed in HD~92945 and HD~206893), but evolves to become more axisymmetric (similar to that in HD~107146) as $e_p(t)\rightarrow0$ with time. We also develop analytic understanding of these findings, finding good quantitative agreement with the outcomes of the $N$-ring calculations. Our results may be used to infer both the dynamical masses of (gapped) debris disks and the dynamical history of the planets interior to them, as we exemplify for HD~206893.
\end{abstract}

\vspace{-2em}

\section{Introduction}
\label{sec:intro}

\subsection{Motivation}

Exoplanetary systems in general are comprised not only of planets, but also of belts of debris similar to Solar System’s asteroid and Kuiper belts \citep{hughes2018review, wyatt19review}. To date, such exo-Kuiper belts, or debris disks, have been detected around $\sim$20$\%$ of nearby main-sequence stars through their excess emission at infrared wavelengths \citep{montesinos2016, sibthorpe2018}.  These observations are explained by the presence of large, km-sized planetesimals that are continually colliding and generating a cascade of  bodies down to $\sim \mu$m-sized dust grains \citep{backman93, wyatt08collisionsreview}. Given that observed disks typically contain $\sim 0.001$--$1 M_{\earth}$ in mm/cm-sized grains \citep{panic2013, holland17}, the reservoir of parent planetesimals is estimated to be anywhere between $\sim 1 M_{\earth}$ and up to $\sim 1000 M_{\earth}$ in mass \citep[e.g.,][]{wyattdent2002, krivovwyatt20}. These mass estimates are obtained by extrapolating the mass of the observed dust to the mass of the unobservable dust-producing planetesimals. As such, they are of course subject to uncertainties related to the size distribution of disk particles, their maximum sizes, as well as the details of the collisional cascade model, both optical and physical \citep[for a detailed discussion, see e.g.][]{krivovwyatt20}. Nevertheless, debris disks are often considered as massive analogues of the current-day Kuiper belt \citep{wyatt19review}.

Recently, thanks to the advent of new generation instruments such as ALMA, it has become possible to image tens of debris disks with high angular resolution at millimeter wavelengths.  Imaging at such wavelengths is particularly fundamental to our understanding of debris disks \citep[e.g.,][and references therein]{hughes2018review, marino2022review}. This is so because millimeter observations trace $\sim$mm-sized grains that are  largely unaffected by non-gravitational radiation forces \citep{burns79}, and thus indirectly trace the spatial distribution of the otherwise undetectable parent planetesimals. High-resolution ALMA images have revealed a large variety of disk structures,  such as  gaps, spirals, warps, and eccentric rings \citep{hughes2018review, wyatt18review, wyatt19review}. Such structures can provide unique insights into the formation and evolution of exoplanetary systems, and may even help to infer the presence and parameters of otherwise-undetectable planets.  The Solar System provides a case in point to this end. Indeed, detailed studies of the asteroid and Kuiper belts have enabled the reconstruction of the Solar System's dynamical history \citep[see e.g. review by][and references therein]{morbi-kbo-review-2020} and, more recently, have led to a suggestion of a ninth planet beyond Neptune (\citealp{KB-p9}; see also \citealp{ST19} for a contrarian view).

Much effort has been put into understanding how planets and/or stellar companions sculpt debris disk morphologies through their gravitational perturbations \citep[e.g.,][to name a few]{wyattetal99, wyattSPIRAL05, leechiang2016, nesvold17, yelverton2018, tim21, farhat-sefilian-23}. However, so far the gravitational effects of debris disks have been largely omitted in studies of planet--debris disk interactions. That is, models often neglect the mass of the debris disk, treating it as a collection of massless planetesimals subject only to the gravity of the host star and (invoked) planets. This omission, however, may not always be justified, especially since observations suggest that debris disks could be as massive as tens -- if not hundreds -- of Earth masses \citep[e.g.,][]{  krivovwyatt20}. It is thus necessary to study the various ways in which the  gravitational perturbations due to both the disk and  planets would affect the evolution and morphology of debris disks.

\subsection{Our Previous Work}
\label{sec:previous_work_intro}

As a first step towards this direction, in \citet[][hereafter `Paper I']{Paper1} we investigated the role of the long-term, secular gravitational effects of debris disks in single-planet systems. To this end, we considered what is arguably the simplest possible setup: a host star surrounded by a radially-extended, massive debris disk situated exterior to, and coplanar with, a single low-eccentricity  planet (see also Figure \ref{fig:diagram_modelsystem_P1}).

A fundamental result of Paper I was that even  when the disk is less massive than the planet, the disk gravity can have a considerable impact on the secular evolution of  the constituent planetesimals -- contrary to what may be naively expected. In particular, we showed that the interaction generally results in the formation of a gapped -- i.e. double-ringed --  structure within the disk, akin to  those seen by ALMA in  \HD~\citep{marino2018}\footnote{Here, we note that during the review process of the current article, a new study by \citet{amaia2023} reported that the previously-thought single gap in the HD 107146 disk could  instead be two narrow gaps close to each other. While intriguing, this possibility is not considered throughout our manuscript.}, \HDN~\citep{marino2019}, \HDnew~\citep{marino2020, nederlander21}, and \HDO~\citep{macgregor2019}.

The physics behind the gap-forming mechanism proposed in Paper I is as follows. The combined gravity of the disk and the planet mediates the establishment of \textit{two} secular apsidal resonances within the disk. These resonances emerge as a result of an equality between the apsidal precession rate of planetesimals due to \textit{both} the disk and the planet, and that of the planet due to the \textit{disk gravity} (see also Figure \ref{fig:A-a-from-matrix-fid-simA-nominal}). At and around the resonant sites, planetesimal eccentricities are excited to relatively large values (i.e., $e \rightarrow 1$) in the course of the evolution. Since planetesimals on eccentric orbits spend more time away from their orbital semimajor axis,  an apparent depletion forms in the disk surface density at and around the resonance location.\footnote{We note that this line of argument has already been pursued previously by \citet{yelverton2018}, who considered a scenario whereby secular resonances arise due to the interactions of two (or more) planets interior to a \textit{massless} disk. We refer the interested reader to Section 1.1 of Paper I for a discussion of this and other mechanisms proposed in the literature for carving gaps in debris disks \citep[e.g.,][]{pearcewyatt15, shannon2016, maryam2016, zheng2017}.}

Paper I  demonstrated that this secular resonance-induced mechanism for sculpting gaps is quite robust. Indeed, it can operate over a wide range of planet-disk parameters: namely, for any given planet (interior to the disk), provided that the disk-to-planet mass ratio is between $\sim 10^{-4}$ and $2$. Additionally, it was found that the gap typically (i) forms over timescales of tens of Myr; (ii) is of $\mathcal{O}(10)$ au width; and (iii) is characterised by a non-axisymmetric, crescent shape with azimuthally varying depth and width. Finally, Paper I showed that one of the resonances always occurs near the disk's inner edge\footnote{This happens due to the divergence of the disk-induced precession rate of planetesimals orbiting near the sharp edges of a razor-thin disk; see Fig. \ref{fig:A-a-from-matrix-fid-simA-nominal}.} such that -- except if the two resonances are very close to each other -- it does not lead to any observable effect.

The work presented in Paper I, however, employed a rather simplified model, in the sense that in addition to the perturbations due to the planet, it only accounted for the axisymmetric component of the disk (self-)gravity, ignoring its non-axisymmetric contribution. This was done by essentially  modeling the disk as being \textit{passive}, i.e., as  providing a \textit{fixed} axisymmetric  gravitational potential. While this treatment allowed us to study the secular planet--disk interactions using analytical methods, it cannot be regarded as fully satisfactory. This is because, as one might expect, the disk can naturally develop some degree of non-axisymmetry due to, e.g., the torque exerted by the eccentric planet. Such a disk would have a non-axisymmetric component of its (self-)gravitational potential, which would in turn affect the orbital evolution of \textit{both}  the planet and the planetesimals. Thus, a more complete and self-consistent treatment of the disk gravity is warranted to uncover the full richness of the dynamics. This is the motivation for our study presented here.

\subsection{The Current Work}
\label{sec:current_work}

In this paper, the second in the series, we go a step further  in our investigation of the secular interaction between  eccentric planets and  external, massive debris disks.

The specific goals of our study are essentially two-fold.  Our first goal is to present a semi-analytic framework that allows for a self-consistent modeling of the secular evolution of  self-gravitating particulate disks and their response to external perturbations in general astrophysical setups. This framework is built upon the continuum version of the classical Laplace--Lagrange theory \citep{mur99}, whereby the disk is modeled as a series of $N {\gg} 1$ massive rings interacting both with each other and with (any) external perturbers  \citep[e.g.,][and references therein]{Touma2002, hahn2003, bat12, SR19}.

\begin{figure*}[ht!] 
\epsscale{0.95}
\plotone{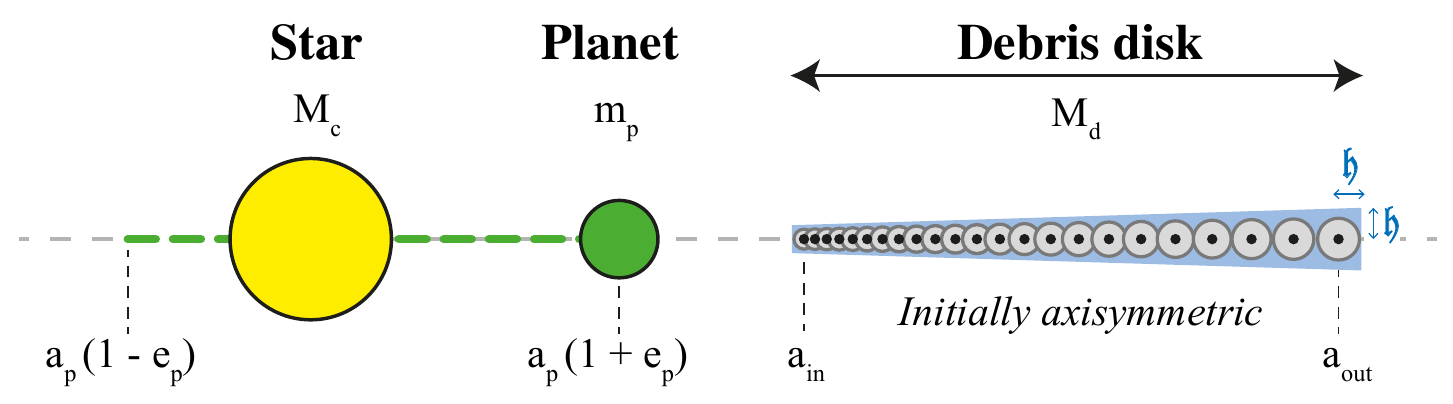}
\caption{A sketch of the initial setup of the model system considered in this and our previous work \citep{Paper1}.
A central star of mass $M_c$ is orbited by a planet of mass $m_p$ and an external debris disk of mass $M_d$ ($M_d, m_p \ll M_c$). 
The entire star-planet-disc system is coplanar. 
The planetary orbit (dashed green line), characterized by its semimajor axis $a_p$, is  slightly eccentric such that it does not cross the disk along its orbit (typically, $e_p \leq 0.1$) . 
The disk, extending from $a\inn$ to $a\out \gg a\inn$, is axisymmetric characterized by a truncated surface density profile given by Equation (\ref{eq:Sigma_d}).
The disk is modeled as a series of $N \gg 1$ massive rings with fixed semimajor axes, each characterized by a small radial and vertical half-thickness $\mathfrak{h}$, such that the aspect ratio $H$ of the disk is constant (Equation \ref{eq:eq_H_softening}). 
See the text (Section \ref{sec:problem_setup}) for further details.
}
\label{fig:diagram_modelsystem_P1}
\end{figure*}

Using this so-called `$N$-ring' model, our second goal is to explore the dynamical consequences of another component of disk gravity within the setup of our toy model (Figure \ref{fig:diagram_modelsystem_P1}): namely, the effects of the non-axisymmetric  torque that the disk would exert on the planet (absent in Paper I). In other words, we now account for all gravitational perturbations between the disk and the planet, with the exception of the non-axisymmetric component of the disk \textit{self-}gravity.\footnote{This assumption shall be relaxed in the next installation of this series, drawing on the PhD dissertation of A. Sefilian \citep{sefilian-phd}.}

As we show below, even at this level of approximation, the planet--debris disk interactions result in a new phenomenon absent in Paper I: namely, the circularization of the planetary orbit over time while, simultaneously, the disk develops a gap around the secular resonance. We find that the damping of the planetary eccentricity is realized through a process known  as \textit{resonant friction} in the literature \citep{tre98, wardhahn1998,  wardhahnprotostars}, which is distinct in origin from the well-known  eccentricity-damping processes such as scattering (e.g., \citealp{levison08, pearcewyatt15}). Indeed, the circularisation of the planetary orbit follows from the long-term, secular exchange of angular momentum between the planet and the planetesimals at and around the secular resonance, with the planet  not intersecting the disk along its orbit (Figure \ref{fig:diagram_modelsystem_P1}).

Our work is organized as follows. We discuss our general setup in Section \ref{sec:problem_setup} and then present the equations governing the evolution of the so-called $N$-ring model in Section \ref{sec:problemframework_P2}. Technical details and tests of the model can be found in Appendix \ref{app:tests_Nring}, where the relation between the present work and  Paper I is also made explicit. In Section \ref{sec:results_p2}, using a fiducial set of planet--disk parameters, we present the main results of our $N$-ring simulations describing the generic evolutionary behavior of both the planet and the debris disk.  Then, in Section \ref{sec:analysis_predictions}, we present and analyze the results of a suite of simulations, providing quantitative explanations for the differences relative to the simplified model of Paper I. We then discuss our results along with their implications and limitations in Section \ref{sec:discussion}. Our main findings are summarized in Section \ref{sec:summary}.

\section{Problem Setup}
\label{sec:problem_setup}

The general setup of the system that we consider in this work  is similar to that explored in Paper I. 
Namely, we consider a central star of mass $M_c$ that hosts a broad debris disk of mass $M_d$ exterior to, and coplanar with, a single planet of mass $m_p$ (with $m_p, M_d \ll M_c$). 
We characterize the planetary orbit by its semimajor axis $a_p$, eccentricity $e_p$, and longitude of pericenter $\varpi_p$. The planet's orbit is taken to be initially eccentric, but typically with $e_p(0) \leq 0.1$ such that it does not cross the disk along its orbit.
We consider the debris disk to be initially axisymmetric, populated by planetesimals on circular orbits with orbital radii of $r = a$. We parameterize the initial disk surface density $\Sigma_d^{t=0}(a)$ by a truncated power-law profile with an exponent $p$ so that  
\begin{equation}
\Sigma_d^{t=0}(a) = \Sigma_0 \bigg( \frac{a\out}{a} \bigg)^p ~~~~~~ \text{for} ~~~~ a\inn \leq a \leq a\out, 
\label{eq:Sigma_d}
\end{equation}
and $\Sigma_d^{t=0}(a) = 0 $ elsewhere; see also Equation (1) of Paper I. Here, $a\inn$ and $a\out$ are the semimajor axes of the innermost and outermost planetesimals in the disk, respectively, and $\Sigma_0$ is the initial surface density at $a = a\out$. The total disk mass $M_d$ contained in such a disk can be estimated with $M_d = 2\pi \int_{a\inn}^{a\out} a \Sigma_d^{t=0}(a) da$, yielding:
\begin{equation}
M_d = \frac{2 \pi}{2-p} \Sigma_0 a\out^2 \bigg[ 1 - \bigg(\frac{a\inn}{a\out}\bigg)^{2-p} \bigg] \approx \frac{2\pi}{2-p} \Sigma_0 a\out^2 , 
\label{eq:disk_mass_p2}
\end{equation}
see also Equation (2, PI) (hereafter, ``PI" means that the equation referred to can be found in Paper I). The approximation in Equation (\ref{eq:disk_mass_p2}) is valid as long as the disk edges are well separated, i.e., $\delta \equiv a\out/a\inn \gg 1$, and $p < 2$ so that most of the disk mass is concentrated in the outer parts.\footnote{We note that the approximation in Equation (\ref{eq:disk_mass_p2}) is provided  for reference only and not used in our analysis, except when the explicit assumption of $a\inn \ll  a\out $ is made for ease of analytical estimates further down in the analysis; see e.g. Equations (\ref{eq:tau_D_p2})--(\ref{eq:tau_D_p2_diskmassMd}).}

\subsection{The \texorpdfstring{$N$}{N}-ring Description of the Disk}

We now introduce a single but important modification to the model system employed in Paper I (for reasons that will be elucidated below): namely, we adopt a  \textit{discretized} description of the disk, rather than a continuous one.  To this end, we model the disk as a series of $N \gg 1$ nested, confocal, massive rings \citep[][]{Touma2002, hahn2003, bat12, SR19}, each characterized by its mass $m_j$, semimajor axis $a_j$, eccentricity $e_j$, and longitude of pericenter  $\varpi_j$ (with $j = 1,.., N$). Within this description, each and every disk ring would evolve (i.e., flex and precess over time) due to its gravitational interactions with the other disk rings, as well as the planet (a mathematical framework for this model is described in Section \ref{sec:problemframework_P2}). Accordingly, the gravitational potential generated by the entire disk would effectively evolve in time, allowing for a fully self-consistent representation of the disk and its (self-)gravity -- provided, of course, that $N \rightarrow \infty$. This is the main (and only) difference between the setups of this and our previous work which,  as described in Section \ref{sec:previous_work_intro}, assumes the disk to be a continuous rigid slab generating a fixed axisymmetric  potential that perturbs the motion of massless planetesimals embedded within.

Conceptually, each individual ring comprising the disk can be envisioned as a swarm of point-mass planetesimals, all sharing a common semimajor axis $a_j$ and the same mean orbital eccentricities $e_j$, apsidal angles $\varpi_j$, and, generally speaking,  inclinations $I_j$ and longitudes of ascending node $\Omega_j$. We  assume that the dispersion of orbital eccentricities and inclinations are small enough throughout the disk that the velocity dispersion $\sigma_v$ of planetesimals  at a given semimajor axis is smaller than the local Keplerian velocity $v_K$, i.e., $\sigma_v / v_K \simeq  [ \langle e^2 \rangle   +  \langle I^2 \rangle ]^{1/2} \lesssim 1$ \citep[e.g.,][]{ida-makino-92, jack93}. This introduces a finite, but non-zero, radial and vertical half-thickness  $\mathfrak{h} \simeq \sigma_v / n$ to each disk ring, where  $n$ is the ring's mean motion.\footnote{This is because e.g. in equilibrium, one has $\sqrt{\langle e^2 \rangle} = 2 \sqrt{\langle I^2 \rangle} $ due to the equipartition between the in-plane and off-plane velocity components \citep{ida-makino-92}, and the inclinations can be directly related to the vertical distribution (e.g. assuming a Gaussian profile provided small inclinations).}
Accordingly, we define the aspect ratio of the disk, $H$, as an intrinsically small parameter, which we assume to be constant throughout the disk  \citep{hahn2003}:
\begin{equation}
    H = \frac{\mathfrak{h}_j}{a_j} = \textrm{const} \ll 1 .
    \label{eq:eq_H_softening}
\end{equation}
Here, it is crucial to note that the dimensionless parameter  $H$ is one of the fundamental parameters of the $N$-ring model, as it represents the magnitude of the  gravitational softening parameter appearing in the ring-ring interaction \citep[Section \ref{sec:problemframework_P2}; see also][]{hahn2003, SR19}.

Qualitatively, the process of replacing particle orbits by massive rings is equivalent to and justified by the so-called \textit{secular approximation} \citep{mur99, SR19}.
This involves averaging the gravitational potential generated by the particles under consideration over the fast-evolving orbital angles, namely, the mean anomalies. 
Thus, the resultant rings -- which have the shape of the particle orbits -- would be of line densities that are inversely proportional to the orbital velocities of the particles along their orbits. 
This orbit-averaging procedure, also known as Gauss' method, renders the Keplerian energy, and so the orbital semimajor axis, of each ring\footnote{Note that hereafter we use the words `planetesimals', `disk rings', and `debris particles' interchangeably.} a conserved quantity.

Based on the discussion above, we assign the semimajor axes of the disk rings in our model such that they are logarithmically spaced between $a\inn$ and $a\out$. That is, the ratio of spacing between any two adjacent rings is constant such that $a_{j+1}/a_j = (a\out/a\inn)^{1/N}$, with $j$ running from the inner to the outer disk edge. 
Constancy of semimajor axis also implies that the disk mass density per unit semimajor axis, defined by $\mu_d(a) = dm(a)/da$, remains invariant in the course of evolution \citep{statler01, irina18}. 
We thus assign the masses of the disk rings $m_j$ by making use of the initial density profile (Equation \ref{eq:Sigma_d}) and the relationship $\mu_d(a) =  2\pi a \Sigma_d^{t=0}(a)$, which ensures that the total disk mass $M_d = \sum_{j} m_j$ is given by Equation (\ref{eq:disk_mass_p2}).

This completes the description of our model system and its initial configuration; a sketch of which is illustrated in Figure \ref{fig:diagram_modelsystem_P1}. We note that this setup is very similar to that explored in \citet{hahn2003} in the context of secular interactions between Neptune and the primordial, massive Kuiper belt.

\subsection{Fiducial Parameters}
\label{sec:fid_par_setupSection}

In this work, unless otherwise stated, we adopt a fiducial system with the following set of parameters: a $M_c = 1.09 M_{\odot}$ star;  a $M_d = 20 M_{\earth}$ disk with a power-law index of $p = 1$ with $a\inn = 30$ au and  $a\out = 150$ au; and a planet of $m_p = 0.6 M_J$  at  $a_p = 20$ au with $e_p(0) = 0.05$ and $\varpi_p(0) = 0$. These parameters correspond to the fiducial model adopted in Paper I; namely, \texttt{Model A} (Table \ref{table:debris_Nring_sims_table}). We remind the reader that according to Paper I, this combination of parameters guarantees that a secular resonance is established within an \HD-like disk at $a\res = 70$ au; see also Appendix \ref{chap:app_table_sim}.

Finally, we model the disk as a series of $N = 5000$ rings with an aspect ratio of $H =0.1$, while the planet is modeled as an (unsoftened) thin ring with $H=0$. 
These choices  are motivated by the study of \citet{SR19}, who determined the minimum number of disk rings that well-approximates the effects of a continuous disk for a given value of softening $H$. As demonstrated by \citet{SR19}, a key condition for this is that the disk rings are physically thick enough and close to each other that their cross sections overlap  (Figure \ref{fig:diagram_modelsystem_P1}).  Further justification and discussion of our choices for $N$ and $H$ is provided in Appendices \ref{sec:p2-vs-p1-section} and \ref{sec:sensitivity_to_N_H_P2}. Here, we note that the choice of the disk's aspect ratio is consistent with the value of $\approx 0.12 \pm 0.05$ measured for the \HD~disk based on ALMA observations \citep{marino2018}
\footnote{Here, it is worthwhile to point out that generally debris disks are expected to have a minimum ``natural'' aspect ratios of $\sim 0.05$, depending on the observational wavelength and the presence, or absence thereof, of embedded large bodies as well as gas \citep[see e.g.][]{thebault2009vertical, daley-aumic-mass, matra-beta-pic-hotcold, Johan22gas, Terrill2023}.}.

\section{Laplace--Lagrange secular theory: continuum version} 
\label{sec:problemframework_P2}

With the $N$-ring description of our model system  in place, we  proceed to describe the equations that govern the long-term, secular evolution of such systems. As in Paper I, we achieve this by working within the framework of orbit-averaged perturbation theory (to second order in eccentricities), ignoring perturbations that occur over short timescales comparable to the orbital periods of involved bodies (e.g., mean-motion resonances and scattering).

\subsection{Potential Softening in Astrophysical Disks}

According to Laplace--Lagrange secular theory, interacting bodies are smeared into massive rings and the resulting orbit-averaged disturbing function $\delta R$ is expressed as a power-series in eccentricities and inclinations and a Fourier series in the orbital angles \citep{plu18, mur99}. Thus, an astrophysical disk may, in principle, be modeled as a continuum of perturbers, i.e., with $N \rightarrow \infty$ rings, each interacting with the others as per the classical disturbing function $\delta R$. 

Unfortunately, however, a direct application of this method to self-gravitating disks is ill-posed from a mathematical point of view, since it would predict an infinite apsidal precession rate at all radii within the disk, which is unphysical  \citep{bat12, SR19}.  
This singularity is a restatement of the fact that the gravitational potential diverges at null inter-particle separations.
In terms of the disturbing function $\delta R$, this translates to the divergence of the Laplace coefficients $b_s^{(m)}(\alpha)$, 
\begin{equation}
    b_s^{(m)}(\alpha) = \frac{2}{\pi} \int_{0}^{\pi} \cos(m\theta) \big[ 1+ \alpha^2- 2\alpha\cos\theta \big]^{-s} d\theta,
\end{equation}
appearing in the expression of $\delta R$: namely, the fact that $b_{3/2}^{(m)} \rightarrow (1-\alpha)^{-2}$ when $\alpha \rightarrow 1$.
Given this, a potential solution would be to set up the disk rings such that e.g. they do not cross each other initially.  This approach, however, is doomed to fail as well, since it would predict a prograde apsidal precession for the disk rings, while in reality the apsidal precession would be \textit{retrograde} \citep[e.g.][]{hep80, ward81, JTgauss, bat12, sil15, SR19}.

These issues are usually overcome in the literature by making use of \textit{softened gravity}, i.e., by spatially smoothing the Newtonian point-mass potential \citep[][and the references therein]{SR19}. This is essentially done by introducing a small, but non-zero, softening length into the calculations, rendering the force between two rings finite -- rather than infinite -- at points of orbit crossings \citep[e.g.,][]{tre98, Touma2002, hahn2003, teygor16softening}. Physically speaking, the process of potential softening can be thought of as spreading the mass of a point-mass object over a Plummer sphere with a radius comparable to the softening length, which, when orbit-averaged, will yield a ring of non-zero thickness. The result of this procedure is a softened analogue of the classical Laplace--Lagrange theory, which would allow for an accurate description of the secular evolution of self-gravitating disks.

Recently, \citet{SR19} investigated the performance of various existing softening prescriptions in reproducing the eccentricity evolution of self-gravitating disks as expected from calculations of the disk  potential that do not rely on any form of softening \citep[e.g.,][]{hep80}. The authors identified softening methods which, in the limit of small (or vanishing) softening parameter, yield results that describe the expected behavior exactly, approximately, or do not converge at all (i.e., as in the un-softened Laplace--Lagrange theory). They also investigated the conditions under which a particular softening method succeeds or fails, both from the mathematical and physical point of views. One of the key physical conditions that the authors found is that the disk rings of small, but non-zero, thickness must be physically overlapping each other \citep[for further details, see][and e.g. Section 6.3 therein]{SR19}.

In our present study, we adopt one of the successful formalisms for potential softening as identified by \citet{SR19}; namely, that of \citet{hahn2003}. 
Physically speaking, the softening formalism of \citet{hahn2003} stems from accounting for the vertical extent of the interacting rings (rather than assuming that they are razor-thin), and vertically averaging the resulting disturbing function over the disk. 
 Mathematically, this process renders the resulting disturbing function a linear combination of \textit{softened Laplace coefficients} defined by
\citep{hahn2003, SR19}:
\begin{equation}
\mathcal{B}_s^{(m)}(\alpha, H) = \frac{2}{\pi} \int\limits_0^{\pi} \cos(m\theta) \bigg[1+\alpha^2-2\alpha\cos\theta +  H^2(1+\alpha^2)  \bigg]^{-s} d\theta , 
\label{eq:Bsm_hahn}
\end{equation}
where $H$ is the disk's aspect ratio (Equation \ref{eq:eq_H_softening}). 
For a detailed discussion of potential softening in disks,  we refer the reader to \citet{SR19}.

\subsection{The Disturbing Function: Softened}
\label{sec:Rj_P2_physical_meanings}

We now present the equations that describe the softened version of the classical Laplace--Lagrange theory, employing the softening formalism of \citet{hahn2003}.

Consider the eccentricity dynamics of an individual ring labeled by index  $j = 1,...,N$ constituting a radially-extended disk. 
The secular disturbing function $R_j$ that governs the evolution of the ring $j$  is then  determined by the perturbations arising due to all other rings labeled by $l \neq j$.
According to \citet{hahn2003}, the disturbing function $R_j$, once expanded to second order in eccentricities, reads as: 
\begin{equation}
    R_j = n_j a_j^2 \bigg[ \frac{1}{2} A_{jj} e_j^2  + \sum\limits_{l=0, l\neq j}^{N} A_{jl} e_j e_l \cos(\varpi_j- \varpi_l) \bigg] ,
    \label{eq:Master_Rj}
\end{equation}
where $n_j = \sqrt{G(M_c + m_j)/a_j^3}$ is the mean  motion of the perturbed $j$-th ring, and the meanings of the coefficients $A_{jj}$ and $A_{jl}$ are provided below. 
Here, we note that in writing Equation (\ref{eq:Master_Rj}), we have considered the planet as the zeroth ring, i.e., indexed as $j = 0$. This is justified by the fact that the mathematical structure of the disturbing function due to a planet is the same as that due to the disk rings, provided the softening is set to zero; see also \citet{SR19}. In doing so, the entire planet--disk system is modeled as a collection of $N+1$ rings. 

The coefficients $A_{jj}$ and $A_{jl}$ appearing in Equation (\ref{eq:Master_Rj}) are related to the perturbations arising due to the axi- and non-axisymmetric components of the $l$-th ring's gravity, respectively. They can be expressed as follows \citep{hahn2003}:
\begin{eqnarray}
A_{jj} &=& \frac{1}{4} n_j \sum\limits_{l=0, l\neq j}^{N} \frac{m_l}{M_c+ m_j} f(\alpha_{jl}, H), 
\label{eq:Ajj}
\\
A_{jl} &=& \frac{1}{4} n_j  \frac{m_l}{M_c+ m_j} g(\alpha_{jl}, H), \qquad j \neq l .
\label{eq:Ajk}
\end{eqnarray}
In Equations (\ref{eq:Ajj}) and (\ref{eq:Ajk}),  $\alpha_{jl}$ is defined such that  $\alpha_{jl} \equiv {a_l}/{a_j}$, and
the functions $f(\alpha, H)$ and $g(\alpha, H)$ which fully characterize the  ring-ring interactions are given by\footnote{In the notation of \citet{SR19}, the functions $f(\alpha,H)$ and $g(\alpha, H)$ read as follows: 
$f = 8 \phi_{11} = 8 \phi_{22}$ and $g = 4 \phi_{12}$, respectively; see their Table 1.}:
\begin{align}
& f(\alpha,H) = \alpha \mathcal{B}_{3/2}^{(1)}(\alpha,H) - 3\alpha^2 H^2 (2+H^2) \mathcal{B}_{5/2}^{(0)}(\alpha,H) , 
\\
& g(\alpha,H) = -\alpha \mathcal{B}_{3/2}^{(2)}(\alpha,H) + 3\alpha^2 H^2 (2+H^2) \mathcal{B}_{5/2}^{(1)}(\alpha,H) , 
\end{align}
where $H$ is the softening parameter (Equation \ref{eq:eq_H_softening}) and $\mathcal{B}_{s}^{(m)}(\alpha, H)$ are the softened Laplace coefficients (Equation \ref{eq:Bsm_hahn}).  Here, we note that the softened Laplace coefficients $\mathcal{B}_s^{(m)}$ can be rapidly evaluated by making use of their relationship to complete elliptic integrals, as outlined in \citet{SR19} (see Appendix C therein).

We point out that the disturbing function of Equation (\ref{eq:Master_Rj}) is valid for cases where the perturbed ring is interior to or exterior to the perturbing ring, i.e., $\alpha_{jl}>1$ and $\alpha_{jl} < 1$, respectively. 
Thus, unlike the expressions of the disturbing function developed in standard textbooks \citep[e.g.][]{mur99}, there is no need for two distinct expressions.  As demonstrated by \citet{SR19}, this follows from the fact that the softened Laplace coefficients satisfy the relationship
$\mathcal{B}_s^{(m)}(\alpha^{-1}) = \alpha^{2s} \mathcal{B}_s^{(m)}(\alpha)$, rendering the functions $f(\alpha, H)$ and $g(\alpha, H)$ symmetric when $\alpha$ is replaced with $\alpha^{-1}$, i.e., 
\begin{equation}
    f(\alpha^{-1}, H) = \alpha f(\alpha, H)  ~~~ \text{and} ~~~ 
    g(\alpha^{-1}, H) = \alpha g(\alpha, H) ,
    \label{eq:f_g_H03_symmetry}
\end{equation}
see also \citet{hahn2003}. Nevertheless, it is trivial to show that the disturbing function (\ref{eq:Master_Rj}) reduces to the classical pairwise expressions in \citet{mur99} (e.g., their equations 7.6 and 7.7) upon setting the softening parameter equal to zero, so that $B_s^{(m)}(\alpha, H = 0) = b_s^{(m)}(\alpha)$.

\subsubsection{Physical meaning of coefficients \texorpdfstring{$A_{jj}$}{Ajj} and \texorpdfstring{$A_{jl}$}{Ajl}}
\label{sec:Rj_Ajj_Ajl_phys_mean}

Finally, we explain the physical meanings of the coefficients $A_{jj}$ and $A_{jl}$ appearing in Equation (\ref{eq:Master_Rj}), highlighting their relationship to their unsoftened counterparts in Paper I wherever possible.  To ease the interpretation, here we remind that the ring indexed by $l$ represents the perturbing ring, while the ring indexed by $j$ is the perturbed one.

To begin with, the coefficient $A_{jj}$ given by Equation (\ref{eq:Ajj}) represents the precession rate of the free eccentricity vector of the $j$-th perturbed ring due to all other perturbing rings in the system with $l \neq j$. 
Thus, the term $A_{jj}$ with $j = 0$, i.e., $A_{00}$, is simply the free precession rate of the planetary orbit due to the entire disk; in Paper I,  this quantity is calculated in the continuum limit (i.e., $N \rightarrow \infty$) without any softening and  is denoted by $A_{d,p}$ (\EQPP{8}). Note that there is a one-to-one correspondence between $A_{00}$ and $A_{d,p}$  simply because the potential does not need to be softened when considering the effect of disks on external objects, e.g., the planet \citep[e.g.,][]{SR19}. Each of the terms $A_{jj}$ with $1 \leq j \leq N$, on the other hand, represents the free precession rate of the $j$-th disk ring due to the combined effects of the planet (i.e., $l = 0$) and the other disk rings (i.e., $l \neq j > 0$). In  terms of Paper I, this corresponds to  $A(a)$, i.e., the sum of the unsoftened planet-induced  and  disk-induced  free precession rates of planetesimals; that is, $A_p$ of \EQPP{4} and $A_d^{H =0}$ of \EQPP{6}, respectively. Note that similar to $A_{d,p}$ and  $A_{00}$, there is a one-to-one correspondence between $A_p(a)$ and $A_{jj}$ with $1 \leq j \leq N$ and $l=0$; this is because $H =0$ by construction in this case.

\begin{figure}[t!]
\epsscale{1.15}
\plotone{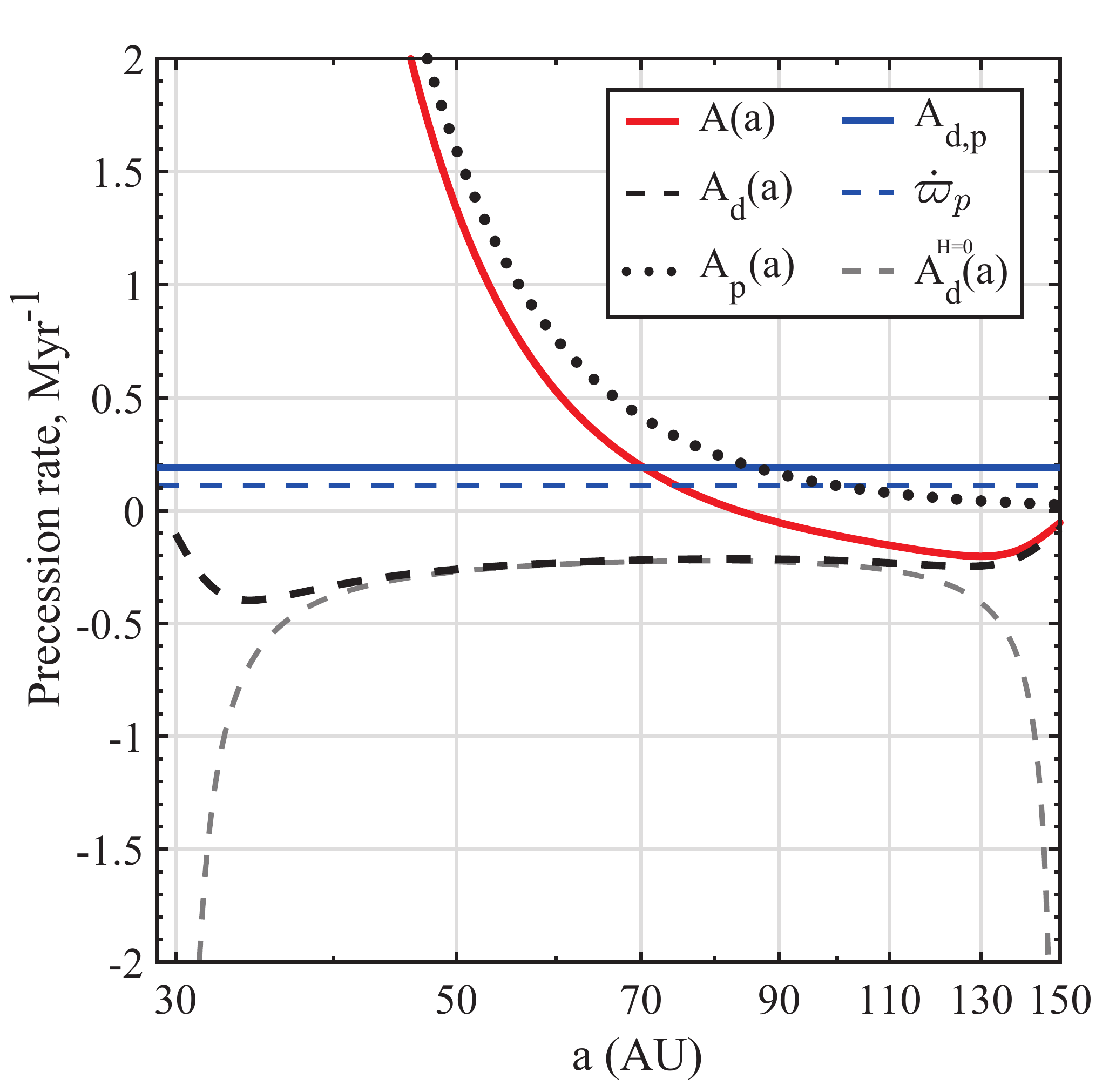}
\caption{Planetesimal free precession rate $A = A_d + A_p$ due to both planet and disk gravity (red curve)  as a function of semimajor axis. Calculations are done using the softened $N$-ring model (Section \ref{sec:problemframework_P2}), assuming the parameters of the fiducial planet--disk model (\texttt{Model A}; Table \ref{table:debris_Nring_sims_table}) with $N = 5000$ and $H=0.1$. The dotted and dashed black curves represent $A_p(a)$ and $A_d(a)$, respectively. The solid and dashed blue lines represent the rate of the free and total planetary precession due to the disk, $A_{d,p}$ and $\dot{\varpi}_p$, respectively. Note that the former is calculated using $A_{jj}$ of Equation (\ref{eq:Ajj}) with $j=0$ (Section \ref{sec:Rj_Ajj_Ajl_phys_mean}), whereas the latter is measured from the simulation of \texttt{Model A} presented in Section \ref{sec:results_p2}.  
This figure is to be interpreted as the softened analogue of Figure 1 of Paper I: for reference, the curve representing the unsoftened version of $A_d(a)$ is shown in dashed gray line and labeled as $A_d^{\rm H=0}(a)$. Note that $A(a) = A_{d,p}$ at $70$ au and $A(a) = \dot{\varpi}_p$ at $\approx 75$ au.}
\label{fig:A-a-from-matrix-fid-simA-nominal}
\end{figure}

For reference, we plot in Figure \ref{fig:A-a-from-matrix-fid-simA-nominal} the radial behavior of the  free precession rate of planetesimal orbits, as well as that of the planet, as computed using the expression of $A_{jj}$ given by Equation (\ref{eq:Ajj}). The calculations are performed for the fiducial planet--disk model, i.e., \texttt{Model A} (Table \ref{table:debris_Nring_sims_table}), using $N = 5000$ disk rings with $H = 0.1$. Figure \ref{fig:A-a-from-matrix-fid-simA-nominal} can thus be interpreted as the softened analogue of Figure 1 in Paper I. 
Note that the softened curve for the disk-induced planetesimal precession rate $A_d(a)$, i.e., $A_{jj}$ of Equation (\ref{eq:Ajj}) with $1 \leq j \leq N$ and $l\neq j >0$, well reproduces the expected behavior from Paper I that does not rely on any form of potential softening. For reference, the latter is shown by a dashed gray curve in Figure \ref{fig:A-a-from-matrix-fid-simA-nominal} and denoted by $A_d^{H=0}(a)$ to distinguish it from its softened counterpart $A_d(a)$.
One can see a good agreement between the softened and unsoftened counterparts throughout the disk, except near the disk edges where the unsoftened definition diverges \citep[see also][and Appendix \ref{sec:p2-vs-p1-section}]{irina18, SR19}. As for the profiles of $A_p(a)$ and $A_{d,p}$ shown in Figure \ref{fig:A-a-from-matrix-fid-simA-nominal}, we remind the reader that they are by definition equivalent to their unsoftened definitions of Paper I; this is simply because the potential need not be softened when considering the effect of the planet (disk) on the disk (planet). Thus, hereafter, we interchangeably refer to the  discretized, softened definitions of the free precession rates of Equation (\ref{eq:Ajj}) using  the same notations of Paper I, i.e., $A_p$, $A_{d,p}$, and $A_d$. Finally, we point out that the total planetary precession rate $\dot{\varpi}_p$ plotted (in dashed blue line) in Figure \ref{fig:A-a-from-matrix-fid-simA-nominal} -- which is different from the free precession rate $A_{d,p}$ -- is obtained from the $N$-ring simulation of \texttt{Model A} presented in Section \ref{sec:results_p2}. The difference between free and total precession rates is discussed in detail in Section \ref{sec:effect_on_wpdot}.

Moving on, we note that the coefficient $A_{jl}$ given by Equation (\ref{eq:Ajk}) characterizes the torque experienced by the $j$-th ring due to the non-axisymmetric component of the $l$-th ring's gravity. 
Thus, the term $A_{jl}$ with $l = 0$ is a measure of the non-axisymmetric perturbations exerted by the planet on the $j$-th disk ring. 
In terms of Paper I, this can be identified as $B_p$ (\EQPP{7}) scaled by $e_p(0)$. Accordingly, the  forced eccentricity of the $j$-th disk ring at $t=0$ due to the planet, considering the disk is massless, can be written as: 
\begin{eqnarray}
 e_{{\rm forced}, p}^{t=0} 
 = \frac{-A_{jl} e_l(0)  }{A_{jj}} \bigg|_{l = 0, H=0}  
 &=& \frac{b_{3/2}^{(2)}(a_p/a_j)}{b_{3/2}^{(1)}(a_p/a_j)} e_p(0) , \nonumber
 \\ 
 &\approx&   \frac{5}{4} \frac{a_p}{a_j} e_p(0),
 \label{eq:eforced_planet_p2_new}
\end{eqnarray}
The above expression is the same as that in Paper I, see Equation (14, PI), and the approximation on the second line of Equation (\ref{eq:eforced_planet_p2_new}) assumes $a_p \ll a_j$ (with $j=1,...,N$).

Finally, we note that the term $A_{jl}$ evaluated at $j = 0$ represents the non-axisymmetric perturbations that the $l$-th disc ring exerts on the planet, and when evaluated at $j \neq 0$, it represents the non-axisymmetric perturbations that the disk rings exert among themselves.   Here, we stress that both of these two contributions to the system's secular evolution were absent in Paper I.
The main goal of our present work is to account for the effects of the former in what we refer to as `nominal' simulations. In such $N$-ring simulations, the disk's non-axisymmetric potential is allowed to operate on the planet, but not on the disk particles themselves, i.e., $A_{jl} = 0$ for all but $j, l = 0$.
The `full' case, whereby the full gravitational perturbations of the disk rings are accounted for, is deferred to the third paper in this series (Sefilian, Rafikov \& Wyatt, in preparation; drawing on \citet{sefilian-phd}).

\subsection{Evolution Equations and Their Solution}

With the expression of the disturbing function in place (Equation \ref{eq:Master_Rj}), the secular evolution of the rings' orbital elements can be determined with the aid of Lagrange's planetary equations.
In particular, when expressed in terms of the eccentricity vector $\mathbf{e} = (k,h)$ where
\begin{equation}
    k = e \cos \varpi
    ~~~~ 
    \text{and} 
    ~~~~
    h = e \sin \varpi ,
    \label{eq:k_h_P2_corrections}
\end{equation}
Lagrange's planetary equations, taken to leading order in eccentricities, read as \citep{mur99}:
\begin{eqnarray}
    \frac{dk_j}{dt} 
    & \approx &  \frac{- 1}{n_j a_j^2} \frac{\partial R_j}{\partial h_j}
    =    - \sum\limits_{l = 0}^{N} A_{jl} h_l
    ,
    \nonumber
    \\
    \frac{dh_j}{dt} 
    & \approx &   \frac{1}{n_j a_j^2} \frac{\partial R_j}{\partial k_j}
    = \sum\limits_{l = 0}^{N} A_{jl} k_l
    .
    \label{eq:EOM_k_h}
\end{eqnarray}
Note that the system of equations (\ref{eq:EOM_k_h}) can be written in a more compact form when expressed in terms of the complex Poincar\'{e} variables, 
\begin{equation}
    \zeta \equiv  e ~ {\rm exp}(i \varpi) = k + i h . 
    \label{eq:zeta_kh}
\end{equation}
Indeed, with a simple application of chain rule, we find that:
\begin{equation}
    \frac{d\zeta_j}{dt}= i \sum\limits_{l=0}^{N} A_{jl} \zeta_l  .
    \label{eq:SimEqn}
\end{equation}
Equation (\ref{eq:SimEqn}) represents the master equation needed for our work, as it fully encapsulates the mutual gravitational interactions among all considered rings, i.e., disk and planet.

Here, we note that the coefficients $A_{jl}$ appearing in Equation (\ref{eq:SimEqn}) can be considered as the time-independent entries of an $(N+1) \times (N+1)$ square matrix $\mathbf{A}$; see Equations (\ref{eq:Ajj}) and (\ref{eq:Ajk}). Accordingly, the master equation (\ref{eq:SimEqn}) constitutes an eigensystem which can be solved  using standard methods, i.e., akin to the problem of $N$ coupled harmonic oscillators \citep[see e.g. Chapter 7 of][]{mur99}. Indeed, the time evolution of $\zeta_j$ can be written in closed form as follows:
\begin{equation}
    \zeta_j(t) = \sum\limits_{l= 0}^{N} T_l E_{jl} ~ \mathrm{exp}[i (\gamma_l t + \delta_l)] ,
    \label{eq:sol_of_EOM}
\end{equation}
where $\gamma_l$ and $E_{jl}$ represent the eigenvalues and eigenvectors of the matrix $\mathbf{A}$, respectively, while $\delta_l$ and $T_l$ are constants of integration determining the phases and relative amplitudes of the eigenvectors, respectively. A handy recipe for determining these constants is given in \citet{mur99}.

Despite the analytic nature of the solution given by Equation (\ref{eq:sol_of_EOM}), we note that its implementation
can become rather involved and inefficient when $N \rightarrow \infty$. 
This is particularly true for our purposes, since we model the disk by a relatively large number of rings, namely,  $N = 5 \times 10^3$, which is a requirement to ensure that the effects of the disk self-gravity are captured properly by the softened $N$-ring model \citep{SR19}; see Appendices \ref{sec:p2-vs-p1-section} and \ref{sec:sensitivity_to_N_H_P2}. 
Accordingly, in our work we instead opt to solve the master equation (\ref{eq:SimEqn}) numerically. We do this by making use of a six-stage, fifth order, Runge--Kutta ODE solver with a variable time-step \citep{NR}, such that the relative error does not exceed $10^{-8}$ per time-step.

Before moving on, we remark that although we employ the $N$-ring model to study the evolution of debris disks in single-planet systems, this framework would work equally well e.g. in the presence of multiple planets, provided mean-motion resonances may be ignored \citep{mur99}. The only obvious caveat is that the outlined $N$-ring model is accurate to lowest order in eccentricities and so the results are more reliable for $e \ll 1$  (see also Section 7.4 of Paper I for a detailed discussion).

\subsection{Implementation and Tests of the \texorpdfstring{$N$}{N}--ring Model}
\label{sec:tests-of-Nring-model}

The softened $N$--ring model described above  was first introduced, implemented, and tested by \citet{hahn2003}, and was further employed to study the secular evolution of the primordial Kuiper belt as well as narrow planetary rings \citep{hahn2003, hahnbending, hahn2008}. 
We implemented our own version of the $N$-ring code\footnote{A copy of the $N$-ring code (written in MATLAB scripts) has been made publicly available on Figshare; see the Data Availability statement at the end of this paper.
} and independently tested various aspects of it to ensure its proper operation.
An extensive report on these tests and their interpretation can be found in Appendix \ref{app:tests_Nring}, which, at first reading, may be skipped.
Below is a brief account of the outcomes of our tests:  
\begin{itemize}

    \item We demonstrate that the $N$-ring model conserves the total angular momentum of the system to second order in eccentricities, i.e., the same level of precision as the disturbing function of Equation (\ref{eq:Master_Rj}); see Appendix \ref{sec:L_conservation_P2}. 
    
    \item We confirm that the $N$-ring model successfully reproduces the analytical results of Paper I -- namely, the equations describing the evolution of planetesimal orbits -- provided that the disk's non-axisymmetric gravitational potential  is switched off; see Appendix \ref{sec:p2-vs-p1-section}. Note that in terms of Section \ref{sec:Rj_P2_physical_meanings}, this translates to  keeping the $A_{jj}$ terms given by Equation (\ref{eq:Ajj}) unmodified (to account for all axisymmetric perturbations), but   setting the terms $A_{jl}$ as given by Equation (\ref{eq:Ajk}) equal to zero  for all but the perturbing ring $l=0$ (i.e., the planet). For ease of discussion, from hereon, we refer to such simulations as `Paper I-like, simplified' $N$-ring simulations. 
    
    \item  We verify the findings of \citet{SR19} that the $N$-ring model is not significantly sensitive to the assumed number of rings $N$ and softening parameter $H$, as long as $H \lesssim 0.1$ and a minimum of $N \sim 10 H^{-2}$ rings are considered;  see Appendix \ref{sec:sensitivity_to_N_H_P2}.
    
    \item We confirm that the $N$-ring model can successfully reproduce the evolution of systems containing only planets (and not disks), provided the softening parameter is set to zero; see Appendix \ref{app:sun-jupiter-saturn}.
    
\end{itemize}
The reader not interested in the details of these tests may skip to the next section without loss of continuity.

\section{Results: Nominal Simulations}
\label{sec:results_p2}

In this section, we present the main results of our ``nominal'' $N$-ring simulations of planet--debris disk systems, in which the non-axisymmetric component of the disk gravity is allowed to operate on the planet, but not onto the disk rings themselves (Section \ref{sec:Rj_P2_physical_meanings}). Our specific aim here is to analyze the dynamical effects of the non-axisymmetric torque acting on the planet (absent in Paper I), with an eye on its potential consequences for the gap-forming mechanism presented in Paper I.

To this end, we ran a set of $67$ nominal $N$-ring simulations using the planet and disk parameters that we had identified as capable of producing a depleted region in an \HD-like disk in Paper I. 
In other words, the combinations of system parameters -- namely, $M_d$, $m_p$, and $a_p$ -- were chosen such that the system would, according to Paper I, guarantee the establishment of a secular resonance within the disk at $a\res = 70$ au (see e.g. Figure 7 of Paper I). 
In doing so, for the sake of some generality, we ignored the constraints from arguments related to the timescale of carving a gap and the width thereof.  
In  terms of Paper I, this means that we selected system parameters both within and outside the allowed portion of the parameter space portrayed in Figure 7 therein. 
For reference, the parameters of the simulated models together with the initial conditions, simulation times and outcomes are listed in Table \ref{table:debris_Nring_sims_table} of Appendix \ref{chap:app_table_sim}.

Despite the broad range of adopted planet--disk parameters, we found that the evolution of all systems followed the same qualitative behavior. Thus, to facilitate the interpretation of the simulations results and to compare them with Paper I, here we present results obtained for \texttt{Model A} -- which, we remind, was the fiducial configuration considered in Paper I.
In what follows, we  first describe  the orbital evolution of the planet and the planetesimals in Section  \ref{sec:typical_evolution_of_e_w_nominal_sim}, before focusing on the evolution of the disk morphology in Section  \ref{sec:typical_evolution_of_disc_morphology_p2}. The effects of parameter variations on the generic behavior are briefly discussed in Section \ref{sec:generality_comment_results}.

\begin{figure*}[ht!]
\epsscale{1}
\plotone{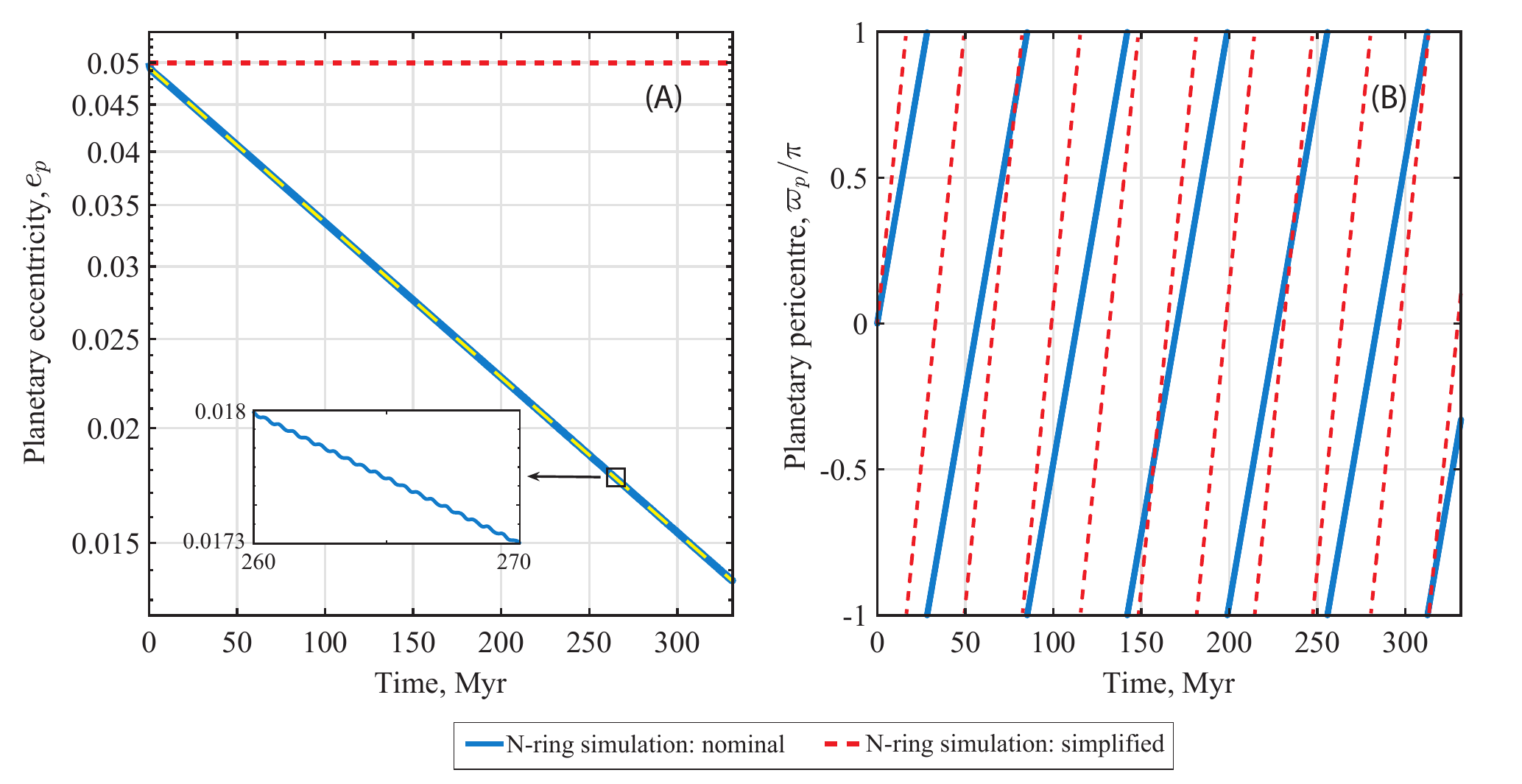}
\caption{The evolution of the planetary eccentricity $e_p$ (panel A) and longitude of pericenter $\varpi_p$ (panel B) in the nominal $N$-ring simulation of the fiducial planet--disk model, i.e., \texttt{Model A}, as an example of the general behavior observed in all other planet--disk systems simulated (Table \ref{table:debris_Nring_sims_table}). 
Results obtained within the simplified, `Paper I-like' $N$-ring simulation of the same planet--disk system are shown in red dashed lines in each panel. 
One can see that in the nominal simulation, the planetary eccentricity does not remain constant at its initial value of $e_p(0) = 0.05$ (as in Paper I), but rather decays  exponentially with  time, 
accompanied by small-amplitude oscillations; see the inset in panel (A). 
The dashed yellow line in panel (A) shows a numerical fit using an exponential model $e_p(t) = e_p(0)~{\rm exp}(-Dt/2)$ (see also Equation \ref{eq:ep_decay}), with $e_p(0) = 0.05$ and  $D \approx 7.7 \times 10^{-3} \rm{Myr}^{-1}$. 
Looking at panel (B), it is also evident that the planet precesses at a slower rate in the nominal simulation compared to the simplified one: indeed, $\tau_{\rm sec} = 2\pi/\dot{\varpi}_p \approx 57$ Myr in the nominal simulation, while $\tau_{\rm sec} \approx 33$ Myr in the simplified one.
See the text (Section \ref{sec:planet_evol_p2}) for more details.
}
\label{fig:ep-wp-fn-time-simA-p2}
\end{figure*}

\subsection{Evolution of Planetary and Planetesimal Orbits}
\label{sec:typical_evolution_of_e_w_nominal_sim}

The nominal $N$-ring simulations differ from the analytical calculations of Paper I by the introduction of the non-axisymmetric torque that the debris disk exerts on the planetary orbit. We thus start by presenting results showing the orbital evolution of the planet (Section \ref{sec:planet_evol_p2}). This will also aid in interpreting many of the dynamical features of the planetesimal evolution described later (Sections \ref{sec:planetesimal_evol_p2} and  \ref{sec:coupling-planet-planetesimal-p2}).

\subsubsection{Evolution of Planetary Orbit}
\label{sec:planet_evol_p2}

Figure \ref{fig:ep-wp-fn-time-simA-p2} summarizes the evolution of the planetary orbit throughout the nominal $N$-ring simulation of \texttt{Model A} (Table \ref{table:debris_Nring_sims_table}): the behavior of its eccentricity $e_p$ and apsidal angle $\varpi_p$ as functions of time are shown in blue curves in the left- and right-hand panels, respectively.
For ease of comparison with Paper I, we also plot in Figure \ref{fig:ep-wp-fn-time-simA-p2}, as dashed red curves, the corresponding results obtained with a simplified $N$-ring simulation of the same system, i.e., with the disk's non-axisymmetric perturbations on the planet switched off.
We remind the reader that such simplified $N$-ring simulations accurately reproduce the analytical results of Paper I; see e.g. Appendix \ref{sec:p2-vs-p1-section}. 
Here, we recall from Paper I that when the disk potential is taken to be axisymmetric, the disk simply causes the planetary orbit to precess at a constant rate so that ${\varpi}_p(t) = A_{d,p} t + \varpi_p(0)$, i.e., $\dot{\varpi}_p = A_{d,p}$ (see \EQPP{8} or $A_{jj}$ of Equation (\ref{eq:Ajj}) with $j=0$; Section \ref{sec:Rj_Ajj_Ajl_phys_mean}), while the planetary eccentricity remains constant, $e_p(t) = e_p(0)$. For further details, see Section 2.2.2 of Paper I.

There are several features to note in Figure \ref{fig:ep-wp-fn-time-simA-p2}. 
Beginning with Figure \ref{fig:ep-wp-fn-time-simA-p2}(A), a striking feature is the behavior of the planetary eccentricity, which, rather than remaining constant as in Paper I, undergoes a long-term decline. Indeed, one can see that in the course of the evolution, the planetary orbit circularizes significantly, with its eccentricity decreasing from the initial value of $e_p(0) = 0.05$ to $\approx 0.013$ by the time that the simulation is stopped at $t = \tau \approx 332 ~ {\rm Myr}$, i.e., approximately a four-fold decrease.
Looking at Figure \ref{fig:ep-wp-fn-time-simA-p2}(A) and the inset therein, it is also evident that the long-term decline is accompanied with additional small-amplitude oscillatory behavior with a short period of $\sim 1 ~{\rm Myr}$.
More interestingly, we find that the decay of the planetary eccentricity follows an exponential behavior rather closely at all times (note that Figure \ref{fig:ep-wp-fn-time-simA-p2}(A) is a semi-log plot). 
Indeed, the eccentricity decline can be well approximated by the exponential function $e_p(t) = e_p(0) ~ {\rm exp}(-D t/2)$ -- wherein the factor of $2$ is retained for later convenience (Equations \ref{eq:D_Tr98_MAIN} and \ref{eq:ep_decay}) -- with $e_p(0) = 0.05$ and $D \approx 7.7 \times 10^{-3} {\rm Myr}^{-1}$. For reference, this is illustrated using the dashed yellow line in Figure \ref{fig:ep-wp-fn-time-simA-p2}(A).

It is worthwhile to note here that for this simulation, the maximum fractional change in the system's total angular momentum deficit is on the order of $|\Delta L_e/L_e|\sim 10^{-8}$; see also Appendix \ref{sec:L_conservation_P2}. 
Thus, the decline of $e_p$ evident in Figure \ref{fig:ep-wp-fn-time-simA-p2}(A) is a real effect and not due to e.g. diffusion of numerical errors within the simulation. 
As a matter of fact, as we shall later show, the circularization of the planetary orbit is a generic phenomenon resulting due to a process known as ``resonant friction'' or ``secular resonant damping'' in the literature \citep{tre98, wardhahn1998, wardhahnprotostars}. This is studied in detail in Section \ref{sec:resonant_friction}.

Looking now at Figure \ref{fig:ep-wp-fn-time-simA-p2}(B), one can see that the planetary longitude of pericenter undergoes a prograde precession at a constant rate (linearly) in time. 
This is in line with the expectations from Paper I, although only on a qualitative level, but not quantitatively. 
Indeed, it is evident that the precessional period of the planet's apsidal angle  $\tau_{\rm sec} = 2\pi/\dot{\varpi}_p$ is longer in the nominal simulation than in the `Paper I'-like simulation: namely, with $\tau_{\rm sec} \approx 57$ Myr and $\approx 33$ Myr in the former and latter cases, respectively. In other words, the value of $\dot{\varpi}_p$ -- which can also be independently inferred from the slope of the numerical $\varpi_p(t)$ curves -- is $\dot{\varpi}_p \approx 0.58 A_{d,p}$ instead of $\dot{\varpi}_p = A_{d,p}$ (as in Paper I and the simplified simulation).
Interestingly, this also indicates that the axi- and non-axisymmetric components of the disk gravity drive planetary precession in opposite senses, with the effect of the latter being smaller than that of the former, resulting in a net prograde precession. 
This effect is characterized in detail both numerically and analytically in Section \ref{sec:effect_on_wpdot}.

To summarize, Figure \ref{fig:ep-wp-fn-time-simA-p2} shows that the back-reaction of the disk upon the planet not only causes the planetary longitude of pericenter to precess, but also leads to the circularization of the planetary orbit over time.
This will have indirect -- but important -- consequences for the evolution of planetesimal orbits  and the development of a gap within the debris disk.

\begin{figure}[ht!]
\epsscale{1.25}
\plotone{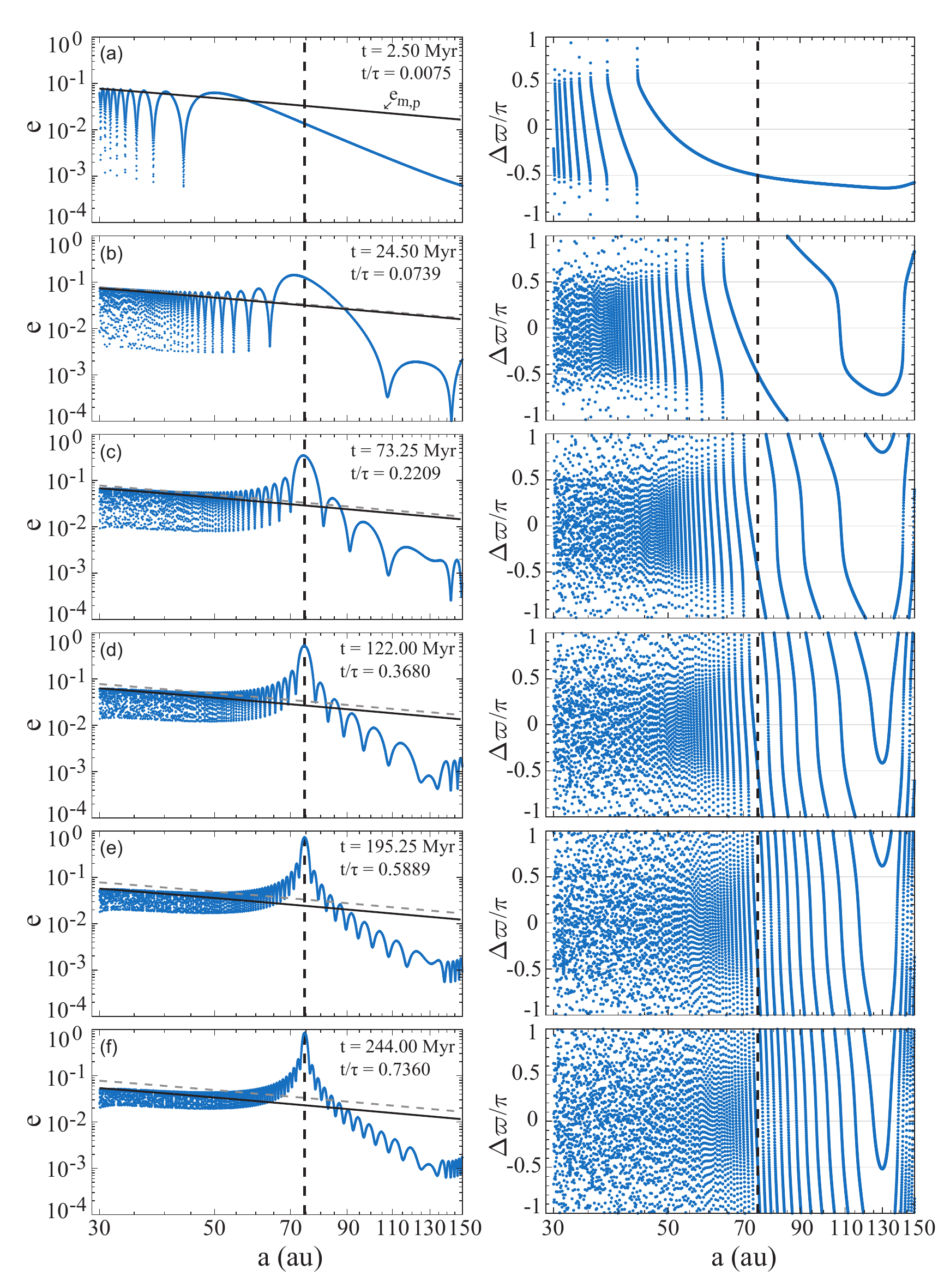}
\caption{Snapshots of the planetesimal eccentricities $e$ (left panels) and apsidal angles $\Delta\varpi$ (right panels, measured relative to that of the precessing planet) as a function of semimajor axis $a$ in the nominal $N$-ring simulation of \texttt{Model A} (Table \ref{table:debris_Nring_sims_table}). 
The snapshots are taken after  $t = 2.5$, $24.5$, $73.25$, $122$, $195.25$, and $244$ Myr of evolution (top to bottom). 
The time $t$ is also indicated relative to $\tau \approx 332$ Myr; the time at which $e(a\res) \rightarrow 1$.
The dashed vertical lines mark the location of the secular resonance, $a\res \approx 75 $ au.
For reference, the solid black lines in the left panels show the maximum planetesimal eccentricities $e_{m,p}(a) = 2 e_{\rm forced, p}(a)$ driven by the planet in the absence of the disk (Equation \ref{eq:eforced_planet_p2_new}).
Note that the curve of $e_{m,p}(a)$ decreases over time (see Equation\ref{eq:eforced_with_D_decay_p2}): for reference, the initial curve at $t = 0$ is shown by dashed gray lines.
This figure is available as an animation in the electronic edition of the journal. The animation runs from $t = 0 $ to $t = \tau = 332 $ Myr with a duration of $36$ seconds.
See the text (Section \ref{sec:planetesimal_evol_p2}) for more details.
}
\label{fig:e-w-fn-a-simA-p2}
\end{figure}

\subsubsection{Evolution of Planetesimal Orbits}
\label{sec:planetesimal_evol_p2}

We now present results describing the evolution of the debris particles. Figure \ref{fig:e-w-fn-a-simA-p2} shows snapshots of the disk rings' eccentricities $e$ and longitudes of pericenter $\Delta\varpi$ (relative to that of the precessing  planet) as a function of their semimajor axes at different times, as indicated in each panel. The times $t$ were chosen such that they correspond roughly to the same ratios of $t/\tau$ as in Figure 2 of Paper I, which, we remind, is a measure of the time relative to the time it takes for the eccentricity at the resonance to grow to unity. There are several notable features in this figure, which we discuss below.

To begin with, Figure \ref{fig:e-w-fn-a-simA-p2} shows that the evolution of the disk rings in the inner and outer parts of the disk proceeds differently, as already expected from Paper I. 
Indeed, at semimajor axes of $a \lesssim 75$ au, the eccentricities of the disk rings are maximized when  they are aligned with the planetary orbit, i.e., $\Delta\varpi = 0$, while at semimajor axes of $a\gtrsim 75$ au, their eccentricities are maximized when the rings are anti-aligned, i.e., $\Delta\varpi = \pi$. 
This is to be expected since in the reported simulation, the disk's non-axisymmetric gravity acts only on the planet and thus does not modify the free precession rates $A(a)$ of the disk particles when compared to Paper I; see e.g. Figure \ref{fig:A-a-from-matrix-fid-simA-nominal}. 
Consequently, the dynamics of disk particles remains planet- and disk-dominated at small and large distances from the planet, respectively (see e.g. Section 2.4 of Paper I). 
This behavior can also be understood by looking at Figure \ref{fig:A-a-from-matrix-fid-simA-nominal}, which shows that  $A(a) \rightarrow A_p(a)$ at $a \lesssim 75$ au, and  $A(a) \rightarrow - |A_d(a)|$ at $a\gtrsim 75$ au. Additionally,  and in line with Paper I, results of Figures \ref{fig:A-a-from-matrix-fid-simA-nominal} and \ref{fig:e-w-fn-a-simA-p2} clearly show that the transition between the two regimes occurs via a secular resonance, where $A(a) = \dot{\varpi}_p$ and the eccentricity of the disk ring grows in time until it reaches unity, $e(a\res) \rightarrow 1$. This, however, happens at $\approx 75$ au, and not at $70$ au as in Paper I; this difference will be discussed later in this section as well as in Section \ref{sec:effect_on_ares_shift}.

\begin{figure}[ht!]
\epsscale{1.25}
\plotone{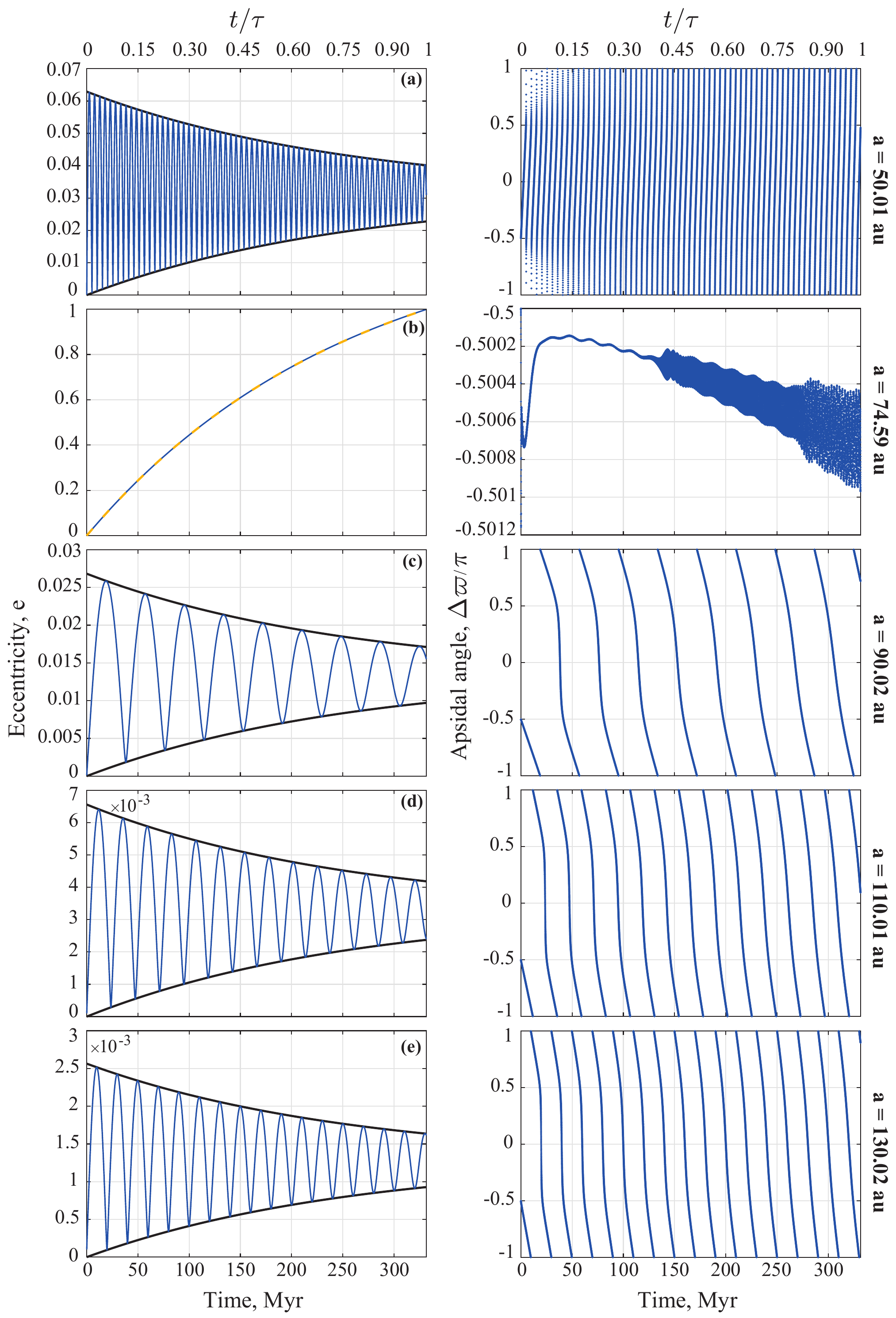}
\caption{The time evolution of planetesimal eccentricities $e$ (left panels) and apsidal angles $\Delta\varpi$ (right panels, relative to the planet) at five different semimajor axes, extracted from the nominal $N$-ring simulation of \texttt{Model A} (Table \ref{table:debris_Nring_sims_table}) shown in Figure \ref{fig:e-w-fn-a-simA-p2}. The values of the probed semimajor axes are indicated on the right side of the figure, and the time $t$ is also indicated in the upper x-axis relative to $\tau \approx 332$ Myr. The envelope of the eccentricity oscillations shown in black curves are obtained using Equation (\ref{eq:master_app_e}) (see also Section \ref{sec:coupling-planet-planetesimal-p2}). Note that at the secular resonance ($a = a\res \approx 75$ au), $\Delta\varpi(t) \approx -\pi/2$ at all times (panel b; note the different scale). It is also evident that the eccentricities at the resonance grow following a quadratic curve, which is perfectly reproduced by Equation (\ref{eq:e_t_at_resonance_p2}) as shown by the dashed yellow curve in panel (b). This figure can be compared to Figure \ref{fig:P1-vs-P2}, which portrays the corresponding results of the same planet--disk model in the simplified, `Paper I'-like, $N$-ring simulations. See the text (Sections \ref{sec:planetesimal_evol_p2} and \ref{sec:coupling-planet-planetesimal-p2}) for more details.}
\label{fig:e-w-fn-time-simA-p2}
\end{figure}

This said, however, we note that there are several differences between the results shown here in Figure \ref{fig:e-w-fn-a-simA-p2} and those of Paper I. 
First, looking at the right column of Figure \ref{fig:e-w-fn-a-simA-p2}, one can see that soon after the evolution starts, the apsidal angles $\Delta\varpi$ of the disk rings, both in the inner and outer disk parts, span the entire range $[-\pi, \pi]$ over time. This is in contrast with the results of Paper I, where $\Delta\varpi$ remained confined at all times within the ranges $[-\pi/2 , \pi/2]$ and $\pm [\pi/2, \pi]$ in the inner and outer disk parts, respectively. 
In our current simulations, this expectation holds true for the majority (but not all) of the disk rings as long as not much time has elapsed from the beginning of the simulation, i.e., $t/\tau \lesssim 0.1 $ or so; see e.g. Figures \ref{fig:e-w-fn-a-simA-p2}(a),(b) and Figure \ref{fig:e-w-fn-time-simA-p2}.

Second, looking at the left column of Figure \ref{fig:e-w-fn-a-simA-p2}, it is evident that the amplitudes of the planetesimal eccentricity oscillations  do not remain constant in time (as expected from Paper I), but rather undergo a slow decline -- see also the animated version of Figure \ref{fig:e-w-fn-a-simA-p2}.
Indeed, one can see that the maximum eccentricities decline by about a factor of $2$ at all semimajor axes, although it is a bit difficult to discern this effect in the outer parts of the disk due to the small eccentricities in that region. 
This is evidenced in Figure \ref{fig:e-w-fn-a-simA-p2}, for instance, for the planetesimals in the inner disk parts by the decreasing upper envelope of $e(t)$ over time; see the full black and dashed gray lines in the left panels.
Upon closer inspection, we also find that this decline in eccentricity amplitudes occurs roughly over the same timescale at all semimajor axes, and that it is not accompanied by any change in the periods of the associated eccentricity oscillations. These  effects can be seen more clearly in Figure \ref{fig:e-w-fn-time-simA-p2}, where we plot the time evolution of the eccentricities and apsidal angles of disk rings at five different semimajor axes. 
Indeed, looking at Figure \ref{fig:e-w-fn-time-simA-p2}, one can see that at a given semimajor axis, the period at which planetesimal eccentricities oscillate does not change as the amplitudes are damped.
We note that Figure \ref{fig:e-w-fn-time-simA-p2} can also be compared with Figure \ref{fig:P1-vs-P2} in Appendix \ref{sec:p2-vs-p1-section}, which depicts the results corresponding to Paper I, i.e., neglecting the disk's non-axisymmetric perturbations on the planet.

Apart from the features discussed above, the results depicted in Figure \ref{fig:e-w-fn-a-simA-p2} indicate that while a secular resonance is established within the disk as expected from Paper I, its location is slightly different than anticipated.  Indeed, one can see that the resonance occurs at around $\approx 75$ au rather than at $70$ au as expected from Paper I; see the dashed vertical lines in Figure \ref{fig:e-w-fn-a-simA-p2}. This said, however, the apsidal angle $\Delta\varpi$ at the resonance remains fixed at $-\pi/2$ throughout the simulation, in line with the expectations from Paper I; see Figures \ref{fig:e-w-fn-a-simA-p2} and \ref{fig:e-w-fn-time-simA-p2}(b). 
Additionally and interestingly, the growth of the eccentricity at the resonance does not occur linearly in time as was the case in Paper I; see e.g. Figure \ref{fig:P1-vs-P2}(b). 
Instead, as can be seen clearly in Figure \ref{fig:e-w-fn-time-simA-p2}(b), it displays a linear growth phase at early times, i.e. $t/\tau \lesssim 0.1$,  after which it smoothly becomes slower, resembling more of a quadratic curve. The change in the growth rate is significant, in the sense that it extends the time needed for the eccentricity at the resonance to grow to unity by about a factor of $\approx 2.5$ relative to the expectations from Paper I: namely, from $\tau \approx 135$ Myr to $\tau \approx 332$ Myr (e.g., see Figure \ref{fig:e-w-fn-time-simA-p2}(b) and the animated version of Figure \ref{fig:e-w-fn-a-simA-p2}).

Finally, we point out that unlike in Paper I, our simulations show no evidence of a secular resonance at $\approx a\inn$ (apart from the one at $\approx 75$ au).
This follows from the fact that the free precession rate of the debris driven by the disk, $A_d(a)$, converges to a finite value  as the disk edges are approached, i.e., $a \rightarrow a\inn, ~a\out$ (rather than diverging as in Paper I); see Figure \ref{fig:A-a-from-matrix-fid-simA-nominal}. 
This is a direct consequence of modeling the disk with a small but non-zero thickness $H$, as already explained in Paper I \citep[see also][]{irina18, SR19}.
It is as a result of this convergence that the resonance condition is no longer satisfied near $a\inn$ and one has $A \approx A_p \gg \dot{\varpi}_p$ as $a \rightarrow a\inn$. The same argument also explains why the apsidal angles $\Delta\varpi(a)$ of debris with $130~{\rm au} \lesssim a \lesssim a\out$ are characterized by a positive slope, i.e., $d\Delta\varpi/da >0 $; see the right-hand panels of Figure \ref{fig:e-w-fn-a-simA-p2}. Indeed, the convergence of $A_d(a)$ to  a finite value as $a \rightarrow a\out$ renders the total free precession rate $A(a)\approx A_d(a)$ in that region to be an increasing function of semimajor axis (rather than decreasing as in Paper I); see Figure \ref{fig:A-a-from-matrix-fid-simA-nominal}.

\subsubsection{The Coupling between the Planet and Planetesimals}
\label{sec:coupling-planet-planetesimal-p2}

Before moving on, it is worthwhile  to pause here and decipher the physics behind the  evolution of planetesimal orbits (Section \ref{sec:planetesimal_evol_p2}). 
Obviously,  planetesimal dynamics is affected only indirectly by the introduction of the disk's non-axisymmetric torque on the planet (absent in Paper I). 
Thus,  the resulting behavior of planetesimals as depicted in Figures \ref{fig:e-w-fn-a-simA-p2} and \ref{fig:e-w-fn-time-simA-p2} should, in principle, be understood as a result of the coupling between the planet and the planetesimals. 


Let us for a moment ignore the evolution of the planet's eccentricity and focus on its apsidal precession (Figure \ref{fig:ep-wp-fn-time-simA-p2}). As described in Section \ref{sec:planet_evol_p2}, the planetary orbit precesses at a rate less than  $A_{d,p}$, i.e., $\dot\varpi_p \lesssim A_{d,p}$, due to the disk's non-axisymmetric gravity acting on it. For reference, the value of $\dot{\varpi}_p$ as extracted from the simulation of \texttt{Model A} is overplotted in Figure \ref{fig:A-a-from-matrix-fid-simA-nominal}, see the blue dashed line therein. Looking at Figure \ref{fig:A-a-from-matrix-fid-simA-nominal}, one can see that while $A(a) = A_{d,p}$ at $a = 70$ au as expected from Paper I, one has  $A(a) = \dot{\varpi}_p$ at around $a \approx 75$ au, which coincides with the exact location of the secular resonance; see Figure \ref{fig:e-w-fn-a-simA-p2} and \ref{fig:e-w-fn-time-simA-p2}(b). Accordingly, this suggests that the shift in the resonance location results from the slower precession rate of the planetary orbit. We will further test and verify this hypothesis later in Section \ref{sec:effect_on_ares_shift}.

\begin{figure*}[ht!]
\epsscale{0.95}
\plotone{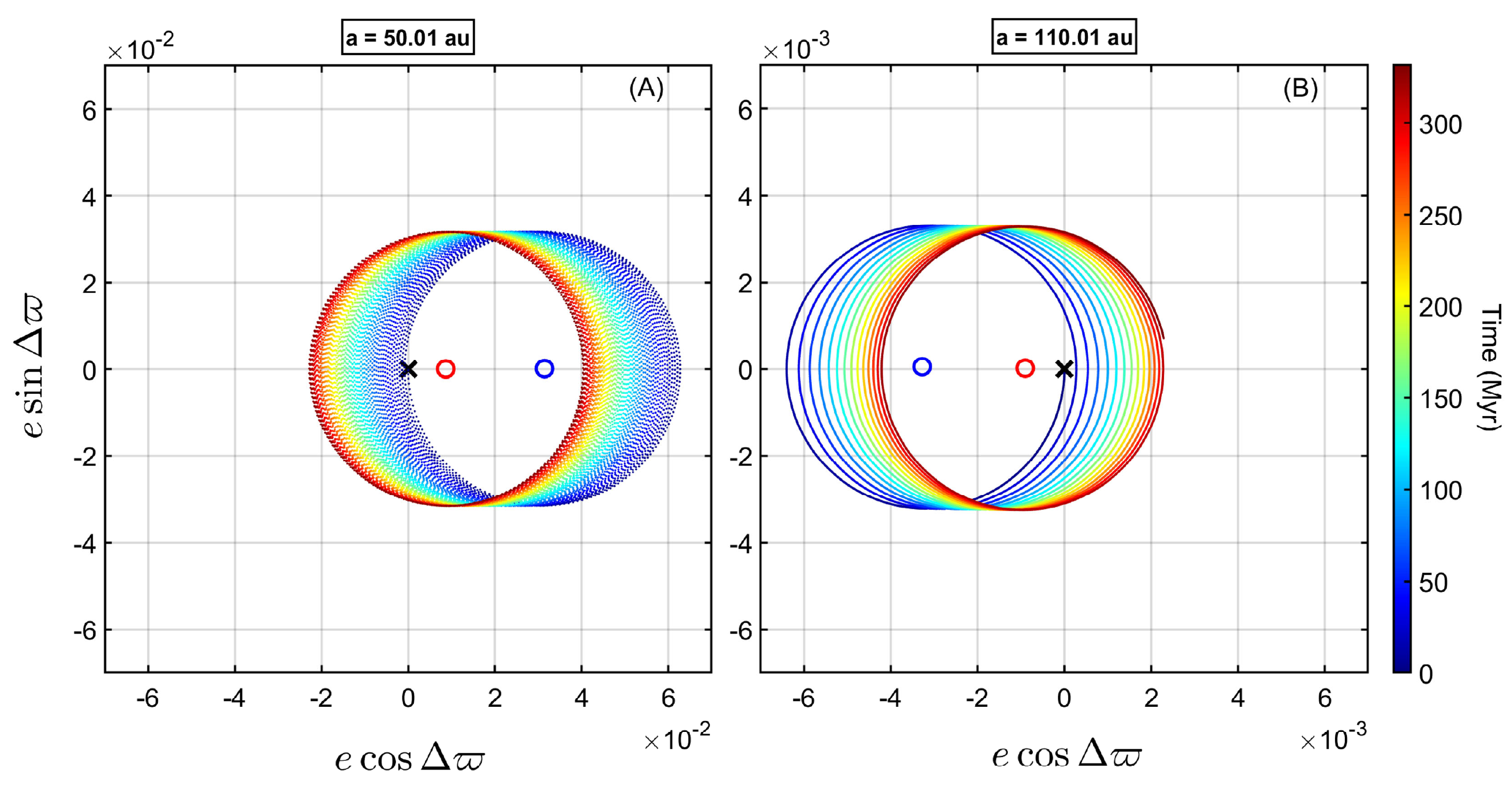}
\caption{The secular evolution of the complex eccentricity,    
$ \zeta =  e ~ {\rm exp}(i \Delta \varpi)$, 
of planetesimals orbiting at semimajor axes of $a = 50$ au (panel A) and $a = 110$ au (panel B) in the  nominal $N$-ring simulation of \texttt{Model A} (Table \ref{table:debris_Nring_sims_table}). 
In each panel, the  blue and red open circles represent the initial and final values of forced eccentricities about which the free eccentricity vector of planetesimals precesses; see also Equations (\ref{eq:e)_free}) and (\ref{eq:e)_forced}). 
Note that the forced eccentricities lie nearly along the x-axis but not exactly; theirs phase angles are  $\varphi_0 \approx 0.18^{\circ}$ and $\varphi_0 \approx 179.16^{\circ}$ in panels (A) and (B), respectively -- see Equation (\ref{eq:phi0}). The black crosses mark the initial planetesimal eccentricities.
Note that the precession is anti-clockwise for the planetesimal at $a = 50$ au, and clockwise for the one orbiting at $a = 110$ au. 
The time $t$ is represented by the colorbar, which runs from $t = 0$ to $t=\tau \approx 332$ Myr. The data is plotted at every $\approx 0.025$ Myr.
See the text (Section \ref{sec:coupling-planet-planetesimal-p2})  for further details. 
}
\label{fig:complex-h-k-scaled}
\end{figure*}

Let us now consider the effects of the decaying planetary eccentricity on the planetesimal dynamics (Figure \ref{fig:ep-wp-fn-time-simA-p2}).
This effect introduces an additional complication to the problem as it renders the gravitational potential of the planet time-dependent: in terms of Paper I, this means that the term $B_p \propto e_p$ is no longer constant in time; see also Section \ref{sec:Rj_P2_physical_meanings} and \EQPP{7}.
To gain additional insights, in Appendix \ref{appendix:analytical_solution_with_ep_decay_D} we derive a full time-dependent solution for the planetesimal eccentricities $e(t)$ and apsidal angles $\Delta\varpi(t)$ in the presence of a circularizing planet  in a frame co-precessing with the planet; see Equations (\ref{eq:master_app_e})--(\ref{eq:tanDw_effect_of_smallD}).
The main takeaway from Appendix \ref{appendix:analytical_solution_with_ep_decay_D} is that the decay of $e_p$ renders the forced component of the planetesimal eccentricity a time-dependent function (Equation  (\ref{eq:e)_forced})), while the free eccentricity -- which, recall, is set by initial conditions -- remains constant (Equation (\ref{eq:e)_free})). Indeed, we find that the forced eccentricity decays following 
%
\begin{equation}
e_{\rm forced}(a,t) \approx e_{\rm forced}^{\rm PI}(a) \times {\rm exp}(-D t /2), 
\label{eq:eforced_with_D_decay_p2}
\end{equation}
%
where $e_{\rm forced}^{\rm PI}(a)$ is given by \EQPP{13}, 
%
\begin{equation}
e_{\rm forced}^{\rm PI}(a) =  \frac{- B_p(a)}{A_d(a) + A_p(a) - A_{d,p}},
\end{equation}
%
and must now be understood as evaluated for $e_p = e_p(0)$. As a result, and unlike in Paper I, the planetesimal eccentricities do not oscillate between their initial values of $0$ and constant maxima of $e_m^{\rm PI} = 2|e_{\rm forced}^{\rm PI}(a)|$. Instead, planetesimal eccentricities now oscillate with a decreasing amplitude, while at the same time the minima attained in the course of oscillations increase over time (Figure \ref{fig:e-w-fn-time-simA-p2}). 
Eventually, as $t \rightarrow \infty$ and $e_{\rm forced}(a,t) \rightarrow 0$, planetesimal eccentricities converge to their free component, $e(a) \rightarrow e_{\rm free} = e_{\rm forced}^{\rm PI}(a) = e_m^{\rm PI}/2$; see Figure \ref{fig:e-w-fn-time-simA-p2}.
This behavior is well captured by Equation (\ref{eq:master_app_e}) of Appendix \ref{appendix:analytical_solution_with_ep_decay_D} which shows that the minimum and maximum eccentricity during the oscillations change in time following $\propto 1-{\rm exp}(-D t/2)$ and $\propto 1+{\rm exp}(-D t/2)$, respectively -- see the black curves in the left panels of Figure \ref{fig:e-w-fn-time-simA-p2}.  For reference, the black lines in Figure \ref{fig:e-w-fn-a-simA-p2} show the maximum eccentricity $e_{m,p}$ predicted by Equation (\ref{eq:master_app_e}) upon neglecting $A_d$ and $\dot{\varpi}_p$, which is valid for the inner disk parts where the dynamics is planet-dominated.  One can see a very good agreement between the simulation results and the curve of $e_{m,p}$ (see also the animated version of Figure \ref{fig:e-w-fn-a-simA-p2}). Equation (\ref{eq:master_app_e}) also explains the behavior of eccentricity growth at the resonance: indeed, taking the limit of $A = A_d + A_p \rightarrow \dot{\varpi}_p$ in Equation (\ref{eq:master_app_e}), one finds that
\begin{equation}
    e(t) = \frac{2 |B_p(0)|}{D} \big[ 1- {\rm exp}(-D t/2) \big] \approx |B_p(0)| t \bigg( 1 - \frac{1}{4} D t \bigg), 
    \label{eq:e_t_at_resonance_p2}
\end{equation}
where the approximation assumes $D \approx 0$, i.e., $D \ll |B_p(0)|$, and all quantities are evaluated at the resonance location $a\res$.  For reference,  Equation  (\ref{eq:e_t_at_resonance_p2}) as evaluated at the values of $D$ and $a\res$ extracted from the simulation of \texttt{Model A} is plotted using a dashed yellow line in Figure \ref{fig:e-w-fn-time-simA-p2}(b): one can see the perfect agreement between the numerical results and Equation  (\ref{eq:e_t_at_resonance_p2}).

Finally, we note that this analysis also clarifies why the disk particles evolve to populate the entire range of $\Delta \varpi = [-\pi , \pi]$ in our simulations (in contrast to Paper I); see e.g. the right panels of Figures \ref{fig:e-w-fn-a-simA-p2} and \ref{fig:e-w-fn-time-simA-p2}. We illustrate this in Figure \ref{fig:complex-h-k-scaled}, where we plot the complex eccentricities of planetesimals orbiting at semimajor axes of $a = 50$ au (panel A) and $a = 110$ au (panel B) in the simulation of \texttt{Model A}, to exemplify the behavior in the inner and outer disk parts, respectively. 
Without loss of generality, let us consider the planetesimals in the inner disk parts, i.e., at $a\inn \lesssim a \lesssim a\res$.  With their orbits being initially circular, their eccentricities start at the origin of the phase space defined by the $e (\cos\Delta\varpi, \sin\Delta\varpi)$ plane; see e.g. Figure \ref{fig:complex-h-k-scaled}(A). Given that $A(a) \approx A_p > 0$ in this planet-dominated region,  planetesimal eccentricities would precess counter-clockwise in a circle around the forced eccentricity vector. 
Since $e_{\rm forced}> 0$ in this case, the circle would initially be restricted to $e \cos\Delta\varpi \geq 0$; meaning that $\Delta\varpi$ can only acquire values between $-\pi/2$ and $+\pi/2$. As time progresses and the forced eccentricity decays, however, the center of the circle slowly shifts towards the origin of the $e (\cos\Delta\varpi, \sin\Delta\varpi)$ plane, causing the circle to cross  all quadrants of the plane and so allowing $\Delta\varpi$ to explore the entire range of $[-\pi, \pi]$. This behavior also explains why  the minimum of the eccentricity oscillations grows to values larger than $0$ with time. 
%
%
Note that as $t \rightarrow \infty$ and $e_{\rm forced} \rightarrow 0$, planetesimal eccentricities precess with a magnitude equal to their free eccentricities around the origin of the $e(\cos\Delta\varpi, \sin\Delta\varpi)$ plane.
%
A similar argument can be applied to the outer disk parts, where the dynamics is in the disk-dominated regime (i.e., $A(a) \approx A_d$) and $e_{\rm forced} <  0$, explaining the results of Figure \ref{fig:complex-h-k-scaled}(B) and Section \ref{sec:planetesimal_evol_p2}. 
In closing, we point out that the decay of the planetary eccentricity gives rise to a negligible misalignment between the forced eccentricity vector of planetesimals and the planetary eccentricity vector (Appendix \ref{appendix:analytical_solution_with_ep_decay_D}). This amounts to a phase shift of $\lesssim 1^{\circ}$ with respect to the x-axis in Figures \ref{fig:complex-h-k-scaled}(A) and (B); see also Equation (\ref{eq:phi0}).

\subsection{Evolution of the Disk Morphology}
\label{sec:typical_evolution_of_disc_morphology_p2}

Having described the secular evolution of the planetesimal and planetary orbits in the fiducial configuration,  we now present and analyze results showing the evolution of the disk morphology.

To do so, we first construct maps of disk surface density distribution by making use of the eccentricity–apsidal angle distribution of planetesimals  as simulated using the $N$-ring model of Section \ref{sec:problemframework_P2} (see e.g. Figure \ref{fig:e-w-fn-a-simA-p2}).
This is done using the same technique adopted in Paper I. To be specific, at each time step, we first populate each of the simulated $N$ rings by $N_{\rm np} = 10^{4}$ new particles; each with mass $m_j/N_{\rm np}$, same orbital elements as the parent ring -- i.e., $a_j$, $e_j$, and $\varpi_j$ -- but with mean anomalies that are randomly distributed between $0$ and $2\pi$. 
We then bin the resulting $N \times N_{\rm np}$ particles in their Cartesian coordinates in the disk plane denoted by ($X$, $Y$), compute the total mass per bin, and divide by its area to arrive at the disk surface density $\Sigma(X,Y)$ at a given time into the evolution.
For further technical details about this procedure, we refer the reader to Appendix C of Paper I.

\begin{figure*}[ht!]
\epsscale{1.0}
\plotone{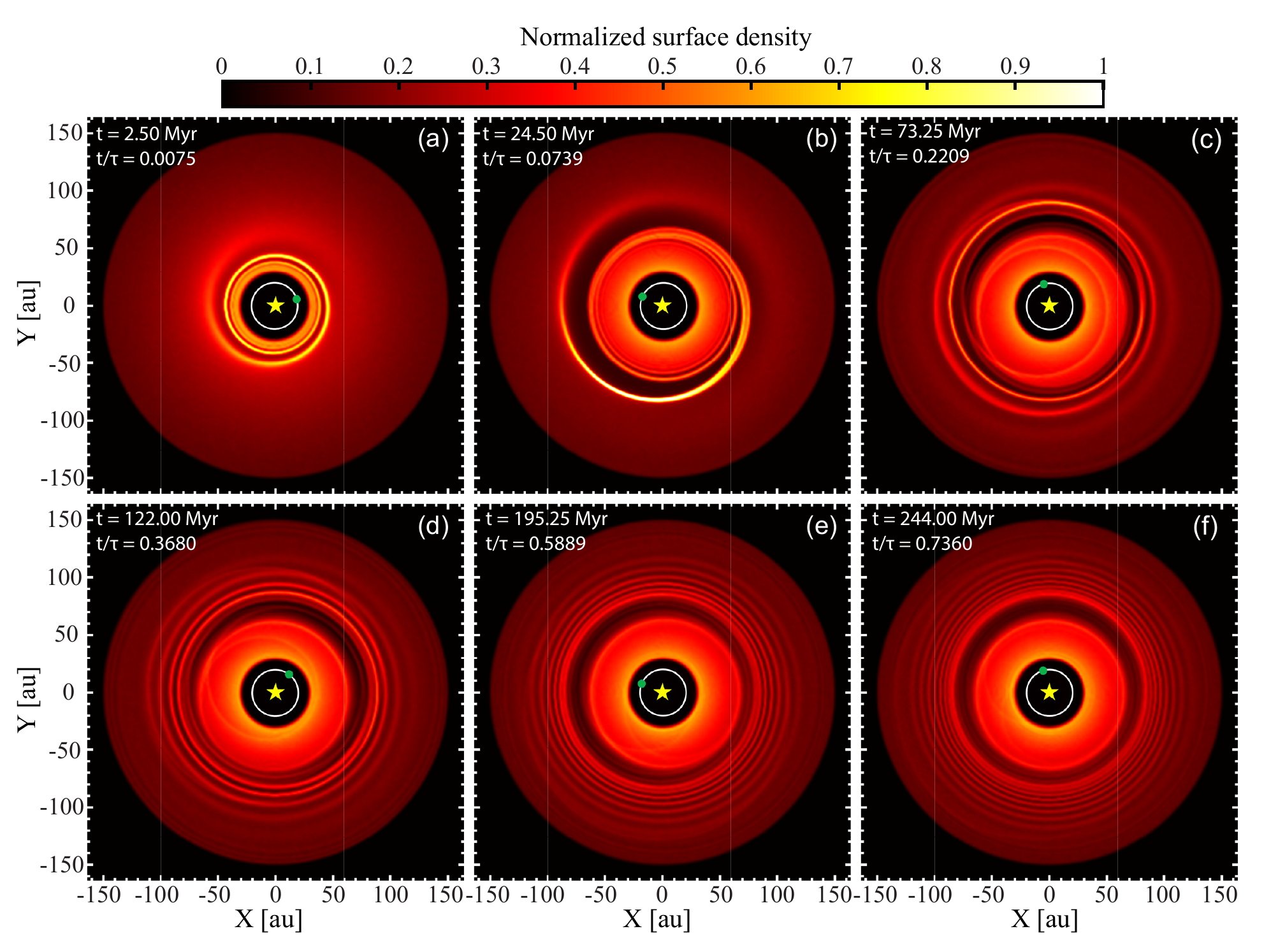}
\caption{Series of two-dimensional snapshots showing the evolution of the (normalized) disk surface density $\Sigma$ in the nominal $N$-ring simulation of the fiducial model (\texttt{Model A}; Table \ref{table:debris_Nring_sims_table}).
The surface density distributions $\Sigma(X,Y)$ are derived from the numerically evolved dynamical state of planetesimals displayed in Figure \ref{fig:e-w-fn-a-simA-p2}, following the same procedure as in Paper I (see Appendix C therein).
The snapshots correspond to the same moments of time $t$ as in Figure \ref{fig:e-w-fn-a-simA-p2}, as indicated in each panel for reference. The time is also indicated relative to $\tau \approx 332$ Myr (Section \ref{sec:typical_evolution_of_e_w_nominal_sim}).
All panels have $400\times 400$ pixels and share the same surface density scale, as well as normalization constant, as shown in the color bar.
In all panels, the stellar position is marked by the yellow star. The planet's orbit and its pericenter position -- which precesses with a period of $\tau_{\rm sec} \approx 57$ Myr (Figure \ref{fig:ep-wp-fn-time-simA-p2}) -- are shown by the white solid line and green circle, respectively. 
To enhance the resolution of images, the orbit of each planetesimal ($N = 5000$ in number) has been populated with $10^4$ particles with the same orbital elements but randomly distributed mean anomalies (see Appendic C of Paper I).
The evolution of the disk morphology occurs over three stages.
\textit{Stage 1} ($t \lesssim \tau_{\rm sec}$): at early times, a trailing spiral arm is launched at the inner disk edge (panel a) which then propagates outward in time as it wraps around the star (panel b).
\textit{Stage 2} ($t \sim \tau_{\rm sec}$): by the time the planet completes one precessional cycle (panel c), a non-axisymmetric gap is sculpted around the location of the secular resonance (i.e., at $a\res \approx 75$ au), which is both wider and deeper in the direction of the planetary pericenter. 
\textit{Stage 3} ($\tau_{\rm sec} \lesssim t \sim \tau$): at late times (panels d--f),  the gap maintains its crescent shape as it co-precesses with the planet's pericenter, slowly becoming more axisymmetric (as $e_p(t) \rightarrow 0$), and a tightly-wound spiral pattern develops beyond the gap. 
See the text (Section \ref{sec:typical_evolution_of_disc_morphology_p2}) for more details. 
This figure is available as an animation in the electronic edition of the journal, running from $t=0$ to $t = \tau$ with a duration of 33 s. 
}
\label{fig:map-simA-paper2}
\end{figure*}

The resulting two-dimensional maps of the (normalized) disk surface density $\Sigma$ corresponding to the same snapshot times as in Figure \ref{fig:e-w-fn-a-simA-p2} are shown in Figure \ref{fig:map-simA-paper2}. For reference, we also highlight in Figure \ref{fig:map-simA-paper2} the planetary orbit and its pericenter position which, in the model considered here (i.e., \texttt{Model A}, see Table \ref{table:debris_Nring_sims_table}), precesses with a period of $\tau_{\rm sec} \approx 57$ Myr; see e.g. Figure \ref{fig:ep-wp-fn-time-simA-p2}. To complement the interpretation of Figure \ref{fig:map-simA-paper2}, we also compute and plot in Figure \ref{fig:av-density-simA-paper2} the radial profiles of the azimuthally averaged disk surface density $\langle \Sigma(r) \rangle$ at times corresponding to those in Figure \ref{fig:map-simA-paper2}. We note that Figures \ref{fig:map-simA-paper2} and \ref{fig:av-density-simA-paper2} can be compared with Figures 8 and 9 of Paper I, respectively, where results of a similar exercise are presented for the same planet--disk model (\texttt{Model A}) but in the absence of the disk's non-axisymmetric perturbations on the planetary orbit.

A close look at Figures \ref{fig:map-simA-paper2} and \ref{fig:av-density-simA-paper2} reveals that the evolution of the disk morphology -- similar to Paper I -- is characterized by three distinct stages that occur on timescales measured relative to the planetary precession period $\tau_{\rm sec}$.
The disk morphology in the first two stages is by and large qualitatively similar to that corresponding to Paper I (see e.g. Section 5.1 therein). For completeness, however, we describe these stages in detail below, pointing out the relevant differences when compared to Paper I.

\subsubsection{Stage 1: \texorpdfstring{$0\leq t \lesssim \tau_{\rm sec}$}{0<t<tau sec}}

\begin{figure*}[t!]
\epsscale{1.15}
\plotone{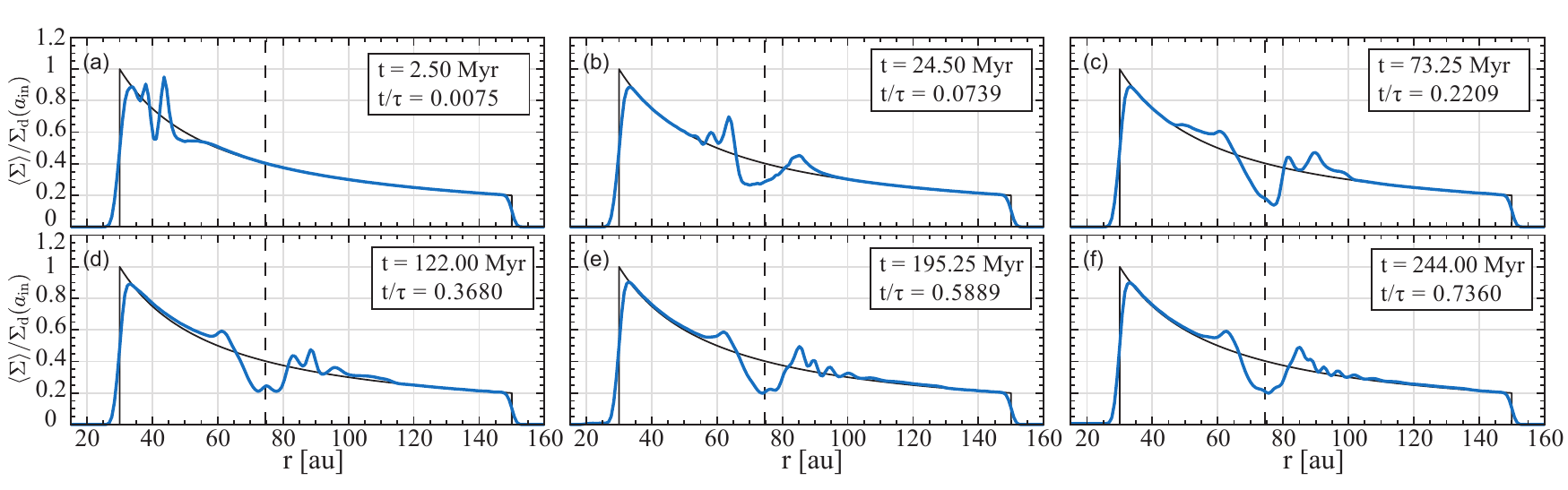}
\caption{The azimuthally averaged surface density of the disk $\langle \Sigma \rangle$ as a function of radial distance $r$ from the central star (solid blue lines).
Each panel corresponds to each of the snapshots of the fiducial configuration shown in Figure \ref{fig:map-simA-paper2}  (i.e., \texttt{Model A}; Table \ref{table:debris_Nring_sims_table}). 
The time $t$ of each snapshot is indicated in each panel, which, for reference, is also shown relative to $\tau \approx 332$ Myr.
The results are obtained by splitting the disk into $200$ annular bins (Appendix C of Paper I), and are all normalized with respect to the initial surface density $\Sigma_d(a)$ (i.e., Equation \ref{eq:Sigma_d} with $p=1$) evaluated at the inner disk edge, $a = a\inn$.  
For reference, the solid black lines show the normalized profile of the initial $\Sigma_d(a)$.
Note the appearance of a clear depletion in the surface density around the location of the secular resonance ($a\res \approx 75$ au, dashed vertical lines), which becomes evident by $\approx 73$ Myr (panel c). It is also evident that at late times (panels d--f), the width and depth of the depletion effectively remain constant in time, while the part exterior to the depletion develops a peak structure in the density profile. 
See the text (Section \ref{sec:typical_evolution_of_disc_morphology_p2}) for more details. 
This figure is available as an animation in the electronic edition of the journal, running from $t = 0$ to $t = \tau$ with a duration of $33$ s.}
\label{fig:av-density-simA-paper2}
\end{figure*}

At early times, i.e., $0\leq t \lesssim \tau_{\rm sec}$, the disk moves away from its axisymmetric initial condition and develops a trailing spiral wave (see Figs. \ref{fig:map-simA-paper2}(a), (b)). The spiral wave, which is initially launched at the inner edge of the disk $a\inn$, propagates radially outwards with time while wrapping completely around the star; see also the animated version of Figure \ref{fig:map-simA-paper2}.  As this happens, the spiral is more tightly wound closer to the planet. This is so much so that in the region interior to the  spiral -- which, for instance, extends out to $\sim$70 au by $t = 24.5$ Myr (Figure \ref{fig:map-simA-paper2}(b)) -- the windings become so difficult to discern that the surface density distribution appears to be roughly axisymmetric. Note that the outermost portion of the spiral moves through the disk at a slower rate as it extends to larger radii: this is simply because the planetesimal precession rate $A(a)$ is a decreasing function of semimajor axis (Figure \ref{fig:A-a-from-matrix-fid-simA-nominal}).
A complementary view of this behavior is provided by panels (a) and (b) of Figure \ref{fig:av-density-simA-paper2} and its animated version.

We note that the behavior described so far is similar to that of `Stage 1' in Paper I, despite the introduction of the disk's non-axisymmetric perturbations on the planet. This is to be expected since, while the latter indirectly affects the evolution of planetesimal orbits (Section \ref{sec:coupling-planet-planetesimal-p2}), the effects remain minor at  $t \lesssim \tau_{\rm sec}$, thus affecting the spatial appearance of the disk only in a subtle way.
For instance, despite the fact that the planetary eccentricity decays by about $20$ per cent from its initial value by $t \lesssim \tau_{\rm sec}$ (Figure \ref{fig:ep-wp-fn-time-simA-p2}), the maximum planetesimal eccentricities are still roughly similar to those in Paper I -- see Figures \ref{fig:e-w-fn-a-simA-p2} and \ref{fig:e-w-fn-time-simA-p2}. 
Relatedly, although planetesimal orbits interior to the spiral's outermost portion become phase-mixed over time such that $\Delta\varpi(a)$ spans the entire $[-\pi, \pi]$ range, the highest concentration remains within the same range as in Paper I, namely, $[-\pi/2, \pi/2]$ -- see Figures \ref{fig:e-w-fn-a-simA-p2}(a), (b). The main difference compared to Paper I, however, is that the outermost portion of the spiral\footnote{We remind the reader that the outermost portion of the spiral is associated with planetesimals that have completed half of their first precession cycle, and thus attained their maximum eccentricities (Figures \ref{fig:e-w-fn-a-simA-p2} and \ref{fig:e-w-fn-time-simA-p2}); see e.g. Sections 2 and 5 of Paper I for further details.} now extends out to about a radius of $\sim 80$ au rather than $\sim 70$ au; see and compare e.g. Figure \ref{fig:map-simA-paper2}(b) here and Figure 8(b) of Paper I.

\subsubsection{Stage 2: \texorpdfstring{$t \sim  \tau_{\rm sec}$}{t~tau sec}}
 
By the time  that the planet has completed approximately one precession cycle, i.e., $t \sim  \tau_{\rm sec}$,  the disk effectively splits into two parts separated by a clear gap in between; see e.g. Figures \ref{fig:map-simA-paper2}(c) and \ref{fig:av-density-simA-paper2}(c) and their animated versions. 
The gap forms in the disk around the location of the secular resonance which, for the system considered here (i.e., \texttt{Model A}), is established at $a\res = 75$ au; see Section \ref{sec:typical_evolution_of_e_w_nominal_sim} and Figure \ref{fig:e-w-fn-a-simA-p2}. Note that this is slightly larger than the case in Paper I, in which case the resonance instead occurs at $a\res = 70$ au. 

We note that the physical features of the gap at this stage are similar to those of `Stage 2' in Paper I. Namely, the gap is crescent-shaped pointing in the direction of the planetary pericenter (Figure \ref{fig:map-simA-paper2}(c)). Thus, both the gap's width and depth vary azimuthally, with the largest (smallest) width and depth  being toward the planetary pericenter (apocenter) -- see also Figure \ref{fig:av-density-simA-paper2}(c). Quantitatively speaking, we find that in an azimuthally averaged sense, the gap has a radial width of $w_g \sim 15$ au, when measured relative to the initial density profile (Figure \ref{fig:av-density-simA-paper2}(c)). This is slightly narrower than that in Paper I for the same model (i.e., \texttt{Model A}), where instead $w_g \sim 20$ au. The depth of the gap, however, is similar to that in Paper I, with about a half of the initial surface density being depleted at the secular resonance; see Figure \ref{fig:av-density-simA-paper2}(c).

Finally, we note that crescent-shape of the gap can be understood using the same reasoning as in Paper I. 
Indeed, by $t \sim \tau_{\rm sec}$, planetesimals interior to the secular resonance have completed at least one full precession cycle, thus settling into a coherent eccentric structure that is apsidally aligned with the planetary orbit and slightly offset relative to the star -- see Paper I for further details.

\subsubsection{Stage 3: \texorpdfstring{$\tau_{\rm sec} \lesssim t \leq \tau$}{tau sec<t<tau}}

At later times, i.e., $\tau_{\rm sec} \lesssim t \leq \tau$, the continued growth of the eccentricity around $a\res$ does not affect the structure of the gap significantly; see panels (d)--(f) of Figures \ref{fig:e-w-fn-a-simA-p2}, \ref{fig:map-simA-paper2} and \ref{fig:av-density-simA-paper2}. 
Indeed, the gap maintains its non-axisymmetric shape while all the time it co-precesses coherently with the planetary longitude of pericenter; see panels (d)--(f) in Figures \ref{fig:map-simA-paper2} and \ref{fig:av-density-simA-paper2}. 
As this happens, and similar to Paper I, the gap's width $w_g$ and depth $d_g$ remain practically invariant. To be specific, we find that in a time-averaged sense, $w_g \approx 15.45 \pm 0.21$ au and $d_g \approx 51 \pm 2 \%$. 
As in Paper I, these variations can be understood by the fact that the inner component of the disk precesses much faster than the outer component (Figure \ref{fig:e-w-fn-a-simA-p2}), causing the offset between them to vary in time.

In addition to this, the disk part exterior to the gap develops a spiral pattern which -- similar to the case in Stage 1 -- wraps onto itself as it propagates radially outwards. The windings are easier to discern at radii closer to the gap than to the outer disk edge, simply because the eccentricities get smaller as $a \rightarrow a\out$ (Figure \ref{fig:e-w-fn-a-simA-p2}). A complementary view of this is provided by Figures \ref{fig:av-density-simA-paper2}(d)--(f), where one can see the emergence of a series of narrow peaks at $r \gtrsim a\res$ in the radial profile of $\langle \Sigma \rangle$. Note that if the disk were evolved for longer, the spiral pattern would fade away once planetesimal orbits become fully phase-mixed -- i.e., with $\Delta\varpi(a)$ spanning the entire $[-\pi, \pi]$ range (Figure \ref{fig:e-w-fn-a-simA-p2}) --  rendering the surface density axisymmetric.

The behavior described thus far is by and large similar to that of `Stage 3' of Paper I. One key difference, however, is that further into the evolution, i.e., as $t \rightarrow \tau$, the gap starts to evolve towards an axisymmetric shape, in the sense that the depletion becomes visible $360^{\circ}$ around the star -- see e.g. panel (f) of Figure \ref{fig:map-simA-paper2} as well as its animated version.
The transition from asymmetry to symmetry is not perfect though, in the sense that one can still discern that the gap is both wider and deeper toward the planetary pericenter, but only to a relatively small degree.

To understand this, it is important to note that by $t \rightarrow \tau$, the planetary eccentricity would have undergone a significant decay relative to its initial value: namely, by as much as $\approx 75 \%$ in \texttt{Model A} -- see Figure \ref{fig:ep-wp-fn-time-simA-p2}(A). As described in Section \ref{sec:coupling-planet-planetesimal-p2}, this forces the maximum planetesimal eccentricities throughout the entire disk to decrease, as well as randomizes the phase angles of planetesimal orbits between $-\pi$ and $\pi$ both interior and exterior to the resonance location (unlike in Paper I; see Figures \ref{fig:e-w-fn-a-simA-p2} and \ref{fig:e-w-fn-time-simA-p2}).
As a result, the disk parts both interior and exterior to the gap become less offset relative to the star individually, and thus, in combination, decrease the asymmetry of the gap in between. This is easier to see in the region interior to the gap, where the eccentricities are naturally larger than in the outer parts. 
Finally, it is noteworthy to mention that the discussed effects of the circularizing planet cannot be reproduced by a planet of constant but smaller initial eccentricity; see e.g. Figure \ref{fig:map-simA-paper2}(f) here and Figure 11 in Paper I which shows results for \texttt{Model A} but with $e_p(0) = 0.025$.

\subsection{Comment on Generality of Results}
\label{sec:generality_comment_results}

We conclude this section by a comment on the generality of the results presented thus far.
As already mentioned at the start of this section, results presented for \texttt{Model A} are generic, in the sense that qualitatively similar behavior is observed in all other simulated planet--disk systems, despite the broad range of parameters explored (Table \ref{table:debris_Nring_sims_table}). 
Quantitatively speaking, on the other hand, results will differ from one system to another depending on the specific parameters of the planet and the disk. 
Nevertheless, several scaling rules can be applied to explain the differences, some of which were already identified and discussed in great detail in Paper I (Section 5.2 therein). We briefly discuss these below.

First, and as in Paper I, varying both the planet and disk masses simultaneously, i.e., in such a way that $M_d/m_p$ remains constant, affects only the secular evolution timescales, but not the details of the ensuing dynamics.
Thus, the very same dynamical end-states will be achieved within e.g. a shorter timescale when both the disk and planet masses are increased, and vice versa -- see also Section 5.2.3 of Paper I.

Second, and as in Paper I, increasing the initial planetary eccentricity causes three effects: 
(i) it decreases the secular evolution timescales (e.g., the timescale $\tau$, see also Section \ref{sec:timescales_paper2_simulations}); 
(ii) it renders the transient spirals, both in the inner and outer disk parts, more open and prominent; 
and (iii) it leads to the sculpting of a wider gap around $a\res$.
The opposite holds true for initially less eccentric planets; see also Section 5.2.2 of Paper I.

Last but not least, and similar to Paper I, the gap that is sculpted at a given resonance location $a\res$ is wider when  the planet is initiated with a semimajor closer to the inner disk edge than to the star, and vice versa -- see also Section 5.2.1 of Paper I.
In addition to this, with increasing planetary semimajor axis, i.e., as $a_p/a\inn \rightarrow 1$, we  find that (i) the planetary orbit precesses at a slower rate than expected from Paper I; (ii) its orbit circularizes at a faster rate compared to the case of $a_p/a\inn \rightarrow 0$; and (iii) the outward shift in the resonance location relative to that expected from Paper I becomes larger. 
The latter three effects and their corresponding scalings, which are of course absent in Paper I, will be characterized next in Section \ref{sec:analysis_predictions}.

Finally, and as in Paper I, we find that the gap depth is not significantly affected by variations in planet--disk parameters: indeed, once the gap is sculpted, we find that $d_g \approx 50\%$ upon time-averaging in all considered systems (Table \ref{table:debris_Nring_sims_table}).

\section{Analysis and Predictions}
\label{sec:analysis_predictions}

As pointed out in the previous section,  there are several qualitative and quantitative differences between the results of our present work and our previous work in Paper I in terms of the evolution of both the planet and the debris disk.
We now characterize these differences in greater detail, focusing first on the behavior of the planet's orbit (Sections \ref{sec:effect_on_wpdot} and \ref{sec:resonant_friction}) and then on the characteristics of the secular resonance (Sections \ref{sec:effect_on_ares_shift}--\ref{sec:reswidth_analysis_section}). 
Wherever possible, we also develop quantitative explanations and predictions for the observed differences using dynamical theory.

In what follows, and for  ease of interpretation and comparability with Paper I,  we also supplement our results with those obtained from what we refer to as `simplified, Paper I-like' $N$-ring simulations, in which we only account for the axisymmetric component of the disk gravity, switching off the non-axisymmetric component (Section \ref{sec:tests-of-Nring-model}).
This is done considering the same $67$ planet--disk models listed in Table \ref{table:debris_Nring_sims_table}.  We remind the reader that such simplified simulations are expected to  accurately reproduce the analytical results of Paper I (Appendix \ref{sec:p2-vs-p1-section}).

\subsection{Precession Rate of the Planetary Orbit}
\label{sec:effect_on_wpdot}

Numerical results of Section \ref{sec:results_p2} indicate that in the same planet--disk model (i.e., \texttt{Model A}, Table \ref{table:debris_Nring_sims_table}), the planetary orbit undergoes prograde precession  at a constant rate that is \textit{slower} than expected from Paper I, i.e., $\dot{\varpi}_p \lesssim A_{d,p}$ (see e.g. Figure \ref{fig:ep-wp-fn-time-simA-p2}(B)). We now analyze this behavior in more detail  and provide a quantitative explanation for it.

\begin{figure}[t!]
\epsscale{1.15}
\plotone{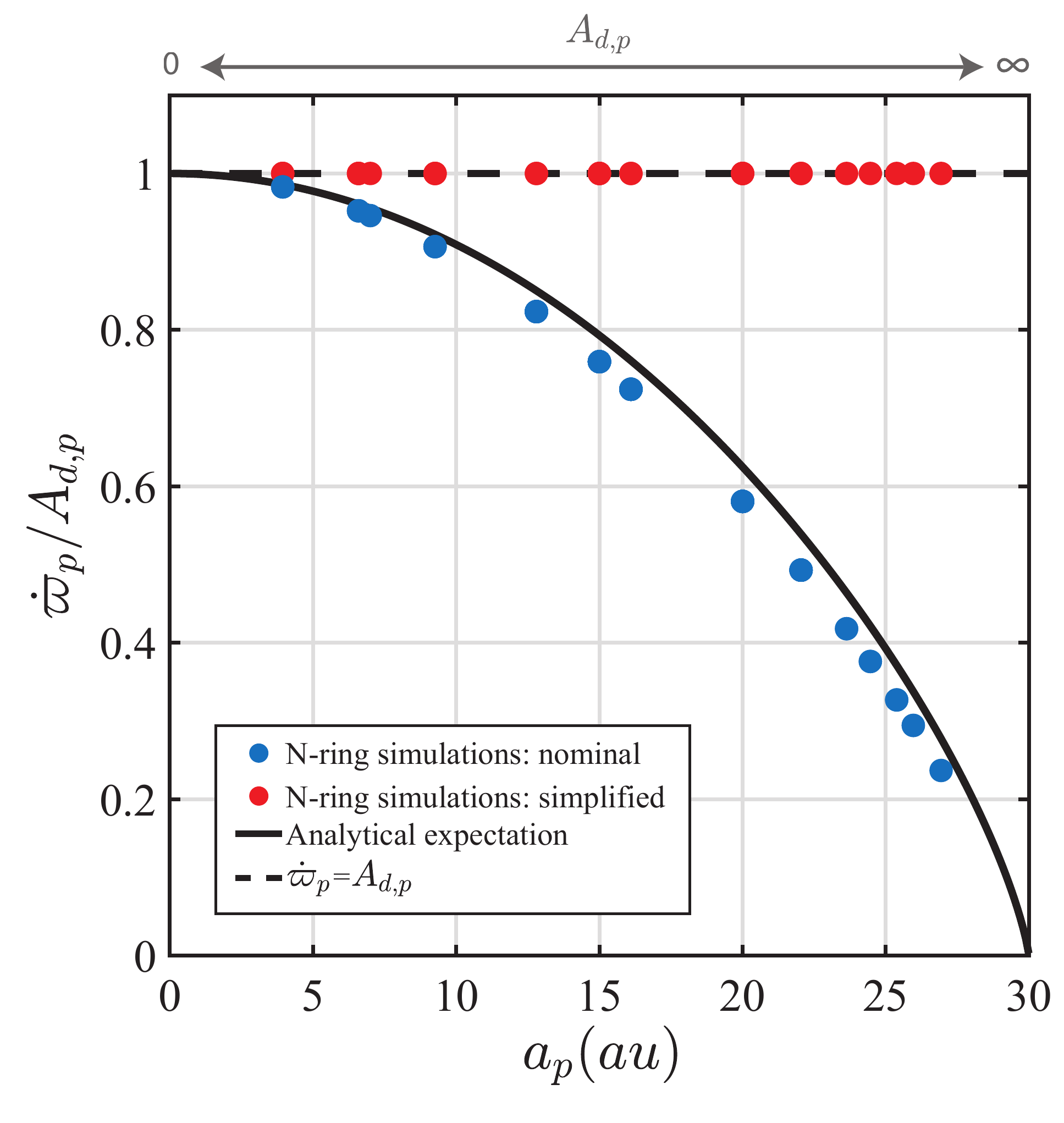}
\caption{The apsidal precession rate of the planetary orbit in each $N$-ring simulation, scaled relative to the theoretical free precession rate (Equation (8,PI)), i.e., $\dot{\varpi}_p/A_{d,p}$, as a function of planetary semimajor axis $a_p$. Results are shown for all planet--disk models listed in Table \ref{table:debris_Nring_sims_table}, each evolved within two sets of simulations: nominal (shown in blue circles) and simplified (i.e., Paper I-like, shown in red circles). Note that results for systems with the same $a_p$ but different combinations of $m_p, M_d$ and $e_p(0)$ (Table \ref{table:debris_Nring_sims_table}) overlap with each other almost perfectly. The solid  black line represents the analytical prediction based on Equation (\ref{eq:wpdot_mod_theory}), which accounts for both axi- and non-axisymmetric perturbations of the disk on the planet (as in the nominal simulations) under two simplifying assumptions about the disk: that its eccentricity is proportional to $1/a$ and its precession is dominated by the planet's gravity within the entire disk (Appendix \ref{app:wpdot_Bdp_derivation}). The dashed black line represents a one-to-one correlation between $\dot{\varpi}_p$ and $A_{d,p}$, as is expected in the simplified simulations.  See the text (Section \ref{sec:effect_on_wpdot}) for more details. 
}
\label{fig:wpdot_fn_ap_P2}
\end{figure}

To this end, we first measure the planetary precession rates in each of the $67$ simulations within both sets of the $N$-ring simulations that were carried out, i.e., nominal and simplified (Table \ref{table:debris_Nring_sims_table}). We do this by simply fitting a straight line to the simulated curves of $\varpi_p(t)$ and measuring the corresponding value of the slope\footnote{In all simulations, a correlation coefficient of $\approx 1$ is obtained, indicating that the planetary apsidal angle advances linearly with time.}.
Additionally, since precession rates scale linearly with masses (at least, in Laplace-Lagrange theory), we normalize the measured values of $\dot{\varpi}_p$ by the theoretical values of $A_{d,p}$  corresponding to each simulated system's  parameters (\EQPP{8}). This essentially should render the results, i.e., $\dot{\varpi}_p/A_{d,p}$, dependent only on the planetary semimajor axis $a_p$ -- except of course if there is a dependency on a non-accounted parameter, namely, the planetary or planetesimal eccentricities.  The results of this exercise are displayed in Figure \ref{fig:wpdot_fn_ap_P2} for both sets of simulations: nominal (blue circles) and simplified (red circles).

There are several notable features in Figure \ref{fig:wpdot_fn_ap_P2}. 
First, it is evident that in all of the simplified simulations,  the planetary orbit precesses at the theoretical rate of $A_{d,p} >0$ as expected from Paper I so that $\dot{\varpi}_p/A_{d,p} \approx 1$, regardless of the system parameters (Table \ref{table:debris_Nring_sims_table}). 
Note that this further confirms the validity of the $N$-ring model outlined in Section \ref{sec:problemframework_P2}, in addition to the tests presented in Appendix \ref{app:tests_Nring}. 
Second, looking at Figure \ref{fig:wpdot_fn_ap_P2}, one can see that for a given planetary semimajor axis, the planetary precession rate  in the nominal simulations is generally smaller than the expectation from Paper I, i.e., $\dot{\varpi}_p(a_p)/A_{d,p}(a_p) \lesssim  1$. 
It is also clear  that the differences between the two sets of simulations become more pronounced for planets orbiting closer to the disk inner edge than to the star, i.e., as $a_p/a\inn \rightarrow 1$. 
Indeed, Figure \ref{fig:wpdot_fn_ap_P2} shows that as the planetary semimajor axis is increased from $a_p/a\inn \approx 0.1$ to $a_p/a\inn \approx 1$ (with $a\inn = 30$ au), the differences grow from a factor of roughly $1$ to a factor of $\approx 5$. 
Another important feature in Figure \ref{fig:wpdot_fn_ap_P2} is that similar to the simplified simulations,  the nominal simulations reveal little or no evidence of scatter in the results at any given value of $a_p$, despite the different initial conditions  (Table \ref{table:debris_Nring_sims_table}). While trivial, what this means is that the effect of the disk's non-axisymmetric potential on the planetary precession rate is  independent of the disk's eccentricity (which, recall, is imposed by the planet's eccentricity; Section \ref{sec:results_p2}).

In order to better understand the results of Figure \ref{fig:wpdot_fn_ap_P2}, it is important to distinguish between the notions of \textit{free} and \textit{forced} precession rates \citep[e.g.,][]{mur99}.
According to Paper I, the planetary orbit precesses at a rate given by $\dot{\varpi}_p = A_{d,p}$; see \EQPP{8}. Strictly speaking, however, this is the \textit{free} precession rate, i.e., the rate at which the planet precesses if the disc potential were axisymmetric -- which is what was assumed in Paper I. In reality, however, there is also a contribution to the planet's precession rate due to the disk eccentricity, which manifests itself as a non-axisymmetric contribution to the disk potential (ignored in Paper I).
This is the \textit{forced} precession rate,  corresponding to the term $i \sum_{l \neq 0}^{N} A_{0l} \zeta_l$ in Equation (\ref{eq:SimEqn}) with $j = 0$. 
In principle then, it is the combination of the free and forced contributions that dictates the total planetary precession rate.

Based on the above argument, in Appendix \ref{app:wpdot_Bdp_derivation} we derive a general analytical expression for the total planetary precession rate  accounting for both the free and the forced components  induced by the disk gravity. 
We find that within a set of reasonable simplifying assumptions, the planetary precession rate $\dot{\varpi}_p$, when scaled by $A_{d,p}$, can be written as follows:
\begin{equation}
\frac{\dot{\varpi}_p}{A_{d,p}} = 1 + \frac{1}{2} \frac{\phi_2}{\phi_1} \frac{b_{3/2}^{(2)}(a_p/a_{\rm in})}{ b_{3/2}^{(1)}(a_p/a_{\rm in})}  
\approx  1 - \frac{13}{16} 
\left(\frac{a_p}{a_{\rm in}} \right)^2 , 
\label{eq:wpdot_mod_theory}
\end{equation}
where the terms $\phi_1>0$  and $\phi_2<0$  govern the strengths of the axisymmetric and non-axisymmetric effects of the disk on the planet, respectively -- see e.g. equations (A5)--(A8) of Paper I. In Equation (\ref{eq:wpdot_mod_theory}), the approximation on the right-hand side is obtained for our fiducial disk model ($p=1$, $\delta=5$) in the limit of $a_p/a\inn \rightarrow 0$, assuming that the disk eccentricity scales as $1/a$; see Appendix \ref{app:wpdot_Bdp_derivation} for details.

Equation (\ref{eq:wpdot_mod_theory}) captures many of the salient features evident in Figure \ref{fig:wpdot_fn_ap_P2}.
First, Equation (\ref{eq:wpdot_mod_theory}) shows that the planetary precession rate is directly proportional to the coefficient $\phi_2$, which is a proxy for the strength of the disk's non-axisymmetric torque on the planet. 
This provides a trivial explanation as to why $\dot{\varpi}_p/ A_{d,p} = 1$ in the simplified, `Paper I-like' $N$-ring simulations, which, by construction, have $\phi_2 = 0$. 
Second, given that generally $\phi_1>0$ and $\phi_2<0$, Equation (\ref{eq:wpdot_mod_theory}) indicates that the disk's non-axisymmetric potential opposes the planetary precession induced by its axisymmetric counterpart. This explains  why the planet precesses at a slower rate in the nominal $N$-ring simulations than in the Paper I-like simulations. 
Note that the correction due to the disk non-axisymmetry depends only on the square of the ratio $a_p/a\inn$, explaining why the differences between the nominal and simplified simulations grow as $a_p \rightarrow a_{\rm in}$.
Third, and more importantly,  Equation (\ref{eq:wpdot_mod_theory}) approximates the results of the nominal $N$-ring simulations very well, even for relatively large values of $a_p$; see the black line in Figure \ref{fig:wpdot_fn_ap_P2}. 
Indeed, the discrepancies between Equation (\ref{eq:wpdot_mod_theory}) and the simulation results are remarkably negligible, despite the oversimplifications regarding e.g. the disk's eccentricity and precession that go into deriving Equation (\ref{eq:wpdot_mod_theory}), namely, that $e_d \propto 1/a$ and $A(a) \approx A_p(a)$ throughout the entire disk (i.e., $a\inn \leq a \leq a\out$); see Appendix \ref{app:wpdot_Bdp_derivation} for further details.

\subsection{Decay of Planetary Eccentricity and Resonant Friction}
\label{sec:resonant_friction}

We now turn to characterizing the behavior of the planetary eccentricity which, as described in Section \ref{sec:results_p2}, exhibits a long-term exponential decline in the nominal $N$-ring simulation of \texttt{Model A}, rather than remaining constant as in Paper I (see Figure \ref{fig:ep-wp-fn-time-simA-p2}).  Our specific aims here are two-fold: to demonstrate that the decay is a generic phenomenon, and that it ensues from a process known as ``resonant friction''. First, however, a brief review of this process is in order.

Dynamical friction is a well-studied process in astrophysics, whereby the gravitational interactions between a massive body (e.g. a planet) and a collection of  lighter objects (e.g. planetesimals) give rise to a net force acting on the former in a way that imitates friction \citep{BT}. \textit{Resonant friction}, sometimes called ``secular resonant damping'',  is a special case of this process that stems from secular, orbit-averaged interactions: namely, due to the coupling of a planet and planetesimals at and around the location of a secular resonance \citep[][]{tre98, wardhahn1998, wardhahnprotostars,  hahnbending, hahn2008}. 
Indeed, as the eccentricities of planetesimals  at and around the resonance are excited and $e(a\res) \rightarrow 1$, they exert a strong torque on the planet. This is because the apsidal angles of planetesimals are shifted to $\Delta\varpi(a\res) = -\pi/2$ (see Figure \ref{fig:e-w-fn-a-simA-p2}), which means that in a frame corotating with the planet, there is a time-steady torque from the planetesimals at the resonance. This torque leads to the redistribution of the system's angular  momentum, without affecting its total budget (Appendix \ref{sec:L_conservation_P2}). In particular, the torque transports angular momentum from the resonant planetesimals to the planet, forcing the planet's eccentricity to damp, while its semimajor axis remains unaffected.
The rate at which this happens is given by \citep[see e.g. equation (20) in][]{tre98}:\footnote{A detailed derivation of Equation (\ref{eq:D_Tr98_MAIN}) can be found in \citet{tre98}, \citet{wardhahn1998}, and \citet{wardhahnprotostars}. For our purposes here, however, we point out that it is derived by neglecting the non-axisymmetric perturbations exerted by the planetesimals among themselves; similar to the assumptions adopted in our nominal $N$-ring simulations.} 
\begin{eqnarray}
\frac{1}{e_p^2} \frac{de_p^2}{dt} 
&=& -D \equiv 
- \frac{\pi^2}{4} 
\frac{m_p}{M_c} 
\frac{\Sigma_d^{t=0}(a) a^2}{ M_c } 
\frac{ n_p  n(a) \alpha^3  }{|dA/d\log a|} 
\left[b_{3/2}^{(2)}(\alpha)\right]^2 
\bigg|_{a_{\rm res}} , 
 \label{eq:D_Tr98_MAIN}
\end{eqnarray}
where $\alpha \equiv a_p/a$,  and, as before, $A(a) = A_d(a) + A_p(a)$, and $n_p \approx \sqrt{G M_c / a_p^3}$ and $n \approx \sqrt{G M_c / a^3}$ are the mean motion of the planetary and planetesimal orbits, respectively. Note that all quantities defining the rate $D > 0$ are evaluated at the secular resonance $a\res$, and so $D = 0$ (and $e_p(t) = {\rm const}$) when e.g. the disk's mass is ignored and there is no secular resonance. 
The solution of Equation (\ref{eq:D_Tr98_MAIN})
is a simple exponential function of time, so that $e_p(t)$ decays
\begin{equation}
e_p(t) = e_p(0) ~ \mathrm{exp}\left(  - D t /2 \right) ,
\label{eq:ep_decay}
\end{equation}
with a characteristic half-life of $\tau_D \equiv 2 \log 2 /D$.

\begin{figure}[t!]
\epsscale{1.15}
\plotone{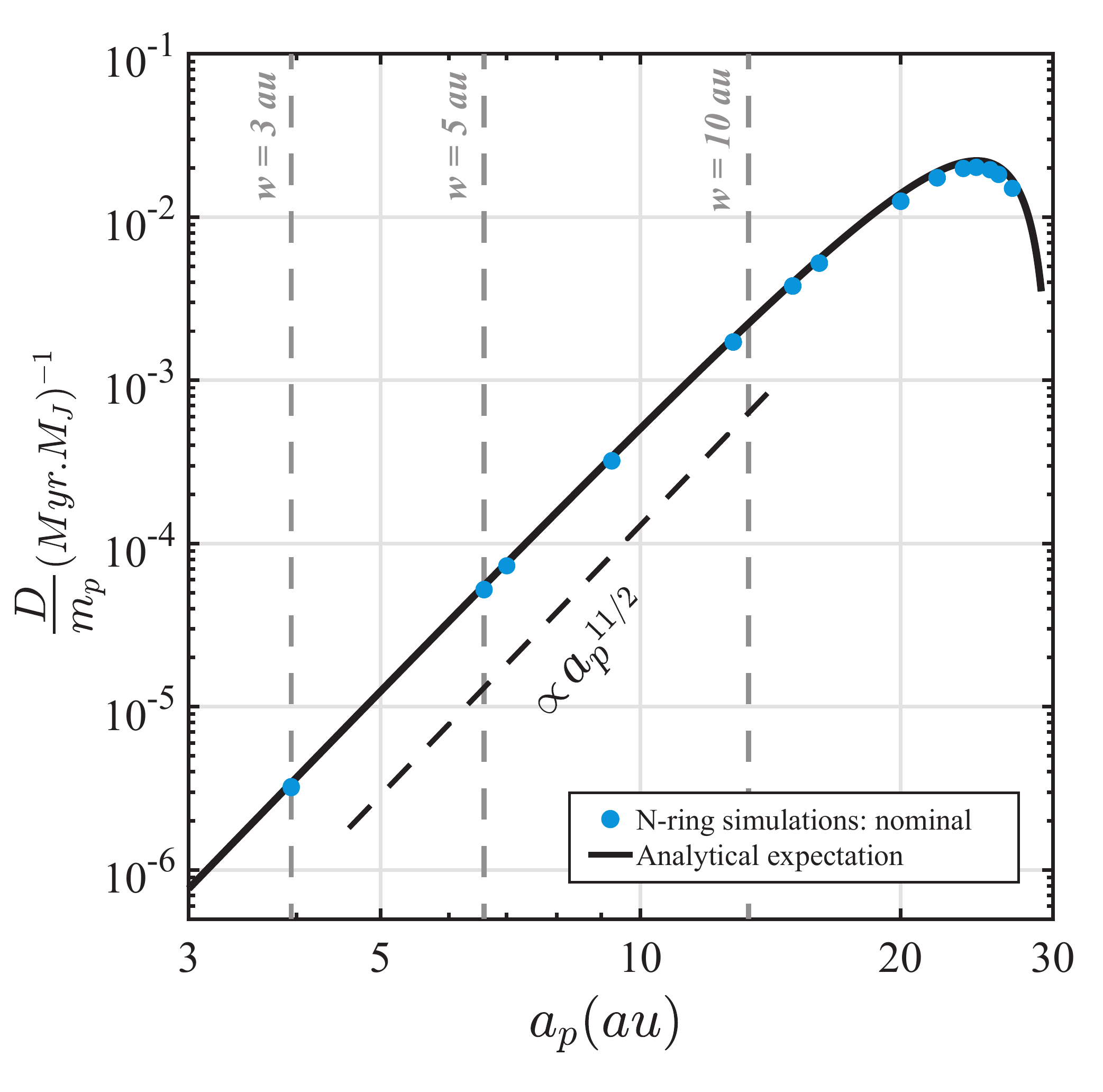}
\caption{The exponential decay rates of planetary eccentricity per unit mass of the planet $D/m_p$ as a function of its semimajor axis $a_p$, measured from the nominal $N$-ring simulations of $67$ different planet--disk models (Table \ref{table:debris_Nring_sims_table}; blue filled circles).
Note that results for systems sharing the same $a_p$ but different $m_p$, $M_d$, and $e_p(0)$ (Table \ref{table:debris_Nring_sims_table}) coincide with each other almost perfectly.
The solid black line represents the theoretical expectation based on Equation (\ref{eq:D_Tr98_MAIN}), evaluated at the system parameters of interest. 
The dashed black line shows the scaling of $D/m_p$ with $a_p$ (Equation \ref{eq:tau_D_p2}). 
The gray dashed lines mark the planetary semimajor axes, for which, based on Paper I, the expected resonance widths are $w = 3$, $5$, and $10$ au (\EQPP{22}).
See the text (Section \ref{sec:resonant_friction}) for more details.
}
\label{fig:D_fn_ap_P2}
\end{figure}

We now show that the exponential damping of planetary eccentricities observed in our nominal $N$-ring simulations is indeed due to resonant friction. 
To this end, we compare the decay rates as measured from our $67$ different simulations (Table \ref{table:debris_Nring_sims_table}) with those expected from Equation (\ref{eq:D_Tr98_MAIN}).
The result of such an exercise is summarized in Figure \ref{fig:D_fn_ap_P2}, where we plot the decay rates per unit planetary mass $D/m_p$ as a function of planetary semimajor axis $a_p$.
We note that in making this figure, results corresponding to our simulations (shown in blue circles) were obtained  by fitting the simulated curves of $e_p(t)$ with the functional form of Equation (\ref{eq:ep_decay}), while keeping both $e_p(0)$ and $D$ as free parameters.\footnote{In each case, the fitting procedure yielded an output $e_p(0)$ that differs only by less than a per cent from the input value for the simulations (Table \ref{table:debris_Nring_sims_table}).}
On the other hand, the theoretical expectations (shown by a solid black curve) were obtained by evaluating Equation (\ref{eq:D_Tr98_MAIN}) at the system parameters of interest (Table \ref{table:debris_Nring_sims_table}), i.e., the combinations of $M_d/m_p$ and $a_p/a\inn$ which, according to Paper I, place a resonance at $a\res = 70$ au (\EQPP{19}).
For a meaningful comparison, we evaluated Equation (\ref{eq:D_Tr98_MAIN}) at the resonance locations that are theoretically expected in the presence of the disk's non-axisymmetric torque on the planet; these are generally larger than the expectations from Paper I as shown in Figure \ref{fig:ares_fn_ap_P2} and further discussed next in Section \ref{sec:effect_on_ares_shift}.

Looking at Figure \ref{fig:D_fn_ap_P2}, one can clearly see that the simulation results agree remarkably well with the theoretical expectations for all considered values of $a_p$, reproducing all the expected salient features. To begin with, one can see that for a given planet semimajor axis,  both theoretical and simulated decay rates scale linearly with the planetary mass, i.e., $D/m_p = {\rm const}$ (recall that some of the simulations share the same $a_p$ but have different $m_p$; Table \ref{table:debris_Nring_sims_table}).  
Additionally, it is also clear that the decay rates become larger with increasing planetary semimajor axis such that $D/m_p \propto a_p^{11/2}$; see the dashed black line in Figure \ref{fig:D_fn_ap_P2}. 
A closer look at Figure \ref{fig:D_fn_ap_P2} also reveals that the agreement between the simulated and theoretical rates is perfect at all values of $a_p$, even when $a_p \rightarrow a\inn$.
This is outstanding, considering the fact that the derivation of Equation (\ref{eq:D_Tr98_MAIN}) assumes that as $t \rightarrow \infty$ and $e(a\res) \rightarrow 1$,  only  the  planetesimals that are \textit{exactly} at the resonance contribute to resonant friction, but not those around it (i.e., $a \approx a\res$). For reference, the values of $a_p$ for which Paper I predicts resonance widths of $3$, $5$, and $10$ au, respectively, are shown in Figure \ref{fig:D_fn_ap_P2}; see the vertical dashed lines (recall from Paper I that $w\propto a_p$; \EQPP{22}).
In summary, Figure \ref{fig:D_fn_ap_P2} provides strong evidence that the decay of planetary eccentricities in the nominal $N$-ring simulations is due to resonant friction.
Note that this process is not observed in Paper I, simply because there the disk's potential is taken to be axisymmetric, shutting off the disk-induced torque on the planet \citep[see also][]{tre98}.

We note that the scalings of $D$  may be understood qualitatively as follows. To begin with, we first recall that simulations with a fixed value of $a_p$ have different input masses $m_p$ and $M_d$ but the same ratio $M_d/m_p$ (to ensure the resonance location remains the same; Paper I). Thus, variations with disk and planet masses (while $M_d/m_p = {\rm const}$) affect only the interaction timescales, but not the nature of the interaction itself.
This explains  why $D \propto m_p$ for a given planetary semimajor axis. 
The scaling $D/m_p \propto a_p^{11/2}$, on the other hand, can be understood by noting that the torque exerted by the disk on the planet, and thus the effect of resonant friction, would be stronger  when the planet and the resonance location are closer together, i.e., when $a_p/a\res \rightarrow 1$. 

These dependencies can also be derived  by considering the approximate form of the decay rate given by Equation (\ref{eq:D_Tr98_MAIN}). Indeed, inserting the condition for secular resonance, i.e. \EQPP{19} or (\ref{eq:resonance_condition_p2}), with $p=1$ into Equation (\ref{eq:D_Tr98_MAIN}),  and taking the limits of $a_p/a\inn \rightarrow 0$ and $a\inn \ll a\res \ll a\out$ (so one can use the asymptotic behaviors of $b_{3/2}^{(m)}(\alpha)$ and  $dA/da$, see \EQPP{B4}), it is straightforward to show that:
\begin{equation}
\tau_D \equiv \frac{2 \log 2}{D} \approx 55 ~ \textrm{Myr} 
\left(\frac{1 M_J}{m_p} \right)
M_{c,1.09}^{1/2} ~  a_{p, 20}^{-11/2} ~ a_{res, 70}^{15/2} , 
 \label{eq:tau_D_p2}
\end{equation}
where the numerical coefficient is obtained for the fiducial disk model (i.e., $p = 1, a\inn = 30$ au, and $a\out = 150$ au), and we have defined $a_{p, 20} \equiv a_p/(20 ~ {\rm au})$, $a_{\res, 70} \equiv a_{\res}/(70 ~ {\rm au})$, and $M_{c,1.09} \equiv M_c/(1.09 M_{\odot})$.

We note that Equation (\ref{eq:tau_D_p2})  can alternatively be expressed in terms of the disk mass. Indeed, using the approximate resonance condition given in the second line of \EQPP{19}, namely,
\begin{equation}
    \frac{M_d}{m_p} \approx 0.15 ~ a_{p, 20}^2 ~  a_{\res, 70}^{-3.5}   , 
    \label{eq:Mdmp_PI_19}
\end{equation}
it is trivial to find that
\begin{equation}
\tau_D \approx 55 ~ \textrm{Myr} 
\left(\frac{48 M_{\earth}}{M_d} \right) 
M_{c,1.09}^{1/2} ~  a_{p, 20}^{-7/2} ~ a_{res, 70}^{4} ,
\label{eq:tau_D_p2_diskmassMd2}
\end{equation}
or, alternatively,
\begin{equation}
\tau_D \approx 55 ~ \textrm{Myr} 
\left(\frac{48 M_{\earth}}{M_d} \right)^{15/7} \left(\frac{m_p}{1 M_J} \right)^{8/7}
M_{c,1.09}^{1/2} ~  a_{p, 20}^{-17/14}  .
\label{eq:tau_D_p2_diskmassMd}
\end{equation}
Equations (\ref{eq:tau_D_p2})--(\ref{eq:tau_D_p2_diskmassMd}) can be used to obtain a crude approximation of the decay rate of $e_p$ in a given planet--disk system, whether hypothetical or observed, without running any simulations provided the resonance occurs within the disk (i.e., $a\inn \leq a\res \leq a\out$). This is illustrated as an example in Figure \ref{fig:tauD_mp_ap_nf} for the fiducial disk model; see Section \ref{sec:disk_gravity_importance} for further details.

\subsection{Location of Secular Resonances}
\label{sec:effect_on_ares_shift}

Section \ref{sec:results_p2} shows that for the same planet--disk model (i.e., \texttt{Model A}), the secular resonance occurs at 
a semimajor axis that is larger in the nominal $N$-ring simulation than in Paper I -- namely, $a\res \approx 75$ au compared to $a\res = 70$ au. We now show that this behavior is generic across our suite of nominal simulations (Table \ref{table:debris_Nring_sims_table}), and interpret it as a consequence of the reduced apsidal precession rate of the planet due to the disk's non-axisymmetric torque on it  (Section \ref{sec:effect_on_wpdot}). 

\begin{figure}[t!]
\epsscale{1.15}
\plotone{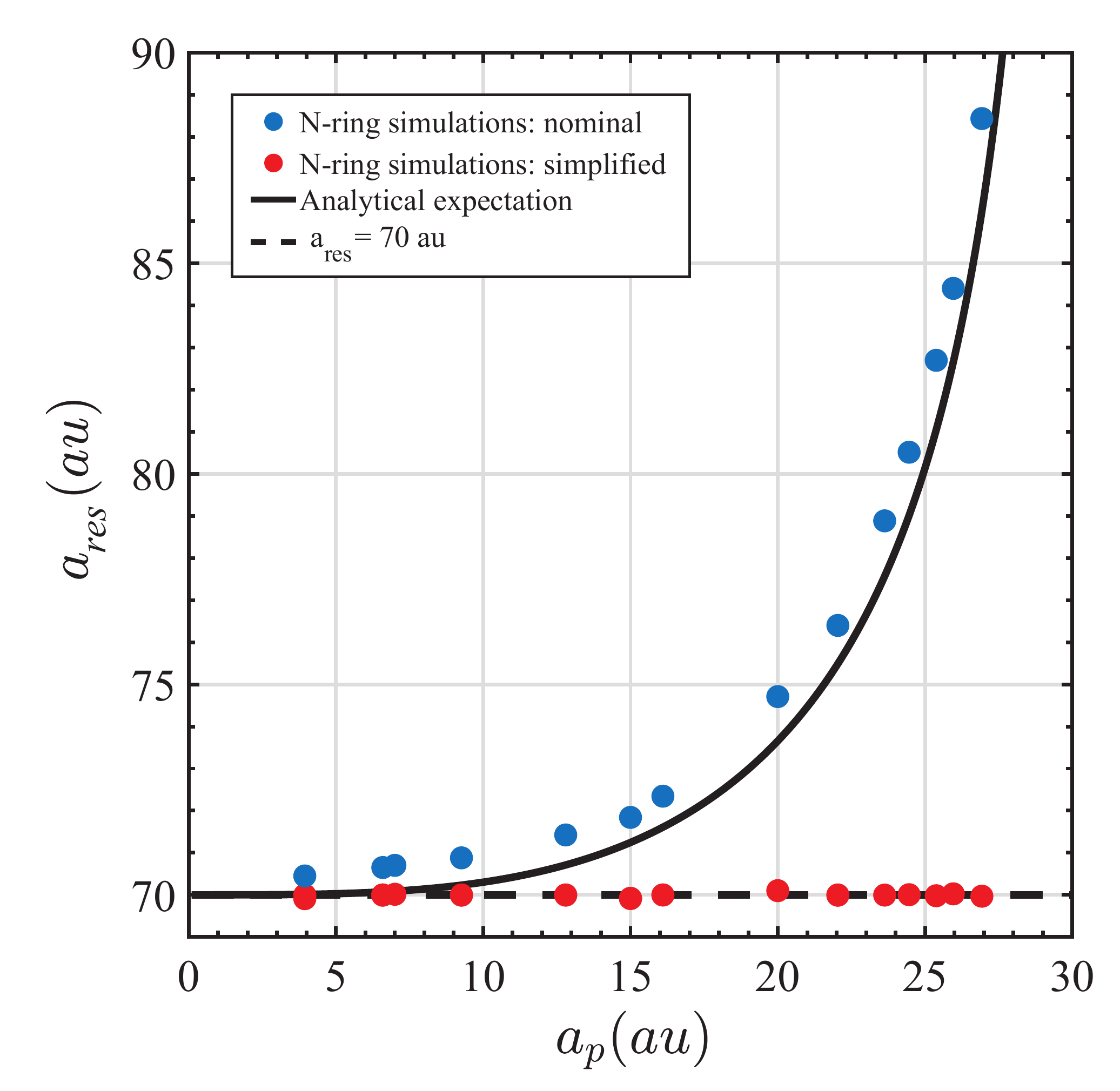}
\caption{The location of the secular resonance $a\res$  in each $N$-ring simulation, plotted as a function of the planetary semimajor axis $a_p$.
Simulations are done for $67$ different planet--disk models (Table \ref{table:debris_Nring_sims_table}), each within two sets:  nominal (blue circles) and simplified (i.e., Paper I-like; red circles). 
Note the overlapping values of $a\res$ for systems sharing the same $a_p$, but differing in $m_p$, $M_d$, and $e_p(0)$ (Table \ref{table:debris_Nring_sims_table}).
One can see that results of simplified simulations lie along the dashed black line: this represents the resonance location expected from Paper I, $a\res = 70$ au. 
It is also evident that results of nominal simulations follow closely the solid black curve: this represents the theoretical expectation based on Equation (\ref{eq:resonance_condition_p2}) that accounts for the reduced planetary precession rate due to the disk's non-axisymmetric torque (Equation \ref{eq:wpdot_mod_theory}).
Clearly, the disk's non-axisymmetric torque on the planet shifts the location of $a\res$ outwards when compared to Paper I. 
See the text (Section \ref{sec:effect_on_ares_shift}) for more details. 
}
\label{fig:ares_fn_ap_P2}
\end{figure}

As already mentioned before,  secular apsidal resonances occur at semimajor axes $a = a\res$ where the precession rate of planetesimal free eccentricities is exactly equal to the apsidal precession rate of the planet, i.e.,
\begin{equation}
    A_d(a\res) + A_p(a\res) = \dot{\varpi}_p ,
    \label{eq:resonance_condition_p2}
\end{equation}
see also \EQPP{17}.
In Paper I, given that the disk's potential was taken to be axisymmetric, $\dot{\varpi}_p$ represented the free precession rate of the planet, i.e., $\dot{\varpi}_p = A_{d,p}$ (see e.g. Section 2.2.2 of Paper I). 
In the nominal $N$-ring simulations, however, this is no longer true since the  non-axisymmetric component of the disk's gravity causes the planet to precess at a slower rate so that $\dot{\varpi}_p \lesssim A_{d,p}$; see Section \ref{sec:effect_on_wpdot} and Equation (\ref{eq:wpdot_mod_theory}).
In terms of the resonance condition, this implies that for a given planet--disk system, the resonance location should be expected to be pushed outwards to a larger semimajor axis in the nominal $N$-ring simulations compared to Paper I.
This could also be understood, for instance, by looking at the curves for $A(a)$, $A_{d,p}$, and $\dot{\varpi}_p$ in Figure \ref{fig:A-a-from-matrix-fid-simA-nominal}.

To verify this explanation, we computed the theoretical locations of the secular resonance based on Equation (\ref{eq:resonance_condition_p2}) using the expression of $\dot\varpi_p$ given by Equation (\ref{eq:wpdot_mod_theory}) -- rather than $\dot{\varpi}_p = A_{d,p}$ --  and compared the outcomes with the numerical results of $a\res$ measured in the suite of $67$ nominal $N$-ring simulations.
For a meaningful comparison, the theoretical calculations assumed the combinations of $M_d/m_p$ and $a_p/a\inn$ that would guarantee a secular resonance at $a\res = 70$ au if $\dot{\varpi}_p$ were equal to $A_{d,p}$, which are the same as those adopted in the numerical simulations (Table \ref{table:debris_Nring_sims_table}).
The results obtained are shown in Figure \ref{fig:ares_fn_ap_P2}, which illustrates how the resonance location varies with planetary semimajor axis both stemming from the theoretical calculations (black curve) and the nominal $N$-ring simulations (blue filled circles). 
One can clearly see that accounting for the disk's non-axisymmetric gravity in the secular resonance condition provides a very good description of the behavior observed in the nominal $N$-ring simulations. Note that the fact that the theoretical expectation underestimates the simulation results is not surprising given that the theoretical values of $\dot{\varpi}_p/A_{d,p}$ overestimate their numerical counterparts (Figure \ref{fig:wpdot_fn_ap_P2}). 
As another check, we also analyzed the set of simplified $N$-ring simulations, in which case $\dot{\varpi}_p = A_{d,p}$ (see Figure \ref{fig:wpdot_fn_ap_P2}), and found that the expected value of $a\res = 70$ au is recovered in every simulation; see the red filled circles in Figure \ref{fig:ares_fn_ap_P2}. %
In summary, the results shown in Figure \ref{fig:ares_fn_ap_P2} confirm the explanation for the shift in resonance locations being due to the non-axisymmetric component of the disk gravity which changes the planetary precession rate.

Before moving on, however, there are other features in Figure \ref{fig:ares_fn_ap_P2} worth noting. 
First, one can see that the resonance location is not shifted significantly for $a_p/a\inn \rightarrow 0$.
This follows from the fact that for $a_p/a\inn \ll 1$, one has $\dot{\varpi}_p \rightarrow A_{d,p} \propto a_p^{3/2} \rightarrow 0$; see Figure \ref{fig:wpdot_fn_ap_P2} and \EQPP{8}. Thus, the planet's precession rate does not contribute much to the resonance condition (Equation \ref{eq:resonance_condition_p2}) in the first place for the corrections due to the disk's non-axisymmetric gravity to have an effect. 
In the opposite case, however, i.e., when $a_p \rightarrow a\inn$, it is evident from Figure \ref{fig:ares_fn_ap_P2} that the resonance locations shift considerably relative to the predictions based on Paper I: for instance, by as much as $\approx 30\%$ for $a_p/a\inn \approx 1$.
Obviously, this shift results from the fact that when $a_p \rightarrow a\inn$, one has $\dot{\varpi}_p \rightarrow 0$  rather than $\dot{\varpi}_p = A_{d,p} \rightarrow \infty$ as in Paper I (see e.g. Figure \ref{fig:wpdot_fn_ap_P2}).

Interestingly, the above discussion  suggests that when the disk's non-axisymmetric gravity is included, as it should, the role of the planetary precession rate $\dot{\varpi}_p$ in the resonance condition of Equation \ref{eq:resonance_condition_p2} is weakened for all $a_p/a\inn$. 
Thus,  to a good degree, the resonance condition can be approximated by $A_d(a\res) + A_p(a\res) \approx 0 $; which is an assumption valid only for $a_p/a\inn \rightarrow 0$ in Paper I.
Accordingly, the scaling of $a\res$ with $M_d/m_p$ and $a_p$ as derived in Paper I by neglecting the planetary precession rate (see \EQPP{19} and Equation (\ref{eq:Mdmp_PI_19})) provides a better explanation for the behavior of the resonance locations in the nominal $N$-ring simulations than in the simplified ones. 
This is so even at large values of $a_p/a\inn$, where previously the condition $A_d(a\res) + A_p(a\res) \approx 0$ given by \EQPP{19} overestimated the resonance location for a given $M_d/m_p$ and $a_p/a\inn$ due to $A_{d,p}$ diverging (see e.g. Figure 4 of Paper I).
For instance, evaluating \EQPP{19}, or Equation (\ref{eq:Mdmp_PI_19}), at the parameters of \texttt{Model A}, one obtains $a\res \approx 77$ au (not so dissimilar from the simulation results in Figures \ref{fig:e-w-fn-a-simA-p2} and \ref{fig:ares_fn_ap_P2}, i.e., $\approx 75$ au), rather than $a\res = 70$ au.

\begin{figure}[t!]
\epsscale{1.15}
\plotone{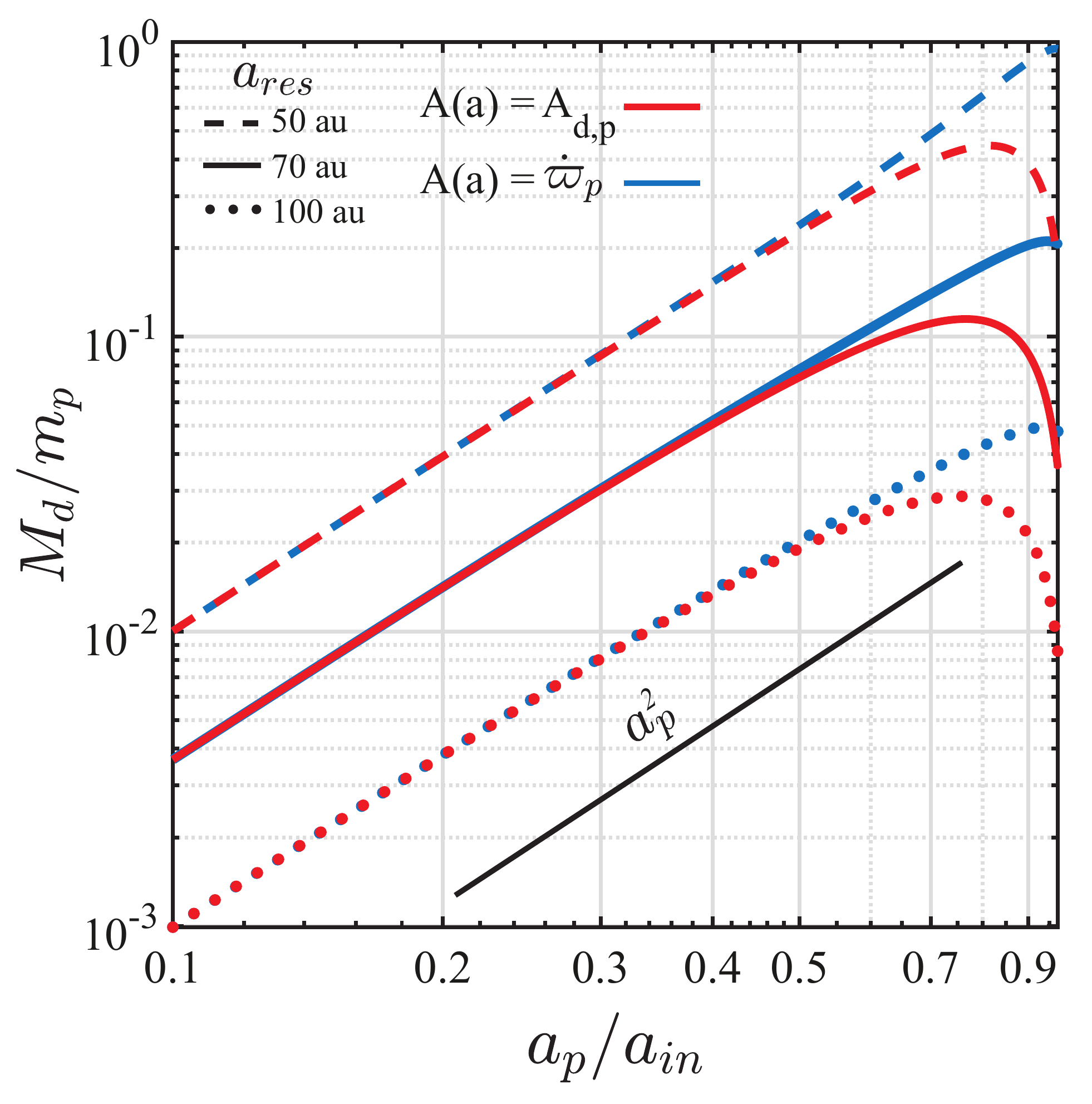}
\caption{The disk-to-planet mass ratio $M_d/m_p$ required, as a function of planetary semimajor axis $a_p/a\inn$, to place a secular resonance within the disk at $a\res$. Calculations are done for the fiducial disk model ($p = 1, a\inn = 30$ au, and $a\out = 150$ au) by solving the resonance condition given by Equation (\ref{eq:resonance_condition_p2}) under two assumptions: once within the context of simplified simulations (i.e., $A(a) = A_{d,p}$; red curves), and once within the context of nominal simulations (i.e., $A(a) = \dot{\varpi}_p$; blue curves). The results obtained for three different values of $a\res$ are shown by different line types, as indicated in the legend.
One can see that for planets orbiting close to the disk, $M_d/m_p$ is larger in the setting of nominal simulations compared to the simplified ones. Note also that the scaling of $M_d/m_p$ with $a_p/a\inn$ in the nominal case is well captured by the black curve representing \EQPP{19}, or (\ref{eq:Mdmp_PI_19}), even in the limit of $a_p\rightarrow a\inn$.  See the text (Section \ref{sec:effect_on_ares_shift}) for more details.
}
\label{fig:md-mp-ares-p1-p2}
\end{figure}

Conversely, this discussion also indicates that for the nominal $N$-ring simulations  to establish a secular resonance at some $a\res$, the ratio $M_d/m_p$ for  a given $a_p/a\inn$ must be equal to or larger than that expected from Paper I. 
This is confirmed in Figure \ref{fig:md-mp-ares-p1-p2}, where we plot the curves of $M_d/m_p$ as a function of $a_p/a\inn$ for three different values of $a\res/a\inn$ using Equation (\ref{eq:resonance_condition_p2}), both in the absence and presence of the disk's non-axisymmetric torque on the planet, i.e., $\dot{\varpi}_p = A_{d,p}$ (\EQPP{8}) and $\dot{\varpi}_p$ given by Equation (\ref{eq:wpdot_mod_theory}), respectively. 
Looking at Figure \ref{fig:md-mp-ares-p1-p2}, one can indeed see that the ratio $M_d/m_p$ must be corrected by increasing its value by at most a factor of $\approx 2 - 3$ for reasonable large values of $a_p/a\inn$.

\subsection{Timescale for Excitation of Planetesimal Eccentricities}
\label{sec:timescales_paper2_simulations}

The results of Section \ref{sec:results_p2} show that for the same planet--disk model (i.e., \texttt{Model A}), the excitation of eccentricity at the location of secular resonance $a\res$ takes a longer  time in the nominal $N$-ring simulation than that expected from Paper I -- namely, $\tau \approx 332$ Myr instead of $\approx 135$ Myr.
We now conduct a detailed analysis of this behavior. 
For ease of discussion, from hereon we use $\tau_{\rm n\mbox{-}axi}$ and $\tau_{\rm axi}$ to refer to the time it takes for $e(a\res) \rightarrow 1$ in the presence and absence of the disk's non-axisymmetric torque on the planet, respectively.

As already mentioned in Section \ref{sec:coupling-planet-planetesimal-p2}, the fact that $e(a\res)$ grows at a slower rate in the nominal $N$-ring simulations (compared to Paper I) is due to the decay of the planetary eccentricity in the course of evolution (Section \ref {sec:planet_evol_p2}). Quantitatively, this can be seen by inverting Equation (\ref{eq:e_t_at_resonance_p2}) to solve for the time $\tau_{\rm n\mbox{-}axi}$ at which $e(\tau_{\rm n\mbox{-}axi}) = 1$, finding that:
\begin{eqnarray}
    \tau_{\rm n\mbox{-}axi} &=& -\frac{2}{D} \log  \bigg( 1  - \frac{D}{2 |B_p(0)|} \bigg)    \bigg|_{a = a\res},
    \label{eq:tau_paper_2_with_D}
    \\
    & \approx &  
    \tau_{\rm axi} \left(1 + \frac{1}{4} \tau_{\rm axi}  D \right) 
    ,
    \nonumber
\end{eqnarray}
where the approximation in the second line assumes $D \ll |B_p(0)|$ and we have written $\tau_{\rm axi} = |1/B_p(0)|$; see \EQPP{16}. Looking at  Equation (\ref{eq:tau_paper_2_with_D}), one can see that  for a given planet--disk system, and depending on the strength of resonant friction (i.e., $D$), the value of $\tau_{\rm n\mbox{-}axi}$ should be expected to be equal to or larger than $\tau_{\rm axi}$ of Paper I.

To test and confirm this explanation, we computed the theoretical values of $\tau_{\rm n\mbox{-}axi}$ using Equation (\ref{eq:tau_paper_2_with_D}) for various values of $e_p(0)$ and compared them with the numerical values of $\tau_{\rm n\mbox{-}axi}$ obtained from the suite of $67$ nominal $N$-ring simulations (Table \ref{table:debris_Nring_sims_table}). 
For a precise comparison, in the theoretical calculations we adopted the numerical values of $D$ and $a\res$ as extracted from the simulations and shown in Figures \ref{fig:D_fn_ap_P2} and \ref{fig:ares_fn_ap_P2}, respectively, rather than using their corresponding theoretical predictions (i.e., Equations (\ref{eq:wpdot_mod_theory}), (\ref{eq:D_Tr98_MAIN}) and (\ref{eq:resonance_condition_p2})).

\begin{figure}[t!]
\epsscale{1.20}
\plotone{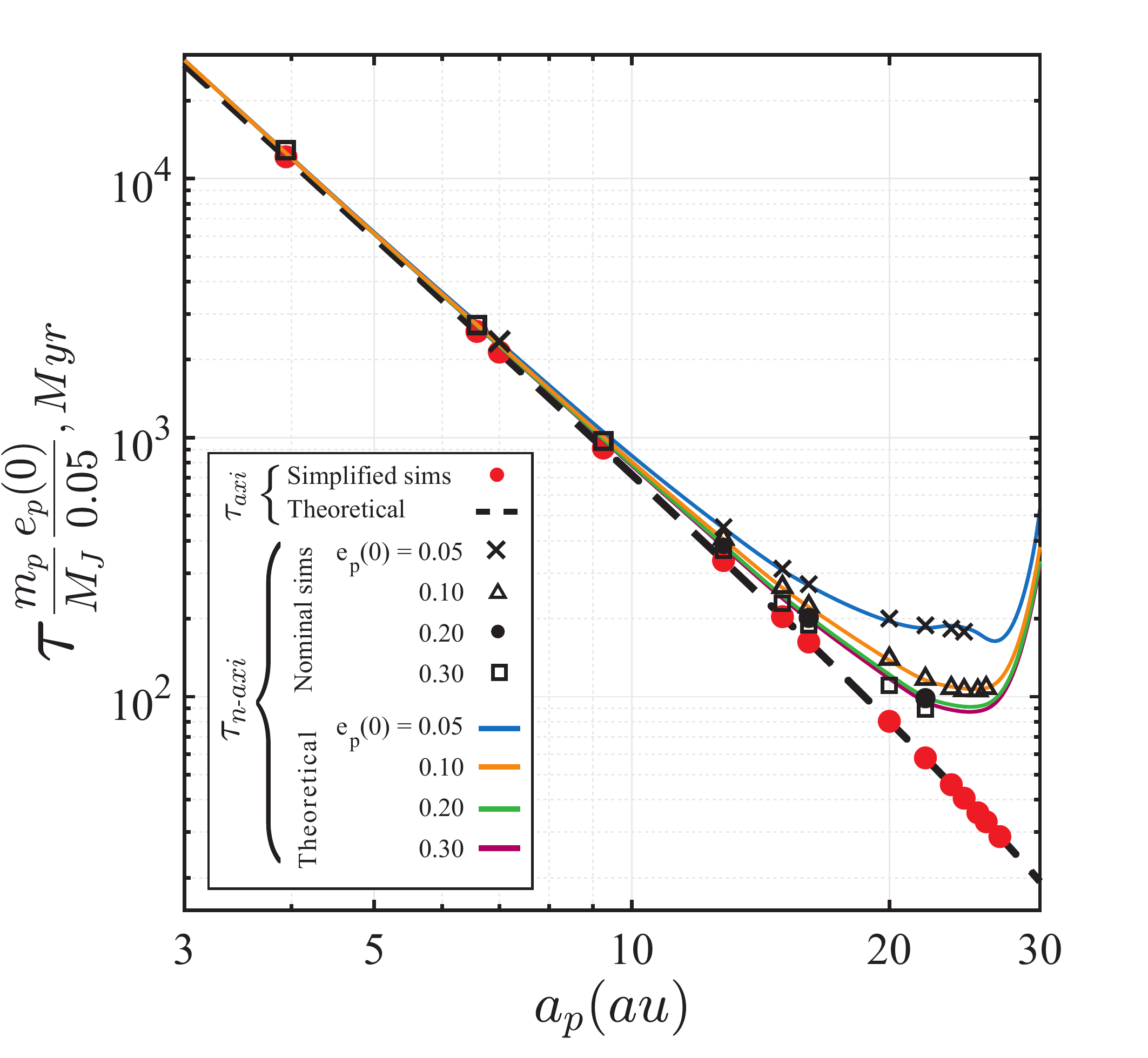}
\caption{The timescale $\tau$ for planetesimal eccentricity excitation at the secular resonance as a function of planetary semimajor axis $a_p$, as measured from $N$-ring simulations. 
Results correspond to $67$ different planet--disk models, each evolved within two sets of simulations (Table \ref{table:debris_Nring_sims_table}): simplified (shown in red circles and denoted by $\tau_{\rm axi}$), and nominal (shown using various black symbols that are explained in the legend; denoted by $\tau_{\rm n\mbox{-}axi}$). 
Note that the timescales are scaled by the planet's mass $m_p$ and initial eccentricity $e_p(0)$. 
One can see that results of simplified simulations lie along the dashed black line, which represents the expectation from Paper I (\EQPP{16}).
It is also evident that results of nominal simulations follow closely the solid curves of different colors (see the legend), which represent the theoretical expectations as a function of $e_p(0)$  upon accounting for the damping of the planetary eccentricity due to resonant friction (Equation \ref{eq:tau_paper_2_with_D}). 
Clearly, the disk's non-axisymmetric torque on the planet causes the timescales to be longer than those expected in Paper I, i.e., $\tau_{\rm n\mbox{-}axi} \gtrsim \tau_{\rm axi}$, with some dependence on $e_p(0)$.
See the text (Section \ref{sec:timescales_paper2_simulations}) for more details. 
}
\label{fig:tau-ap-withD-paper2}
\end{figure}

The results of this exercise are summarized in  Figure \ref{fig:tau-ap-withD-paper2}, where we plot  both the theoretical and numerical values of $\tau_{\rm n\mbox{-}axi}$ as a function of planetary semimajor axis $a_p$ for four different values of $e_p(0)$, as indicated in the legend. 
Note that the results are normalized by the planet's mass and initial eccentricity motivated by the fact that according to Paper I,   $\tau_{\rm axi}  \propto [m_p e_p(0)]^{-1}$ (\EQPP{16}).
For reference, Figure \ref{fig:tau-ap-withD-paper2} also shows the normalized values of $\tau_{\rm axi}$ as extracted from the simplified simulations (see the red circles): as expected, they all lie perfectly along the  
black dashed curve representing \EQPP{16}, irrespective of $m_p$ and $e_p(0)$.
More importantly, looking at Figure \ref{fig:tau-ap-withD-paper2}, one can clearly see that Equation (\ref{eq:tau_paper_2_with_D}) (shown in various solid lines; see the legend) provides a very good description of the numerical values of $\tau_{\rm n\mbox{-}axi}$   (shown in various black symbols; see the legend) for all considered values of $e_p(0)$ and $a_p$.

Results of Figure \ref{fig:tau-ap-withD-paper2} 
confirm that resonant friction on the planet is indeed the cause for the delay in exciting the planetesimal eccentricities at the resonance. 
Indeed, looking at Figure \ref{fig:tau-ap-withD-paper2} one can see that the values of $\tau_{\rm n\mbox{-}axi}$ -- regardless of $e_p(0)$ -- converge to the curve of $\tau_{\rm axi}$ at small values of $a_p$. This is because in this limit, resonant friction on the planet is so weak, i.e., $D \rightarrow 0$, that $e_p(t)$ remains roughly constant, and $\tau_{\rm n\mbox{-}axi} \rightarrow \tau_{\rm axi}$; see also Equation (\ref{eq:tau_paper_2_with_D}). 
In the opposite limit, however, i.e., as $a_p \rightarrow a\inn$ and the effects of resonant friction on the planet become significant, the values of $\tau_{\rm n\mbox{-}axi}$ diverge away from the curve of $\tau_{\rm axi}$, attaining values larger than $\tau_{\rm axi}$ by as much as a factor of $\sim 5$. 
This factor depends on the specific value of $e_p(0)$, with initially more eccentric planets leading to relatively smaller shifts in $\tau_{\rm n\mbox{-}axi}$ than nearly circular planets. This makes intuitive sense since the decay rate of $e_p$ does not depend on $e_p(0)$ (Section \ref{sec:resonant_friction}) and by and large, $B_p \propto e_p$ controls the amplitude of planetesimal eccentricities (see e.g. Equations (\ref{eq:e_t_at_resonance_p2}) and  (\ref{eq:master_app_e})).

\subsection{Resonance widths}
\label{sec:reswidth_analysis_section}

The results of Section \ref{sec:results_p2} show that the gap width in the nominal $N$-ring simulation of \texttt{Model A} is only slightly narrower than that in Paper I:  namely, $w_g \sim 15$ au here compared to $\sim 20$ au in Paper I. We find that this is generally the case for all systems simulated (Table \ref{table:debris_Nring_sims_table}), despite the process of resonant friction acting to damp the orbital eccentricities. 
We now explain why this is the case.

At the outset, we clarify that deriving an analytical expression for the gap widths is beyond the scope of this paper. Instead, what we do here is to follow the same methodology of Paper I and  use  the resonance widths $w$ as proxy for the gap widths $w_g$ (see e.g. Section 3.3 therein). Note that while these two quantities are not exactly equal, they are expected to be of the same order, $w \sim w_g$ (R. Rafikov \& A. Sefilian, in preparation).

The resonance width $w$ is defined as the radial range over which planetesimal eccentricities $e(a)$ are excited `significantly'. Defining what is meant by `significantly' is not straightforward, mainly because in linear Laplace--Lagrange theory, as in this work, the eccentricities formally diverge at the resonance.  In Paper I, following \citet{yelverton2018}, we measured the resonance widths by computing the distance over which the \textit{forced} planetesimal eccentricities, representing  the time-averaged  values of $e(a)$, exceed a \textit{constant} threshold value $\tilde{e}$, i.e., 
\begin{equation}
    |e_{\rm forced}(a)| \geq \tilde{e}        ,
    \label{eq:res-width-def-p1}
\end{equation}
or, equivalently, $\mathrm{max}[e(a)] \geq 2 ~ \tilde{e}$; see \EQPP{20}. In Paper I,  the threshold eccentricity was chosen to be that of the planet, i.e., $\tilde{e} = e_p$, as this represents the maximal forced eccentricity attained away from the resonance. 

One can use Equation (\ref{eq:res-width-def-p1}) to show that for a given planet--disk system, the resulting values of $w$ will be similar both in our previous and current works. This requires some care though,  because neither the forced eccentricities nor the planetary eccentricity (set equal to $\tilde{e}$ in Paper I) are constant in our work. There are two ways to approach this; namely, by considering the initial and final states:

\begin{itemize}

\item First, since planetesimals are initiated on circular orbits,   their free eccentricities  are given by their initial forced eccentricities, i.e., $e_{\rm free} = e_{\rm forced}(0)$. Thus, for the purposes of Paper I, the resonance width condition of Equation (\ref{eq:res-width-def-p1}) could be written using either $e_{\rm free}(a)$ or $e_{\rm forced}(a)$, since these are the same at all times. Nevertheless, the use of the free eccentricities -- rather than the forced -- in the left-hand side of Equation (\ref{eq:res-width-def-p1}) is better justified in more general setups such as in this work. This is because unlike the forced eccentricity, $e_{\rm free}$ is a conserved quantity set by initial conditions. Doing so for the purposes of this work, and noting that by definition $e_{\rm free} = e_{\rm forced}(0)  = e_{\rm forced}^{\rm PI}(a)$ (Equation \ref{eq:eforced_with_D_decay_p2}), it is trivial to see that the resulting resonance widths will be equal to those in Paper I upon using the same  $\tilde{e} = e_p(0)$ condition as in that work.

    \item A complementary explanation can be proposed  by considering the limit of $t \rightarrow \infty$. In this case, resonant friction would completely damp the forced planetesimal eccentricities as $e_p(t) \rightarrow 0$. As a result, eccentricity oscillations will reach a steady state so that $e(a) \rightarrow e_{\rm free}(a) \equiv \mathrm{max}[e(a)]$  (Figures \ref{fig:e-w-fn-a-simA-p2} and \ref{fig:e-w-fn-time-simA-p2}),  which amounts to exactly half  of the maximum amplitude of eccentricity oscillations in Paper I, i.e., $e(a)  \rightarrow e_{\rm max}^{\rm PI}/2 = e_{\rm forced}^{\rm PI}(a)$. Accordingly, setting the  threshold eccentricity  equal to $\tilde{e} = e_p(0)/2$ -- which is what the maximum eccentricity will be away from the resonance as $t \rightarrow \infty$ (Figure \ref{fig:e-w-fn-a-simA-p2}) --  the same resonance width as in Paper I will be recovered. 
\end{itemize}

In summary, the above discussion implies that the resonance widths $w$ -- and thus the gap widths $w_g$ -- will be very similar in Paper I and our current work. Indeed, only slight deviations between the results are to be expected, if any, due to the shifting of the resonance locations found in this work  (Section \ref{sec:effect_on_ares_shift}). This is because our calculations in Paper I show that for a fixed value of $a_p$, the resonance widths scale as $w \propto a\res^{-1/2}$; see \EQPP{21} and Figure 6 of Paper I.

\section{Discussion}  
\label{sec:discussion}

The results of previous sections show that secular planet--debris disk interactions can lead to the damping of the planet's orbital eccentricity, without affecting its semimajor axis (Sections \ref{sec:planet_evol_p2} and \ref{sec:resonant_friction}). 
This process, which we identify as \textit{resonant friction}, ensues from the gravitational coupling between the planet and the disk  via secular apsidal resonances \citep{tre98, wardhahn1998}.
These resonances are established within the disk at the site where the apsidal precession rate of the planetesimal orbit (due to both the planet and disk gravity) matches that of the planet (due to the disk gravity).\footnote{We remind the reader that in this paper, we ignore the non-axisymmetric perturbations among the disk rings; see Section \ref{sec:limitations_und_future_work} for detailed discussion.}
Accordingly, the circularization of the planetary orbit occurs simultaneously with the sculpting of a depletion, i.e., a gap, within the debris disk.

In what follows, we discus the implications of this work for the predictions made in Paper I for the debris disks  around \HD~and \HDN~(Section \ref{sec:implicatins_for_paper1_discussion}). We also discuss the implications firstly for the dynamical modeling of debris disks in general (Section \ref{sec:disk_gravity_importance}), and  secondly for indirectly measuring the  masses of debris disks as well the primordial eccentricities of exoplanets (Section \ref{sec:Md_ep0_estimations}),  taking the \HDnew~system as an example. We then discuss the major limitation of our current work, namely, the neglect of the non-axisymmetric perturbations among the disk rings (Section \ref{sec:limitations_und_future_work}), before commenting on a related previous work (Section \ref{sec:comment_on_pw15}).

\subsection{Implications for the predictions made in Paper I}
\label{sec:implicatins_for_paper1_discussion}

Results of $N$-ring simulations in Sections \ref{sec:results_p2} and \ref{sec:analysis_predictions} show that the dependence of the characteristics of the secular resonances -- namely, their locations, associated timescales, and widths -- on the planet--disk parameters is modified by the disk's non-axisymmetric torque on the planet.
It is thus necessary to re-evaluate the parameter spaces -- i.e., the combinations of $m_p$, $a_p$, and $M_d$ -- identified in Paper I as capable of explaining the gapped debris disks observed around \HD~and \HDN; see Figures 7 and 12 of Paper I, respectively. 
This can be readily achieved with the aid of the analytical estimates which we derived in Sections \ref{sec:effect_on_ares_shift}--\ref{sec:reswidth_analysis_section} for the characteristics of the secular resonances. 
However, we stress that our goal in this work was not to model the specifics of the observed gapped debris disks in detail, but rather to take an additional (but not final) step towards a full understanding of the dynamical effects of disk gravity in planet--debris disk interactions. 
We thus refrain from a detailed re-evaluation of the parameter spaces identified in Paper I, which we think is best relegated to the third paper in this series, where the full gravitational effects of the disk will be considered (Section \ref{sec:limitations_und_future_work}).

For now, we note that our nominal simulations suggest that the allowed portions of the parameter spaces in Paper I for both \HD~and \HDN~will not be strongly modified for $ a_p/a\inn \ll 1$. 
This is because, in this case, the disk's non-axisymmetric torque on the planet does not affect the properties of the secular resonances; see e.g. Figures \ref{fig:ares_fn_ap_P2}--\ref{fig:tau-ap-withD-paper2}.
In the opposite limit, however, i.e., when $ a_p/a\inn \lesssim 1$, we find that the resonance location shifts to larger separations from the star. 
To correct for this shift, the ratio of masses $M_d/m_p$ for a given value of $a_p$ should be larger than that in Paper I by a factor of $\sim 2 - 3$; see Figure \ref{fig:md-mp-ares-p1-p2}. 
This could be done, for instance, by keeping $M_d$ unchanged, and adopting values of $m_p$ that are smaller by a factor of $\sim 2 - 3$ than in Paper I.
This obviously will prolong the timescales $\tau$ for eccentricity excitation at the resonance (since $\tau \propto 1/m_p$, \EQPP{16}), in addition to the delay arising due to  resonant friction (Figure \ref{fig:tau-ap-withD-paper2} and Equation \ref{eq:tau_paper_2_with_D}).
However, it is unlikely that this would affect the parameter spaces of Figures 7 and 12 of Paper I any further, since the resulting gaps are already well-developed by $t \sim \tau_{\rm sec}  = 2\pi/\dot{\varpi}_p$ (i.e., when $e(a\res) \lesssim 1$), which, we remind the reader, are generally much smaller than $\tau$ (i.e., when $e(a\res) \approx 1$); see Section \ref{sec:typical_evolution_of_disc_morphology_p2}. 
We note that this also implies that the timescale $\tau_{\rm sec}$ is better suited as a metric for constraining the planet--disk parameters, as it also does not depend on the planet's eccentricity (in contrast to $\tau$; Equation (\ref{eq:tau_paper_2_with_D})).

\subsection{The important role of debris disk (self)-gravity}
\label{sec:disk_gravity_importance}

\begin{figure}[t!]
\epsscale{1.25}
\plotone{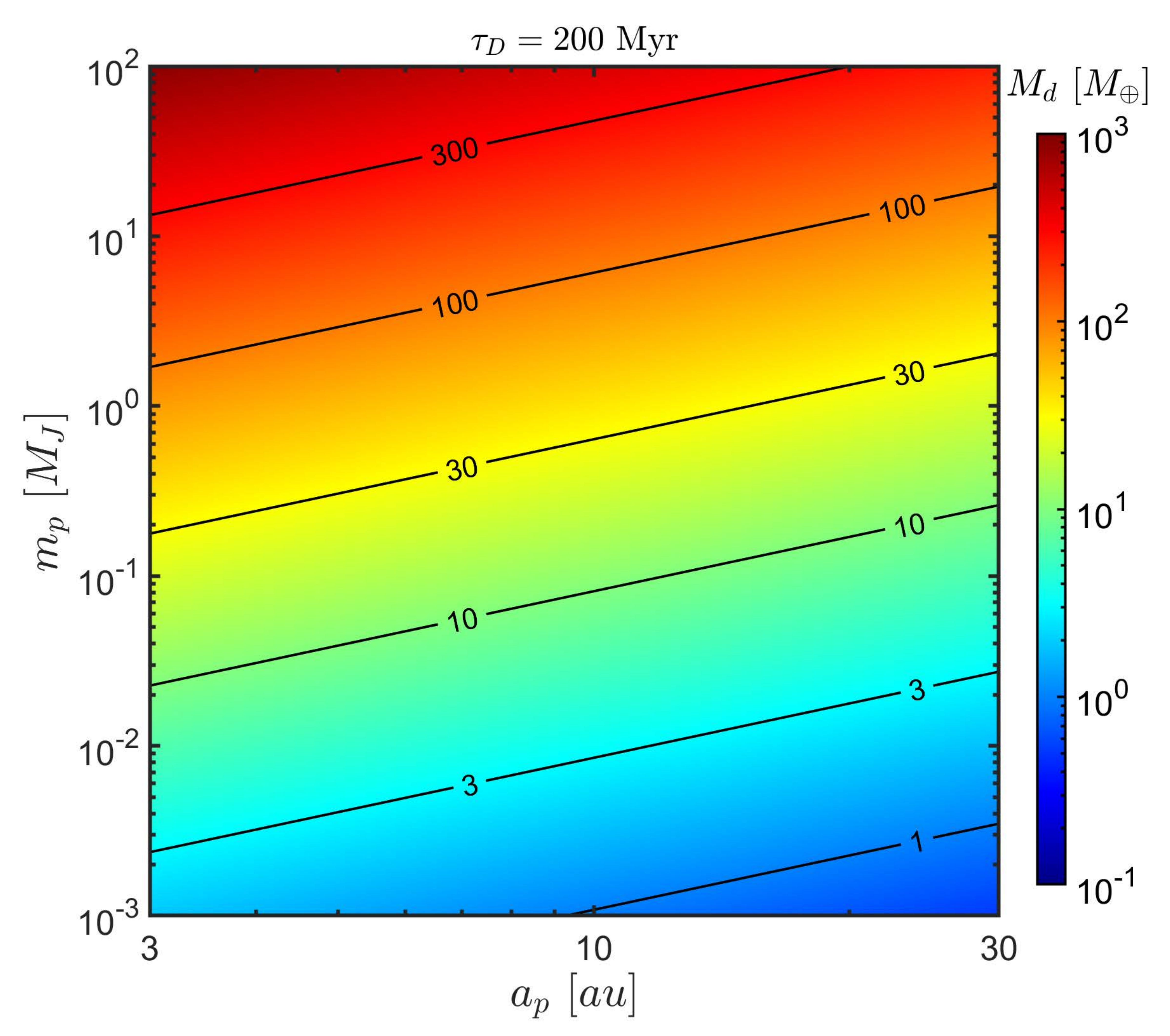}
\caption{Contour plot of the disk mass $M_d$  for damping the planetary eccentricity due to resonant friction with a half-life time of $\tau_D = 200$ Myr, plotted in the space of planetary semimajor axis $a_p$ and mass $m_p$. Calculations are done using the approximate equation for $\tau_D$, i.e., Equation (\ref{eq:tau_D_p2_diskmassMd}), assuming the parameters of the fiducial disk model (i.e., $p=1, \delta = 5$) around a solar-mass star of age $t_{\rm age} = \tau_D$. For a given combination of $a_p$ and $m_p$, increasing (decreasing) $M_d$ leads to a shorter (longer) $e_p$-damping timescale than that stipulated (Equation (\ref{eq:tau_D_p2_diskmassMd})). Note that for a fixed value of $M_d$, the planet's mass scales approximately linearly with its semimajor axis, namely, $m_p \propto a_p^{17/16}$; see Equation (\ref{eq:tau_D_p2_diskmassMd}).
See the text (Sections \ref{sec:resonant_friction} and \ref{sec:disk_gravity_importance}) for more details.
}
\label{fig:tauD_mp_ap_nf}
\end{figure}

The study presented here highlights an important caveat related to studies of  planet--debris disk interactions in general.  One of our main results is that resonant friction is a robust and effective process, in that it only requires the presence of a single planet orbiting interior to a debris disk of modest mass with  $M_d/m_p \lesssim 1$ (Section \ref{sec:resonant_friction} and Figure \ref{fig:md-mp-ares-p1-p2}). Additionally, the resulting eccentricity-damping timescale $\tau_D$ can be comparable to or less than the typical ages of debris disks over a broad range of planet--disk parameters; see Equations (\ref{eq:tau_D_p2}), (\ref{eq:tau_D_p2_diskmassMd}), and Figure \ref{fig:D_fn_ap_P2}.  
As an example, this is further illustrated in Figure \ref{fig:tauD_mp_ap_nf}, where, using the approximate equation for $\tau_D$ (i.e., Equation (\ref{eq:tau_D_p2_diskmassMd})), we plot the contours of the disk mass ($M_d$) in the planetary mass ($m_p$)--semimajor axis ($a_p$) plane corresponding to a particular choice of $\tau_D = 200$ Myr. Looking at Figure \ref{fig:tauD_mp_ap_nf}, it is evident that disk masses of $M_d \sim 1 - 100 M_\oplus$ can sufficiently damp the orbital eccentricities of planets of reasonable masses orbiting interior to the disk. Here, we note that for a given planet, the values of $M_d$ reported  in Figure \ref{fig:tauD_mp_ap_nf} represent the minima required so that $e_p$ decays to half its initial value by the age of the system $t_{\rm age}$, which we have set equal to $\tau_D$. Thus, at least the same amount of $e_p$-decay could occur with, for instance, lower disk masses (and thus longer $\tau_D$) in older systems. Conversely, resonant friction would be weak for a given planet that lies above some constant-$M_d$ contour in Figure \ref{fig:tauD_mp_ap_nf}.

The extreme scenario that follows from this process of resonant friction is  a planet orbiting interior to a gapped debris disk which, at the time of detection, would be on a circular orbit with $e_p = 0$; an end-state that can be easily attained for planets closer to the disk than to the star, in which case $D/m_p$ tends to be maximized (Figure \ref{fig:D_fn_ap_P2}). If not for the disk gravity in the model, such a configuration would otherwise not only be unexpected, but also deemed to be incompatible with the imaged radial structure of the debris disk -- at least not without invoking additional processes.  We thus caution that effects of resonant friction should be considered when modeling and interpreting debris disk structures.

We note that this realization could potentially be important when interpreting radial substructures of debris disks in general, i.e., even when a gap is not detected. This is because while the process of resonant friction requires the establishment of a secular resonance within the disk for it to operate, it does not necessarily require that the secular resonance is wide enough for it to produce an observable gap. A narrow secular resonance could also lead to significant damping of the planet's eccentricity, provided the system is old enough. This could occur e.g. when the secular resonance is established near $\approx a\inn$ or $\approx a\out$, which tends to be narrower than the case when $a\inn \lesssim a\res \lesssim a\out$ (see e.g. \EQPP{22} and Figure 6 of Paper I).\footnote{Here, we note in passing that the occurrence of secular resonances (and thus excitation of eccentricities, $e(a\res) \rightarrow 1$) near the innermost radius of a given planetesimal population could be responsible for truncating and shaping the observed inner disk edge.  We defer a detailed investigation of this process to a future paper which is currently in preparation; referring the reader for now to the work of \citet{jeremy2023} for a step in this direction.} 
Although beyond the scope of our current work, it is worthwhile to note that this discussion would equally apply to configurations where the planet is situated exterior to a self-gravitating debris disk \citep[e.g.][]{wardhahn2003}, both in coplanar and inclined configurations\footnote{In cases where the disk particles and the planet are initially misaligned, the system may establish a secular \textit{inclination resonance}, leading to the formation of a warp within the disk. This occurs at the site where the rate at which the planetesimal longitude of ascending node  precesses is the same as that of the planet. This then gives rise to resonant friction, which, somewhat akin to the coplanar case, acts to damp the planetary inclination \citep[e.g.][]{wardhahn2003, hahnbending}. We defer a detailed investigation of this phenomenon to a future work.} 
\citep[e.g.][]{hahnbending}. In summary, the discussion here lends further support to our statement in Paper I: including the effects of disk gravity should be considered in studies of planet--debris disk interactions.

A word of caution is necessary at this stage though. The $N$-ring model of Section \ref{sec:problemframework_P2} is valid only to second order in eccentricities; see e.g. Equation (\ref{eq:Master_Rj}). Thus, one concern is that the strength of resonant friction would in reality be limited when $e(a\res)$ grows to values large enough that the effects of higher-order terms in the secular disturbing function become important. This can be easily assessed in a future work, either numerically or analytically, by making use of a higher-order expansion of the disk potential \citep[for work in this direction, see][]{JTgauss,touma-sridhar-2012, ST19}. 
Nevertheless, such details are unlikely to negate the main message here: a planet could sculpt an observable radial structure (e.g., a gap) in a debris disk, and yet have its orbital eccentricity damped to below its initial value, provided the disk is massive enough to establish a secular resonance (see also Section \ref{sec:resonant_friction}).
We stress that this generally translates to  a requirement of $10^{-4} \lesssim M_d/m_p \lesssim 1$ for all $a_p \lesssim a\inn$ (see also Figure\ref{fig:tauD_mp_ap_nf}), which ensures that a secular resonance is established at some location within the disk, i.e., $a\inn \lesssim a\res \lesssim a\out$; see Equation (\ref{eq:Mdmp_PI_19}), Figure \ref{fig:md-mp-ares-p1-p2}, and Figure 4 of Paper I.

\subsection{Implications for disk mass estimates and the evolutionary history of planets}
\label{sec:Md_ep0_estimations}

Taking resonant friction for granted, our results may be used to set constraints not only on the total masses of debris disks, but also the dynamical past of the planets orbiting interior to them,  as follows. For a given debris disk characterized by a depletion in its surface density, the location of the gap can be used to set constraints on the possible mass and semimajor axis of a yet-undetected planet interior to the disk; see e.g. Section 4 of Paper I. These inferences, however, are not unique, as they depend on the assumed value of disk mass $M_d$. Thus, and as already discussed in Paper I, the detection of planets with the inferred parameters (namely, $m_p$ and $a_p$) would indirectly measure the debris disk's total mass; see e.g. Equation (\ref{eq:Mdmp_PI_19}) and \EQPP{19}. This is particularly appealing, as the disk mass is a parameter which is fundamentally  unknown \citep{krivovwyatt20}. Additionally, however, our current results imply that the planetary eccentricity at the time of detection can be translated to its primordial value, $e_p(0)$, assuming that e.g. resonant friction has been active over the system's age; see Equations (\ref{eq:tau_D_p2}) and (\ref{eq:tau_D_p2_diskmassMd}). A proper valuation of the primordial planetary eccentricity, as we would demonstrate next for \HDnew, would require the knowledge of the current planetary parameters, namely, $m_p$, $a_p$, and $e_p$, as well as the observed location of the gap within the system's debris disk. This could be useful to better understand the conditions at the end of planetary formation processes. 
This is a particularly intriguing result, considering that combined JWST and ALMA observations in the near future are expected to reveal a wealth of information on exoplanets and debris disks around the same stars (including \HD, \HDN, and \HDnew~as part of a JWST Cycle 1 GO program; \citet{seba-jwst}).

A case in point is provided by the \HDnew~system. This $1.24 M_{\odot}$ F5V star hosts a double-ringed debris disk \citep{marino2020, nederlander21}, extending from $\sim$ 30 au to $\sim 180$ au, situated exterior to a directly imaged brown dwarf companion  \citep{delorme17, milli17}. The orbital semimajor axis of the brown dwarf \HDnew~B  is $a_p \sim 11$ au, and its minimum possible mass is estimated to be $\sim 12 M_J$. In Paper I,  using these parameters, we had arrived at the conclusion that the $\sim$27 au wide gap centered at $a\res \approx$75 au within this disk could be reproduced by its interaction with \HDnew~B, provided that $M_d \approx 170 M_{\earth}$.
If our hypothesis is true\footnote{Note that our inference for the disk mass of \HDnew~in Paper I, i.e., $M_d \approx 170 M_{\earth}$, remains unchanged in light of introducing the disk's non-axisymmetric torque on the inner companion. This is because $a_p/a\inn \approx  1/3$, and thus the resulting $a\res$ would be the same as in Paper I (i.e., $a\res \approx 75$ au).},
our calculations imply an eccentricity decay half-life of $\tau_D \approx 190$ Myr (Equations (\ref{eq:D_Tr98_MAIN}) and (\ref{eq:tau_D_p2})).
Considering the age of the system of $t_{\rm age} = 50 - 700$ Myr and the current eccentricity $e_p(t_{\rm age}) \sim 0.15$ of \HDnew~B \citep{marino2020}, we then infer initial orbital eccentricities of $e_p(0) \gtrsim 0.18$ (for $t_{\rm age} = 50$ Myr), and disfavor ages larger than $\approx 506$ Myr (so that $e_p(0) \leq 1$).
We stress, however, that these values must not be taken at face value, but rather as proof-of-concept, not least because of our ignorance of  the recently discovered second inner companion, \HDnew~c, at $\sim 3$ au \citep{hinkley_HD206893}; see also Section \ref{sec:limitations_und_future_work}.

Moving on,  we note that our results may also be used to derive upper limits on the total masses of debris disks in known planet-hosting systems, whether the disk is resolved or not.  Indeed, noting that $e_p(0)$ could not exceed unity, the current planetary eccentricity, if known, establishes an upper estimate on the total disk mass; provided a secular resonance is established somewhere within the disk -- see Equations (\ref{eq:tau_D_p2}) to (\ref{eq:tau_D_p2_diskmassMd}) as well as Figure \ref{fig:tauD_mp_ap_nf}. Here, we point out that such an analysis has already been performed by \citet{ward-neptune-98} considering Neptune and the Kuiper belt. Indeed, \citet{ward-neptune-98} derived an upper limit of $\sim 2 M_{\earth}$ for the total mass of the Kuiper belt, assuming that the current eccentricity of Neptune, $e_{N} \approx 0.009$, results from resonant friction. It would be interesting to use this approach for planet--hosting systems to derive independent estimates of debris disk masses, a task for which Figure \ref{fig:tauD_mp_ap_nf} could serve as a starting point -- see also Section \ref{sec:resonant_friction} and Equation (\ref{eq:D_Tr98_MAIN}).

\subsection{Limitations and Future Work }
\label{sec:limitations_und_future_work}

The major limitation of our current work is that in addition to the axisymmetric contribution of the disk gravity, we accounted for only one aspect of its non-axisymmetric component: namely, its effect on the planet. 
That is to say, we ignored the non-axisymmetric perturbations that the disk rings would exert among themselves. 
We stress, however, that this neglect is not due to an inherent limitation of the softened $N$-ring model outlined in Section \ref{sec:problemframework_P2}, which, on the contrary, can be employed to address the full self-consistent problem.
Instead, this omission was done on purpose in order to facilitate building up a more systematic understanding of the effects of disk (self-)gravity in planet--debris disk interactions, both generally and specifically when secular resonances are established. 
Indeed, this incremental approach allowed us to isolate and focus on the process of resonant friction, as well as to derive semi-analytical equations that well reproduce the simulated orbital evolution of the planet and planetesimals. 
In a forthcoming paper (Paper III, in preparation, based on \citet{sefilian-phd}),  we shall relax our assumptions regarding the disk gravity and consider the fully self-consistent case, i.e., by accounting for the disk's complete gravitational potential.
For now, we note that the introduction of the non-axisymmetric perturbations among the disk rings will lead to a rich phenomenology of additional effects (as also discussed in Section 7.1.2 of Paper I).

Another major limitation of our work is the fact that we ignore the collisional activity of planetesimals within the disk.
This, together with other assumptions already discussed in detail in Paper I (see Section 7 therein), such as the system's coplanarity, our adherence to single-planet systems, and limiting the analysis to second order in eccentricities, can be easily relaxed by introducing slight modifications to the $N$-ring model outlined in Section \ref{sec:problemframework_P2}.    
We defer addressing these limitations to future papers of this series, referring the interested reader to Section 7 of Paper I for a discussion of the potential implications of relaxing some of them.

\subsection{Comment on the work of Pearce \& Wyatt (2015)}
\label{sec:comment_on_pw15}

Finally, we comment on the work of \citet{pearcewyatt15}, who employed $N$-body simulations to study the evolution of planet--debris disk systems that are akin to ours in terms of architecture. In their simulations, the authors accounted for the full back-reaction of the disk on the planet (i.e., both axi- and non-axisymmetric components, similar to our current work), but neglected the disk self-gravity (in contrast to our work). Despite this, however, they reported on the sculpting of a gap within the disk, coupled with the circularization of the planetary orbit in the absence of any significant orbital migration -- a result reminiscent of our nominal $N$-ring simulations. We thus suspect that \citet{pearcewyatt15} may have stumbled upon evidence of resonant friction, although not identified as such by the authors.

Our reasoning for this is as follows. According to Figure 5 of \citet{pearcewyatt15}, the  decay of the planet's eccentricity follows by and large an exponential behavior, especially at late times (see `Stage 3' in their figure). \citet{pearcewyatt15} interpreted this as a direct consequence of planetesimal scattering. This might indeed be the dominant cause at early times into the evolution, i.e., when the planet that they consider is highly eccentric ($e_p \sim 0.6$), and thus crosses the disk along its orbit. However, once all debris particles that initially cross the planetary orbit have been scattered, it is likely that the continued decay is instead mainly due to resonant friction, a process which the authors did not identify. To test this, one could compare the simulation data of \citet{pearcewyatt15} with the decay rate of Equation (\ref{eq:D_Tr98_MAIN}). In doing so, we find that the analytical expression of $D$ given by Equation (\ref{eq:D_Tr98_MAIN}) provides a good first-order approximation to the simulation data of \citet{pearcewyatt15}, although underestimating the time to damp the planetary eccentricity by about a factor of $\sim 3$. This difference could in principle be attributed to the fact that in our linear calculations planetesimal eccentricities at the resonance reach unity (enhancing the effect of resonant friction), whereas this might be limited to lower values (i.e., $e \ll 1$) in the more accurate, direct $N$-body simulations of \citet{pearcewyatt15}. It is also important to  note that the scattering  of planetesimals evident in \citet{pearcewyatt15} further complicates a direct comparison with our analytical theory, not least because scattering renders the disk mass, and thus the strength of resonant  friction, time dependent. Nevertheless, further support to our reasoning can be sought in the fact that this discussion also implies the occurrence of a secular resonance within the  disks of \citet{pearcewyatt15}, without which, we remind, resonant friction does not occur (Section \ref{sec:resonant_friction}).  This is indeed the case in their work, providing further support to our supposition, although the authors do not mention it explicitly. Indeed, the resonance condition that is satisfied in the work of \citet{pearcewyatt15} is $A_p(a) = \dot{\varpi}_p$, i.e., Equation (\ref{eq:resonance_condition_p2}) with $A_d(a) = 0$; see e.g. their Figure 2.  This also complements the explanation provided in \citet{pearcewyatt15} for why a gap is sculpted within the debris disks that they simulate.

\section{Summary}  
\label{sec:summary}

In this paper, the second in the series, we have investigated the secular interaction between a single low-eccentricity planet and an external, self-gravitating debris disk coplanar to it. In doing so, we have gone further than the simplified model of Paper I, having accounted for the full back-reaction of the disk on the planet, i.e., not only the corresponding axisymmetric but also the non-axisymmetric component of the disk gravity (absent in Paper I). For simplicity, however, we have ignored the non-axisymmetric perturbations among the disk rings.

We achieved this by employing a discretized description of the disk, the so-called $N$-ring model (Section \ref{sec:problemframework_P2}), whereby the disk is modeled as a series of geometrically spaced $N \gg 1$ massive rings that interact both with each other and the planet. This framework allows for a self-consistent treatment of the secular evolution of gravitating disk--planet systems to second order in eccentricities using  Lagrange's planetary equations (Equations \ref{eq:EOM_k_h}). This is, however, subject to one condition: that the interaction potential between the disk rings is softened appropriately; namely, by accounting for the disk's small, but non-zero, aspect ratio $H$ \citep{hahn2003, SR19}. Accordingly, the result is a softened analogue of the classical Laplace--Lagrange perturbation theory. While the softened $N$-ring model is not novel per se \citep[see e.g. references in][]{SR19}, this is the first time (to our knowledge) that it has been utilized to examine the  eccentricity evolution of massive debris disks in exoplanetary systems.\footnote{We note that a similar $N$-ring approach has been utilized in studying the inclination evolution of self-gravitating astrophysical disks; see e.g. \citet{bat11, bat12, batygin2018}.}

We employed the $N$-ring model to study how the introduction of the non-axisymmetric torque that the disk exerts on the planet modifies results of Paper I: namely, that secular apsidal resonances can lead to the formation of a gap within the disk. Our main findings can be summarized as follows. 
\begin{enumerate}[label=(\roman*)]
\item  The disk gravity does not only drive prograde apsidal precession of the planetary orbit (i.e., $\dot{\varpi}_p >0$), but also induces significant evolution of its orbital eccentricity $e_p$. In particular, we find that the planetary eccentricity can undergo a long-term exponential decline, accompanied by small-amplitude  oscillatory behavior. 

\item The circularization of the planetary orbit results from a process known as ``resonant friction'' (or ``secular resonant damping'') in the literature. This is a direct consequence of angular momentum exchange between the disk and the planet, which is mediated by the secular apsidal resonance within the disk itself. Thus, it does not require the planet to cross the disk along its orbit.

\item   Despite modifying the planetary precession rate, the inclusion of the disk's non-axisymmetric torque on the planet  does not significantly affect the condition for secular resonance (and thus the formation of a gap within the disk). That is, we generally find the same requirement as in Paper I, that the debris disk is less massive than the planet, $M_d /m_p \lesssim 1$, when $a_p\lesssim a\inn$.

\item  The specific behavior of the secular resonances as a function of disk and planet parameters, however, does get affected: namely, their locations $a\res$, associated timescales $\tau$, and widths $w$. By means of dynamical theory, we derive semi-analytic formulae that well reproduce the simulation results; thus updating the equations derived in Paper I.

\item By investigating the evolution of planet--disk systems that are motivated by \HD~and \HDN, we find that by and large the evolution of the disk morphology proceeds in the same three stages as in Paper I. The gap forms over timescales comparable to the planetary precession period, typically tens of Myr. 

\item We find that the resulting gap has features that are akin to those in Paper I: asymmetric, being wider and deeper in the direction of the planetary pericenter, and of fractional depth of about $0.5$. The degree of asymmetry, however, is reduced with time due to resonant friction.

\item More generally, our results may be used to  constrain the properties of yet-undetected planets interior to gapped, i.e., double-ringed, debris disks. The detection (or non-detection) of such planets sets indirect constraints on their primordial eccentricity, as well the total masses of their debris disks.

\end{enumerate}

In closing, we note that the methodology and results of this work will be used in the third paper of this series (Paper III), where we investigate the fully self-consistent problem, i.e.,  by incorporating the full gravitational effects of the disk. In the future, we also plan to exploit the flexibility of the $N$-ring code to model a range of other scenarios: e.g., collisionally active debris disks, disks perturbed by planets on inclined orbits, and disks in multi-planet systems.

\section*{Acknowledgments}

\noindent   
This work was supported, in whole or in part, by the Bill \& Melinda Gates Foundation [OPP1144]. Under the grant conditions of the Foundation, a Creative Commons Attribution 4.0 Generic License has already been assigned to the Author Accepted Manuscript version that might arise from this submission. 
This work derives from the PhD research of A. Sefilian at the University of Cambridge, made possible by the support of the Gates Cambridge Trust.
A.A.S. acknowledges support by the Alexander von Humboldt Foundation through a Humboldt Research Fellowship for postdoctoral researchers. R.R.R. acknowledges financial support through the Ambrose Monell Foundation, and STFC grant ST/T00049X/1. A.A.S. expresses his utmost gratitude to Jihad Touma for hosting him at the American University of Beirut (AUB), in Lebanon, during the last-year of his PhD studies.  A first draft of this manuscript was prepared at AUB, which offered a safe haven amidst the collapsing situation in Lebanon. The authors thank the anonymous referee for a positive report and
constructive comments on the manuscript, and Natasha Simonian for her help in composing Figure 1.

\software{MATLAB \citep{MATLAB}.}

\section*{Data availability}
\noindent The data that support the findings of this study, including a copy of the $N$-ring code developed as part of this work, are made publicly available in Figshare at doi: \href{http://doi.org/10.6084/m9.figshare.22598377}{10.6084/m9.figshare.22598377}.


\newpage
\appendix

\section{Tests of the \texorpdfstring{$N$}{N}-ring model}
\label{app:tests_Nring}
\restartappendixnumbering

In this Appendix, we describe several tests which are used for verifying the performance of the $N$-ring model outlined in Section \ref{sec:problemframework_P2}.

\subsection{Conservation of Angular Momentum}
\label{sec:L_conservation_P2}

We first present an analytic proof that the $N$-ring model conserves  the system's total angular momentum, $L$, given by:
\begin{equation}
   L =  \sum\limits_{j=0}^{N} L_j =  \sum\limits_{j=0}^{N} m_j n_j a_j^2 \sqrt{1-e_j^2} ,
   \label{eq:L_system}
\end{equation}
%
see \citet{mur99}. To this end, we retain terms up to second order in eccentricities in Equation (\ref{eq:L_system}) -- i.e., to the same degree of precision as the secular disturbing function (Equation \ref{eq:Master_Rj}) -- and write $L$ as $L \approx L_0 - L_e$, where
\begin{equation}
    L_0 =   \sum\limits_{j=0}^{N}  m_j n_j a_j^2
    , 
    ~~~\text{and}~~~
    L_e = \sum\limits_{j=0}^{N} \frac{1}{2} m_j n_j a_j^2 e_j^2 . 
    \label{eq:L0_and_Le}
\end{equation}
Here, $L_0$ is the system's total circular angular momentum which, by virtue of orbit-averaging, is a conserved quantity. This is simply because secular perturbations do not alter the individual orbital energies, or semimajor axes, of interacting bodies \citep{mur99}. The term $L_e$, on the other hand, is the system's total angular momentum deficit (AMD) which, physically, quantifies the amount of angular momentum that needs to be injected into the system to circularize the orbits of all interacting bodies. While the individual AMD of each ring is not a conserved quantity and can evolve significantly over secular timescales, the system's total AMD, $L_e$, is conserved. This can be demonstrated as follows. First, by making use of the relationship $ de_j^2/ dt = 2 [ k_j  (dk_j/dt) +  h_j (dh_j/dt) ]$ and the equations of motion given by Equation (\ref{eq:EOM_k_h}),
one can write $dL_e/dt$ as:
\begin{eqnarray}
    \frac{dL_e}{dt} &=& \sum\limits_{j=0}^{N} \sum\limits_{l = 0, l \neq j}^{N} m_j n_j a_j^2 A_{jl} (h_j k_l - h_l k_j) , 
    \label{eq:dLedt_sum}
    \\
    &=& \sum\limits_{j=0}^{N} \sum\limits_{l = 0, l \neq j}^{N} \frac{1}{4} \frac{m_j m_l}{M_c + m_j} n_j^2 a_j^2 g(\alpha_{jl}, H) (h_j k_l - h_l k_j) \nonumber .
\end{eqnarray}
Next, one can write $dL_e/dt \equiv S_1 - S_2$, with $S_1$ and $S_2$ being the terms that involve summing over the $h_j k_l$ and $h_l k_j$ terms in the second line of Equation (\ref{eq:dLedt_sum}), respectively. Since $j$ and $l$ are dummy indices, they can be interchanged e.g. in the expression of $S_1$, and upon using the  relationships $(n_l / n_j)^2 = \alpha_{jl}^{-3} (M_c + m_l)/(M_c+m_j)$ and  $g(\alpha_{lj}, H) = \alpha_{jl} g(\alpha_{jl}, H)$ (Section \ref{sec:problemframework_P2}),  it can be shown after some algebra that $S_1 = S_2$. This concludes our analytic proof that $dL/dt = 0$ \citep[see also][]{hahn2003}.

The conservation of the system's total angular momentum can be used as a  reliable diagnostic for the quality of the numerical scheme adopted in evolving a given  $N$-ring system (Equation \ref{eq:SimEqn}). Based on this, we analyzed all of the simulations carried out for the purposes of this work (see Table \ref{table:debris_Nring_sims_table}), finding that the total AMD is conserved  to within a fractional error of $|\Delta L_e / L_e | \sim 10^{-9} - 10^{-8}$ in all runs (see Equation \ref{eq:L0_and_Le}). As can be expected, we found that this is the case at all times as long as no ring attains an eccentricity of $e(t) \gtrsim 1$  in the course of the evolution (e.g., at and around $a\res$).\footnote{We remind the reader that eccentricities larger than unity are not physical in our model; they simply result from adopting a second-order Laplace--Lagrange theory  which is valid for $e \ll 1$ (Section \ref{sec:problemframework_P2}); see e.g. \citet{malhotra98, mur99}. For a detailed discussion of higher-order eccentricity terms, the reader is referred to Section 7.4 of Paper I.} We also repeated some of the simulations listed in Table \ref{table:debris_Nring_sims_table}  by modeling the planet as a thick, softened ring -- rather than razor-thin with $H = 0$ -- to find that it does not affect the conservation of the system's total angular momentum. Finally, it is important to note that the analysis presented here is general: that is, it holds true whether the non-axisymmetric perturbations among the disk rings are ignored (as in the current work) or not (as planned for future works).

\subsection{Tests Against Known Analytic Solutions}
\label{sec:p2-vs-p1-section}

We now demonstrate that  the $N$-ring model can successfully reproduce the analytical solutions derived in Paper I  for the evolution of  planetesimal eccentricities $e(t)$ and apsidal angles $\Delta\varpi(t)$  -- namely,   Equations (11) and (12) therein.

\begin{figure}[t!]
\epsscale{1.2}
\plotone{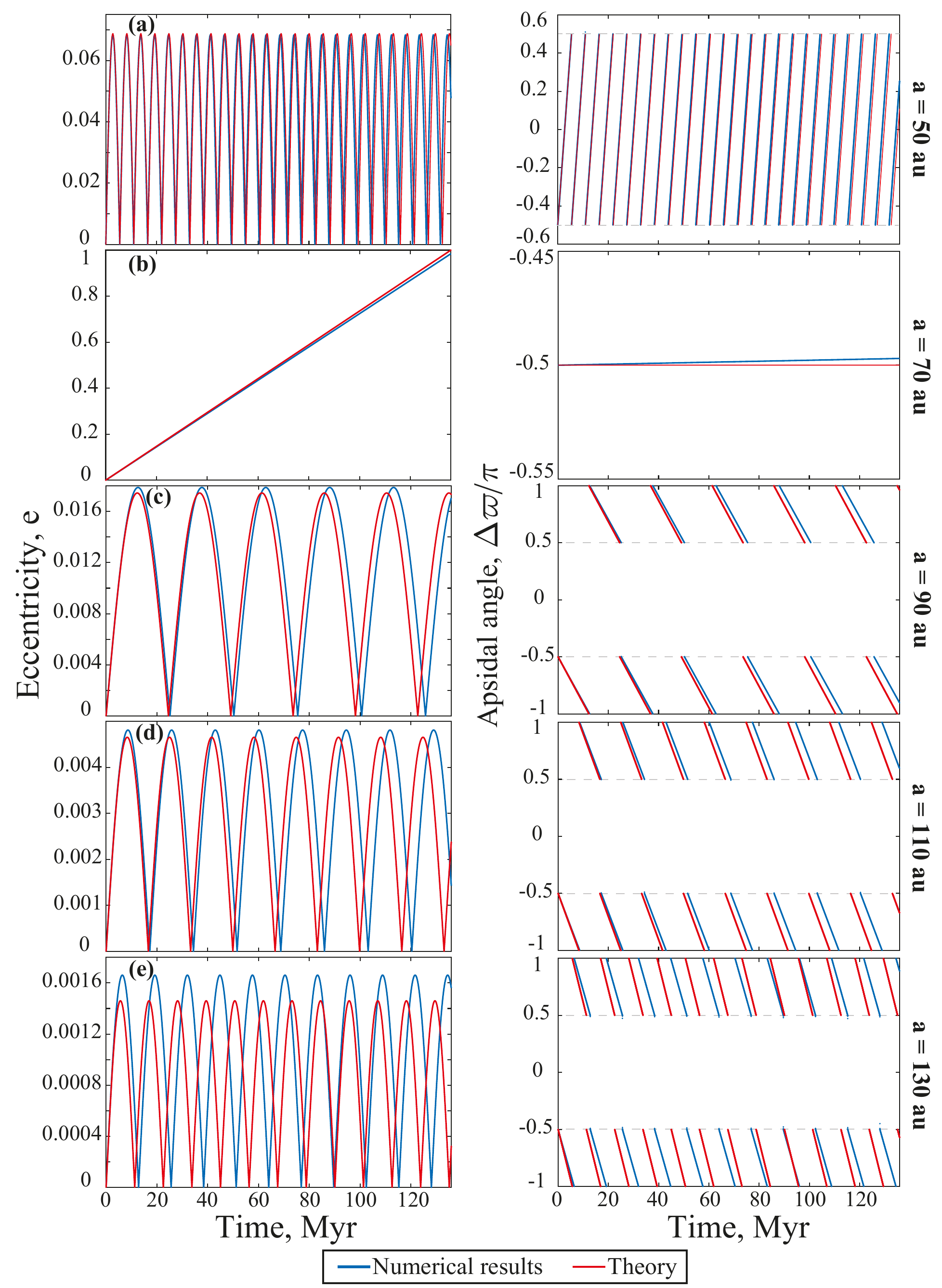}
\caption{Test of the $N$-ring model of Section \ref{sec:problemframework_P2}: a comparison of the planetesimal orbital evolution in the softened $N$-ring numerical simulation (blue curves) versus the unsoftened analytical results of Paper I (Equations (11) and (12) therein; red curves). Each row displays the time evolution of the planetesimal eccentricity $e$ (left panels) and apsidal angle $\Delta \varpi$ (right panels, measured relative to that of the precessing planet) at a given semimajor axis $a$, as indicated on the right side of the figure. The planetesimals were initiated on circular orbits ($e(0) = 10^{-5}$) in the fiducial planet--disk model (\texttt{Model A}, Table \ref{table:debris_Nring_sims_table}). The $N$-ring simulation assumed $N = 5\times 10^{3}$ disk rings, each with a softening parameter of $H=0.1$, while leaving the planet's potential unsoftened. 
One can see a very good agreement between the numerical results of the $N$-ring model and the theoretical results of Paper I at all semimajor axes. See the text (Appendix \ref{sec:p2-vs-p1-section}) for more details.}
\label{fig:P1-vs-P2}
\end{figure}

At the outset, we remind the reader that results of Paper I were derived under the assumption that the debris disk contributes to the system's evolution through only its \textit{axisymmetric} (self-)gravity, i.e., by ignoring its non-axisymmetric component. Thus, in order to perform a direct comparison between the results of the $N$-ring model and Paper I, we switched off the  terms $A_{j,l\neq 0}$ appearing in the interaction matrix $\mathbf{A}$ (see Equations (\ref{eq:Ajk}), (\ref{eq:SimEqn}), and Section \ref{sec:Rj_P2_physical_meanings}), and simulated the evolution of \texttt{Model A}  as an example (Table \ref{table:debris_Nring_sims_table}).  We did so by modeling the disk as a collection of $N = 5 \times 10^{3}$ softened rings with  $H  = 0.1$, and the planet as an unsoftened ring. Additionally, the disk rings were initiated with $e(0) = 10^{-5}$ and $\varpi(0) = -\pi/2$ to mimic the initial conditions of Paper I's analytical results.

The results of such a simulation are shown in Figure \ref{fig:P1-vs-P2}, where we plot the time evolution of planetesimal eccentricities  (left panels) and apsidal angles  (relative to the precessing planet, right panels)  at different semimajor axes, as indicated to the right of each panel.
For comparison, Figure \ref{fig:P1-vs-P2} also shows the corresponding results obtained using Equations (11) and (12) of Paper I. Looking at Figure \ref{fig:P1-vs-P2}, it is evident that the agreement between the $N$-ring simulation results (shown in blue curves) and the analytical results of Paper I (shown in red curves) is very good at all semimajor axes. Indeed, the simulated curves follow closely those of Paper I, with agreement to within several percent in both the amplitudes of eccentricity variations as well as the periods associated with the oscillations of $e(t)$ and $\Delta\varpi(t)$. Generally, the discrepancies  are negligible throughout nearly the entire disk, i.e., $a\inn \lesssim a \lesssim a\out$, with differences being about $2 - 3 \%$ when compared to the analytical solutions; see panels (a)--(d) of Figure \ref{fig:P1-vs-P2}. Note, however, that the discrepancies become more pronounced, i.e., on the order of $\approx 10\%$, for planetesimals near the outer disk edge, $a \approx a\out$; see e.g. Figure \ref{fig:P1-vs-P2}(e). Nevertheless, such deviations are of no practical importance for our purposes, in the sense that they do not significantly affect e.g. the disk's spatial appearance.

We note that the discrepancies in Figure \ref{fig:P1-vs-P2} can be  explained based on the study of \citet{SR19}. There, it was shown that for the softening prescription of \citet{hahn2003} (as in this paper), a discretized disk must be modeled with $N \sim C H^{-2}$ rings to accurately capture the secular effects of a continuous disk. This scaling was found to hold for particles far from the disk edges, i.e., $a\inn \ll a\ll a\out$, with $C$ being roughly inversely proportional to the desired level of accuracy \citep[for more details, see Section 6.3 of][]{SR19}.  The deviations evident in Figures \ref{fig:P1-vs-P2}(a)--(d) are consistent with those findings: indeed, for our adopted  values of $N$ and $H$, results of \citet{SR19} predict a discrepancy of $\sim 1$ per cent at $a\inn \ll a\ll a\out$. Additionally, results of \citet{SR19} show that for particles orbiting near the disk edges (i.e., $a \approx a\inn, a\out$), accurate representation of the disk-induced dynamics requires the use of a rather small value of softening, i.e., $H \lesssim 10^{-3}$ (see e.g. their Section 5). This explains why the discrepancies in panel (e) of Figure \ref{fig:P1-vs-P2}, where $a \lesssim a\out$,  are larger than in the other panels. Relatedly, we remark that if not for the planet in our simulation, which, recall, dominates the planetesimal free precession rate in the inner disk parts (Figure \ref{fig:A-a-from-matrix-fid-simA-nominal}), relatively large deviations would have also been observed in e.g. panel (a) of  Figure \ref{fig:P1-vs-P2}, where $a \gtrsim a\inn$. Finally we note that, although not shown in Figure \ref{fig:P1-vs-P2},  the $N$-ring simulation -- in contrast to Paper I -- does not feature a secular resonance at $a \approx a\inn$, i.e., apart from the one at $70$ au (or, more generally,  at $\gtrsim a\inn$ in other systems; see e.g. Figure \ref{fig:e-w-fn-a-simA-p2}). This is a simple restatement of the fact that  a softening value of $H=0.1$ is not sufficient to capture the divergence of the disk-induced precession near the disk edges; see Figure \ref{fig:A-a-from-matrix-fid-simA-nominal} and \citet{SR19}.

To summarize, the general conclusion from Figure \ref{fig:P1-vs-P2} is that the softened $N$-ring model of Section \ref{sec:problemframework_P2}, when taken  in the appropriate limit,  reproduces the analytical results of Paper I to within an acceptable degree of accuracy.  This conclusion is further supported by the results of Section \ref{sec:analysis_predictions}, where, for ease of distinguishing between our current and previous works, we also report on the outcomes of a suite of Paper I-like $N$-ring simulations; see e.g. Figures \ref{fig:wpdot_fn_ap_P2}, \ref{fig:ares_fn_ap_P2}, and \ref{fig:tau-ap-withD-paper2}.

\subsection{Sensitivity to input parameters \texorpdfstring{$N$}{N} and \texorpdfstring{$H$}{H}}
\label{sec:sensitivity_to_N_H_P2}

Apart from the physical parameters describing a given system (i.e., masses and initial orbital elements), the $N$-ring model takes two other key numerical parameters as inputs: the number of disk rings $N$ and their aspect ratio $H$.  To investigate the sensitivity of the model to these parameters, we performed two new sets of simulations of \texttt{Model A} (Table \ref{table:debris_Nring_sims_table}) in which  $N$ and $H$ are varied  away from their fiducial values, i.e., $N=5 \times 10^{3}$ and $H=0.1$ (Section \ref{sec:fid_par_setupSection}).

In the first set of simulations, we varied the number of disk rings within the range $5\times 10^{2} \leq N \leq 10^{4}$, while holding $H$ constant at $H=0.1$. In doing so, we found that all simulations have nearly perfect agreement with each other, except for some slight quantitative deviations towards the smallest end of $N$. Indeed, simulations with $N \sim 5\times 10^{2}$ resulted in e.g. the resonance location being shifted inwards at most by a few percent compared to the nominal location (typically, by $\lesssim 2\%$). No significant differences were observed, however, for $N\gtrsim 2\times 10^{3}$, implying that results obtained with the fiducial value of $N$ can be considered to be converged (assuming $H=0.1$). For future use in Paper III, we also confirmed that this is the case regardless of whether the non-axisymmetric self-gravity of the disk is included or not. We note that the behavior described here is in line with \citet{SR19}, who determined the minimum $N$ that well-approximates the effects of a continuous disk (i.e., with $N \rightarrow \infty$) for a given value of softening -- see e.g. their Section 6.3 and Figure 9.

In the second set of simulations, we varied the aspect ratio of the disk rings  $H$ between $0.05$ and $0.2$, while keeping their number constant at $N = 5\times10^{3}$. In doing so, we found that the results could be sensitive to the value of $H$, depending on whether the disk's non-axisymmetric self-gravity is (i) ignored or (ii) included.  In the first case, i.e., as in the present paper, we measured only negligible differences  between the simulations adopting different values of $H$ -- similar to the case of variations in $N$.  In the second case, however, which is the subject of Paper III, we observed some quantitative differences in terms of e.g. the eccentricity oscillations and precession periods of planetesimals in the outer disk parts, i.e.,  $ a\res \lesssim a \lesssim a\out$.  These findings can be understood based on results of \citet{SR19} concerning the softening prescription of \citet{hahn2003}. Indeed, looking at their Figure 1(A), one can see that for a $p=1$ disk, the axisymmetric component of the softened disk self-gravity  matches the unsoftened results for all $H \lesssim 0.2$. However, as can be seen from their Figure 1(B), this is not the case for the non-axisymmetric component of the disk self-gravity, which fully converges to the unsoftened results only when $H \lesssim 0.01$.

In summary, the $N$-ring model can, in general, be sensitive to the adopted number of disk rings and their aspect ratio. Nevertheless, results presented in this work do not vary strongly provided that $H\lesssim 0.1$ and $N\gtrsim 2 \times 10^{3}$, justifying our particular choices of their fiducial values (Section \ref{sec:fid_par_setupSection}). Note that, in principle, we could have instead adopted e.g. $H \lesssim 0.01$ in our simulations. This, however, would have necessitated a much larger value of $N$ than that adopted \citep[see e.g.][]{SR19}, rendering the $N$-ring model computationally expensive and slow to operate. Finally, we stress that the constraints on $N$ and $H$ discussed here apply strictly to power-law disks with $p=1$ (Equation \ref{eq:Sigma_d}), and thus might differ for different disk models;  see e.g. Figures 1 and 9 of \citet{SR19}.

\subsection{The Sun-Jupiter-Saturn System}
\label{app:sun-jupiter-saturn}

As mentioned in Section \ref{sec:problemframework_P2}, the softened $N$-ring model represents the continuum version of the classical Laplace--Lagrange planetary theory. Thus, as a final and simple test, we investigated the performance  of the $N$-ring model in reproducing the secular dynamics of systems harboring planets, but not disks. To do so, we considered the well-known examples provided in \citet{mur99} related to the Sun-Jupiter-Saturn system (see their Sections 7.3 and 7.5). We first verified that the $N$-ring model, with $N=2$,  perfectly reproduces the  Laplace--Lagrange planetary solutions for Jupiter and Saturn as reported in Figure 7.1 of  \citet{mur99}, provided that we set the softening exactly to zero. We then verified that in the same limit, i.e., $H=0$, the $N$-ring model also accurately reproduces the evolution of massless test particles in the Sun-Jupiter-Saturn system as reported in e.g. Figure 7.5 of \citet{mur99}. This said, however, we found that stringent conservation of the system's angular momentum requires defining the mean-motion as $n_j = [G (M_c + m_j)/a_j^3]^{1/2}$, rather than as $n_j = [G M_c/a_j^3]^{1/2}$ as done in \citet{mur99}. This is in agreement with Appendix \ref{sec:L_conservation_P2}, as well as previous results on the same subject  \citep[][]{hahn2003}.

\section{Simulation Parameters}
\label{chap:app_table_sim}

\restartappendixnumbering

The results of this work are based on the outcomes of a set of $67$ simulations carried out using the softened $N$-ring model outlined in Section \ref{sec:problemframework_P2}. The parameters of the simulated planet--debris disk systems, along with their outcomes, are listed in Table \ref{table:debris_Nring_sims_table}. In each simulation, unless otherwise stated here or in the text itself, (i) the central star's mass is $M_c = 1.09 M_{\odot}$; (ii)  the planet is modeled as a thin, unsoftened ring; (iii) the planet's initial apsidal angle is set to $\varpi_p(0) = 0$; (iv) the disk is modeled as a collection of $N = 5\times10^{3}$ softened rings, each with a softening parameter of $H = 0.1$; (v) the disk rings are initiated with eccentricities $e(0) = 10^{-4}$  and apsidal angles $\varpi(0) = -\pi/2$, respectively; and (vi) integrations are stopped after $200$ Myr.

\startlongtable
\begin{deluxetable}{c c c c c c c}
\centerwidetable
\setlength{\tabcolsep}{2pt} 
\renewcommand{\arraystretch}{0.95} 
\tabletypesize{\scriptsize} 
\tablecaption{Parameters and outcomes of the planet--debris disk systems considered in Sections \ref{sec:results_p2} and \ref{sec:analysis_predictions}.}
\label{table:debris_Nring_sims_table}
\tablehead{
\colhead{Sim.} & \colhead{$M_d [M_{\oplus}]$}  & \colhead{$m_p [M_J]$}     & \colhead{$a_p [au]$}  & \colhead{$e_p(0)$}  &  \colhead{Outcome} & \colhead{Notes}}
\startdata
%
1 &  10  & 0.31 & 25.96 &  0.05 & 0 &   
\\ 
2 &  ...  & ... & ... &  0.10 & 1  &  
\\ 
3 &  20 & 0.59 & 25.38 &  0.05  &  0 &  
\\ 
4 &  ... & ... & ... &  0.10 & 1  &  
\\ 
5 &  50 & 1.40 & 24.46  & 0.05 & 1   &  
\\ 
6 &  ... & ... & ... & 0.10 &  1 &  
\\ 
7 &  100 & 2.74 & 23.63  & 0.05 & 1 &  
\\  
8 &  ... & ... & ...  & 0.10  &  1  &  
\\ 
%
9 &  10 & 0.43 & 15  & 0.05 & 0 &  
\\  %
10 &  ... & ... & ...  & 0.10 & 0 & 
\\  %
11 &  ... & ... & ...  & 0.30 & 1 & Stopped at $150$ Myr
\\  %
12 &  20 & 0.86 & ...  & 0.05 & 0  &  
\\  %
13 &  ... & ... & ...  & 0.10 & 1 &  
\\  %
14 &  ... & ... & ...  & 0.30 & 1 &   Stopped at $150$ Myr
\\  %
15 &  50 & 2.15 & ...  & 0.05 & 1 &   
\\  %
16 &  ... & ... & ...  & 0.10  & 1  &   
\\  %
17 &  ... & ... & ...  & 0.30 & 1 &   Stopped at $150$ Myr
\\  %
18 &  100 & 4.30 & ...  & 0.05 & 1  &   
\\  %
19 &  ... & ... & ...  & 0.10 & 1 &   
\\  %
20 &  ... & ... & ...  & 0.30 & 1 &    Stopped at $150$ Myr
\\  %
21 &  10 & 0.3 & 20  & 0.05 & 0  &  
\\  %
22 &  ... & ... & ...  & 0.10 & 0 &   
\\  %
23 &  ... & ... & ...  & 0.30 & 1 &   Stopped at $100$ Myr
\\  
\textbf{24} &  20 & 0.6 & ...  & 0.05 & 1  &    \textbf{\texttt{Model A} of Paper I}.
\\  
 &   &  &   &  &   & Stopped at $400$ Myr
\\  %
25 &  ... & ... & ...  & 0.10 &  1 &   
\\  %
26 &  ... & ... & ...  & 0.30 & 1 &   Stopped at $100$ Myr
\\  %
27 &  50 & 1.5 & ...  & 0.05 & 1  &   
\\  %
28 &  ... & ... & ...  & 0.10  & 1 &  
\\  %
29 &  ... & ... & ...  & 0.30 & 1  &   Stopped at $100$ Myr
\\  %
30 &  100 & 3 & ...  & 0.05 & 1 &   
\\  %
31 &  ... & ... & ...  & 0.10 & 1 &   Stopped at $150$ Myr
\\  %
32 &  ... & ... & ...  & 0.30 & 1  &   Stopped at $100$ Myr
\\  %
33 &  30 & 1.68 & 12.8  & 0.05 & 0 &   
\\  %
34 &  ... & ... & ...  & 0.10 & 1 & Stopped at $150$ Myr 
\\  %
35 &  ... & ... & ...  & 0.20 & 1 & Stopped at $150$ Myr 
\\  %
36 &  ... & ... & ...  & 0.30 & 1 & Stopped at $150$ Myr   
\\  %
37 &  100 & 5.59 & ...  & 0.05 & 1 &   
\\  %
38 &  ... & ... & ...  & 0.10 & 1 &   
\\  %
39 &  ... & ... & ...  & 0.20 & 1 &   
\\  %
40 &  ... & ... & ...  & 0.30 & 1 &   
\\  %
41 &  21 & 0.81 & 16.1  & 0.05 & 0 &  
\\  %
42 &  ... & ... & ...  & 0.10 & 1  &  Stopped at $150$ Myr   
\\  %
43 &  ... & ... & ...  & 0.20 & 1 &  Stopped at $150$ Myr   
\\  %
44 &  ... & ... & ...  & 0.30 & 1 &  Stopped at $150$ Myr   
\\  %
45 &  70 & 2.72 & ...  & 0.05  & 1 &  
\\  %
46 &  ... & ... & ...  & 0.10 & 1 &  
\\  %
47 &  ... & ... & ...  & 0.20 & 1  &  
\\  %
48 &  ... & ... & ...  & 0.30 & 1 &  
\\  %
49 &  10.5 & 0.29 & 22.04  & 0.05 & 0 &  
\\  %
50 &  ... & ... & ...  & 0.10 & 0 &   Stopped at $150$ Myr 
\\  %
51 &  ... & ... & ...  & 0.20 & 1 & Stopped at $150$ Myr 
\\  %
52 &  ... & ... & ...  & 0.30 & 1 &  Stopped at $150$ Myr
\\  %
53 &  35 & 0.97 & ...  & 0.05 & 1 &  
\\  %
54 &  ... & ... & ...  & 0.10 & 1 &  
\\  %
55 &  ... & ... & ...  & 0.20 & 1 &  
\\  %
56 &  ... & ... & ...  & 0.30 & 1  &  
\\  %
57 & 121.2 & 12 & 9.26 & ... & 1 &  Stopped at $100$ Myr
\\  %
58 & 50.5 & 5 & ... & ...  & 1 &   Stopped at $100$ Myr
\\  %
59 & 10 & 0.99 & ... & ...  & 1 &  Stopped at $300$ Myr
\\  %
60 & 64.3 & 12 & 6.59 & ...  & 1 &  
\\  %
61 & 20 & 3.73 & ... & ...  & 1  &  
\\  %
62 & 10 & 1.87 & ... & ...  & 1 & Stopped at $400$ Myr 
\\  %
63 & 50 & 25.1 & 3.94 & ...  & 1 &  Stopped at $300$ Myr 
\\  %
64 & 24 & 12 & ... & ...  & 1 &  Stopped at $300$ Myr 
\\  %
65 & 10 & 5.02 & ... & ...  & 1 & Stopped at $500$ Myr 
\\ %
\textbf{66} & 95 & 15.8 & 7 & 0.05 & 1  &   \textbf{\texttt{Model B} of Paper I}. 
\\ %
 &  &  &  &  &   &  
Stopped at $400$ Myr 
\\  %
\textbf{67} & 6 & 0.2 & 26.93 & ... & 0  &  \textbf{\texttt{Model C} of Paper I}.
\\  %
 &  &  &  &  &   & 
Stopped at $400$ Myr
\\
\enddata
\parshape 1 0em 1\columnwidth 
\tablecomments{ 
{\footnotesize The combinations of $M_d$, $m_p$, and $a_p$ (columns 2--4) are chosen such that, according to Paper I, the system is expected to produce a gap in a \HD-like disk at $a\res = 70$ au; see Figure 7 therein. Note, however, that here we have discarded all but the resonance location constraint considered in Paper I (see Section 4 therein). That is, the chosen parameters lie both within and outside the ``allowed'' region of Figure 7 in Paper I. Column 5 represents the planet's initial eccentricity.  Outcomes 1 and 0 in column 6 signify whether  planetesimal eccentricities had achieved a minimum value of $1$ at the resonance (i.e., $e(a_{\rm res}) \geq 1$) or were in the process of doing so, respectively, by the time the simulation was stopped. Simulation $24$ (i.e., $\texttt{Model A}$) is the fiducial model adopted in this work, same as in Paper I.}}
\end{deluxetable}

\section{Planetesimal dynamics in the Case of a Circularizing Planet}
\label{appendix:analytical_solution_with_ep_decay_D}
\restartappendixnumbering

We now update the analytical solutions derived in Paper I for the secular evolution of planetesimals, i.e., \EQPP{11} and \EQPP{12}, which were derived assuming a non-evolving, axisymmetric surface density (Equation (\ref{eq:Sigma_d})).  To do so, we adopt the analytic expression for the secular disturbing function  of planetesimals, $R$, given by \EQPP{9}, and introduce two \textit{ad hoc}  modifications that imitate the dynamical effects due to  the disk's non-axisymmetric torque on the planet (absent in Paper I). First, we multiply the  term $B_p \propto e_p$ appearing in $R$ -- which, recall, represents the non-axisymmetric component of the planetary potential --  by an exponential factor of ${\rm exp}(-Dt/2)$. This essentially ensures that the decay of the planetary eccentricity due to resonant friction  is taken into account properly (Section \ref{sec:resonant_friction}). Second, we replace the term $A_{d,p}$ appearing in $R$ -- which, recall, represents the planetary \textit{free} precession rate  --  by $\dot{\varpi}_p$ of Equation (\ref{eq:wpdot_mod_theory}) which accounts for both the free and forced precession rates (see Section \ref{sec:effect_on_wpdot} and Appendix \ref{app:wpdot_Bdp_derivation}). This guarantees that we are still working in a frame co-precessing with the planet, an assumption inherent in the derivation of $R$ in Paper I.

As a result, the disturbing function $R$ now reads as:
\begin{align}
    R &  =  n a^2 \Bigl[ \frac{1}{2} \left( A_d + A_p - \dot{\varpi}_p \right) e^2 
    \nonumber\\
 & \hspace*{1.5cm} +   B_p(0) ~ \mathrm{exp}(-D t / 2 ) e \cos(\varpi-\varpi_p) \Bigr]. 
    \label{eq:Rgeneral_with_res_friction}
\end{align}
In Equation (\ref{eq:Rgeneral_with_res_friction}),  $B_p(0)$ must be understood as $B_p$ of Paper I (\EQPP{7}) evaluated using the planet's initial eccentricity, $e_p(0)$, and the meaning of all other terms is the same as in Paper I -- see e.g. Section 2.2 therein. Equipped with $R$ of Equation (\ref{eq:Rgeneral_with_res_friction}), the evolution of the planetesimal eccentricity vector in a frame corotating with the planet's apsidal line, i.e., $\mathbf{e} = (K,H) = e (\cos\Delta\varpi, \sin \Delta\varpi)$, can be determined by Lagrange's planetary equations such that \citep[][see also Section 2.3 of Paper I]{mur99}:
\begin{align}
  &  \frac{dK}{dt} \approx \frac{- 1}{n a^2} \frac{\partial R}{\partial H}  =    - (A - \dot{\varpi}_p ) H    , \nonumber
\\
 & \frac{dH}{dt} \approx  \frac{1}{n a^2} \frac{\partial R}{\partial K}  =   (A - \dot{\varpi}_p) K + B_p(0) ~ \mathrm{exp}(-D t /2 ) ,
\label{eq:EOM_analytical_with_res_friction}
\end{align}
where, as before, $A = A_d + A_p$. Note that when the non-axisymmetric torque of the disk on the planet is ignored (as in Paper I), one has $\dot{\varpi}_p \rightarrow A_{d,p}$ and $D = 0$ (see Sections \ref{sec:effect_on_wpdot} and \ref{sec:resonant_friction}), so that the system of equations (\ref{eq:EOM_analytical_with_res_friction}) reduce to those in Paper I -- namely, Equations (10) therein.

The system of equations (\ref{eq:EOM_analytical_with_res_friction}) can be solved using standard methods \citep{mur99}. In particular, when planetesimals  are initiated on circular orbits, i.e., $K(0) = H(0) =0$, the evolution of their eccentricity vector can be represented as follows:
\begin{eqnarray}
{\bf e}(t)&=&
\left\{
\begin{array}{l}
K(t) \\
H(t)
\end{array}
\right\}=
{\bf e}_{\rm free}(t)+{\bf e}_{\rm forced}(t),
\label{eq:vec_sol}\\
{\bf e}_{\rm free}(t)&=&
\frac{B_p(0)}{\sqrt{(A-\dot{\varpi}_p)^2+(D/2)^2}}
\nonumber\\
&\times &
\left\{
\begin{array}{l}
\cos[(A-\dot{\varpi}_p) t+\varphi_0]\\
\sin[(A-\dot{\varpi}_p) t+\varphi_0]
\end{array}
\right\},
\label{eq:e)_free}\\
{\bf e}_{\rm forced}(t)&=& -
\frac{B_p(0)\exp(-D t / 2 )}{\sqrt{(A-\dot{\varpi}_p)^2+(D/2)^2}}
\left\{
\begin{array}{l}
\cos\varphi_0\\
\sin\varphi_0
\end{array}
\right\},
\label{eq:e)_forced}
\end{eqnarray}
and the phase shift $\varphi_0$ is such that
\begin{eqnarray}
\left\{
\cos\varphi_0\, , \,
\sin\varphi_0
\right\}=
\frac{\left\{
A-\dot{\varpi}_p\, , \,
D/2
\right\}}{\sqrt{(A-\dot{\varpi}_p)^2+(D/2)^2}}.
\label{eq:phi0}
\end{eqnarray}
This solution for vector ${\bf e}(t)$ is illustrated in Figure \ref{fig:complex-h-k-scaled}. Note that when the planetary eccentricity decays, i.e. $D\neq 0$, ${\bf e}_{\rm forced}$ is misaligned with the planetary eccentricity vector. However, this misalignment is small when $D\ll (A-\dot{\varpi}_p)$, which is the case in practice (see e.g. Figures \ref{fig:A-a-from-matrix-fid-simA-nominal} and \ref{fig:D_fn_ap_P2}).

We can use the solution (\ref{eq:vec_sol})--(\ref{eq:phi0}) to find
\begin{align}
& e(t) = 
 \sqrt{ 
 \frac{4B_p^2(0) F_1(t)} {4(A - \dot{\varpi}_P)^2 + D^2}  
 } , 
\label{eq:master_app_e}
 \\
& \tan\Delta\varpi(t)  = 
\frac{- 2 (A - \dot{\varpi}_p) \sin [(A-\dot{\varpi}_p) t] + D F_2(t)}{ 2 (A-\dot{\varpi}_p) F_2(t) + D \sin [ (A - \dot{\varpi}_p) t] } ,
\label{eq:master_app_w}
\end{align}
where, for conciseness, we have defined:
\begin{align}
    F_1(t) & \equiv   1 + \mathrm{exp}(-Dt) -2 ~ \mathrm{exp}(-D t / 2 ) \cos[(A-\dot{\varpi}_P)t]     
    ,
    \\
    F_2(t) & \equiv \mathrm{exp}(-Dt/2) - \cos [(A-\dot{\varpi}_p)t] 
    .
\end{align}
This solution for ${e}(t)$ is illustrated in Figure \ref{fig:e-w-fn-time-simA-p2}. Finally, note that Equations (\ref{eq:master_app_e}) and (\ref{eq:master_app_w})  can be expressed in a more appealing and comparable form to their analogues in Paper I when expanded to lowest order in $D$, so that:
\begin{align}
 &  e(t)   =  \bigg| \frac{2 B_p(0)}{A-\dot{\varpi}_p} \bigg( 1- \frac{1}{4} D  t\bigg) \sin\bigg( \frac{A- \dot{\varpi}_p}{2}t \bigg) \bigg|   
    + \mathcal{O}(D^2)   
    ,
 \\
&  \tan\Delta\varpi(t)   =  \tan\bigg( \frac{A - \dot{\varpi}_p}{2} t - \frac{\pi}{2}   \bigg)
+ \mathcal{O}(D) 
.
\label{eq:tanDw_effect_of_smallD}
\end{align}
This completes our development of the general solutions describing the evolution of planetesimal orbits in the case of a circularizing planet.

\section{Planetary Precession Rates due to  eccentric disks}
\label{app:wpdot_Bdp_derivation}
\restartappendixnumbering

In this work, we have presented a new expression for the apsidal precession rate of a planet that is subject to both the axi- and non-axisymmetric perturbations of an external disk; see Equation (\ref{eq:wpdot_mod_theory}).  We now present a derivation of this equation. To do so, we consider the debris disk to be a continuous entity -- rather than a collection of $N$ discrete rings -- characterized by the mass distribution $\mu_d(a)=dm/da \approx 2\pi a\Sigma_d(a)$.  We also assume that the disk has some eccentricity $e_d(a)$ within its radial extent, i.e., for $a\inn \leq a \leq a\out$. Plugging these ansatzes into Equations (\ref{eq:Master_Rj})--(\ref{eq:Ajk}),  the disturbing function of the planet due to the disk can be written as \citep[see also Eq. (4)--(6) in][]{SR19}:
\begin{equation}
    R_{d,p} = n_p a_p^2 \bigg[ \frac{1}{2} A_{d,p} e_p^2  + \int\limits_{a_{\rm in}}^{a_{\rm out}} \delta B_{d,p}(a) ~ {\bf e}_p\cdot {\bf e}_d(a) da\bigg] ,
    \label{eq:Master_Rp1}
\end{equation}
where ${\bf e}_p$ and ${\bf e}_d(a)$ are the eccentricity vectors of the planet and of the disk elements, respectively, and 
\begin{eqnarray}
A_{d,p} &=& \frac{1}{4} n_p \int\limits_{a_{\rm in}}^{a_{\rm out}} \frac{\mu_d(a)}{M_c} \bigg(\frac{a_p}{a}\bigg)^2 b_{3/2}^{(1)}(a_p/a)da, 
\label{eq:Ajj-new}
\\
\delta B_{d,p}(a) &=& -\frac{1}{4} n_p  \frac{\mu_d(a)}{M_c} \bigg(\frac{a_p}{a}\bigg)^2 b_{3/2}^{(2)}(a_p/a) .
\label{eq:Ajk-new}
\end{eqnarray}
Note that the softening is set to zero in these expressions since the planet orbits completely interior to the disk and the potential does not need to be softened \citep{SR19}.  Here, we point out that with some algebra, Equation (\ref{eq:Ajj-new}) can be expressed in the form of \EQPP{A3} for power-law surface density profiles $\Sigma_d(a)$, see Equation (\ref{eq:Sigma_d}).

Next, we identify the eccentricity of the disk elements with that of the planetesimals in our simulations so that, using Equations (\ref{eq:vec_sol})--(\ref{eq:phi0}), we can write:
\begin{equation}
{\bf e}_p\cdot{\bf e}_d(a)={\bf e}_p\cdot{\bf e}_{\rm free}(a)+{\bf e}_p\cdot{\bf e}_{\rm forced}(a).
\label{eq:terms}
\end{equation}
Plugging the first term of Equation (\ref{eq:terms}), i.e., ${\bf e}_p\cdot{\bf e}_{\rm free}(a)=e_p e_{\rm free}(a)\cos[(A(a)-\dot\varpi_p)t+\varphi_0]$, into Equation (\ref{eq:Master_Rp1}) would result in an integral over a function rapidly oscillating in $a$,   simply because $|A(a)-\dot\varpi_p|t\gg 1$ at late times. Because of the rapidly oscillating integrand, this integral will be very small and can be neglected. As a result, the free eccentricities of  debris particles in the disk do not contribute to the planetary disturbing function, and thus to the planetary precession.
On the other hand, the second term in Equation (\ref{eq:terms}) is just 
\begin{equation}
{\bf e}_p\cdot{\bf e}_{\rm forced}(a) = e_p e_{\rm forced}\cos\varphi_0\approx e_p e_{\rm forced},
\label{eq:termsff}
\end{equation}
where the last approximation assumes $D/2\ll (A(a)-\dot\varpi_p)$, see Equation (\ref{eq:phi0}). Plugging Equation (\ref{eq:termsff}) into Equation (\ref{eq:Master_Rp1}) would result in a non-vanishing contribution such that\footnote{One can show that the slight misalignment between the vectors ${\bf e}_{\rm forced}$ and ${\bf e}_p$ as described by Equation (\ref{eq:phi0}) leads to only a small correction to $\dot e_p$ compared to Equation (\ref{eq:D_Tr98_MAIN}) as long as $D\ll (A-\dot\varpi_p)$, which is the case in practice (see e.g. Figures \ref{fig:A-a-from-matrix-fid-simA-nominal} and \ref{fig:D_fn_ap_P2}).} 
\begin{equation}
    R_{d,p} \approx n_p a_p^2 \bigg[ \frac{1}{2} A_{d,p} e_p^2  + e_p\int\limits_{a_{\rm in}}^{a_{\rm out}} \delta B_{d,p}(a) e_{\rm forced}(a) da\bigg] .
    \label{eq:Master_Rp2}
\end{equation}

Equipped with Equation (\ref{eq:Master_Rp2}), we can then compute the time evolution of the planetary apsidal angle using Lagrange's planetary equations which, to lowest order in $e_p$, yield \citep{mur99}:
\begin{equation}
    \dot{\varpi}_p \approx \frac{1}{n_p a_p^2 e_p} \frac{\partial R_{d,p}}{\partial e_p}  = A_{d,p} + \int\limits_{a_{\rm in}}^{a_{\rm out}} \delta B_{d,p}(a) \frac{e_{\rm forced}(a)}{e_p(t)} da, 
    \label{eq:varpi_dot_app_1r}
\end{equation}
In Equation (\ref{eq:varpi_dot_app_1r}), and as before, $A_{d,p}$ represents the free precession rate of the planetary orbit (\EQPP{8}). On the other hand,  the second term represents the forced precession rate resulting due to the non-axisymmetric component of the disk gravity, which we had ignored in Paper I. 

We now note that, according to Equation (\ref{eq:e)_forced}), 
\begin{eqnarray}
\frac{e_{\rm forced}(a)}{e_p(t)} &=&\frac{ - B_p(0)\exp(-D t / 2 )}{e_p(t) \sqrt{(A-\dot{\varpi}_p)^2+(D/2)^2}}    . 
\label{eforced_over_ep_rR}
\end{eqnarray}
Next, recalling that $e_p(t) = e_p(0) ~ {\rm exp}(-D t /2 )$ (Equation \ref{eq:ep_decay}), and noting that the largest contribution to the planetary precession rate would arise from the inner disk parts where $A \approx A_p \gg \dot{\varpi}_p, D$ (Figure \ref{fig:A-a-from-matrix-fid-simA-nominal}), Equation (\ref{eforced_over_ep_rR}) can be further simplified to read as follows:
\begin{align}
& \frac{e_{\rm forced}(a)}{e_p(t)} \approx
\frac{-B_p(0)}{A_p e_p(0)} =
\frac{b_{3/2}^{(2)}(a_p/a)}{b_{3/2}^{(1)}(a_p/a)}
\approx \frac{a\inn}{a} \frac{b_{3/2}^{(2)}(a_p/a_{\rm in})}{b_{3/2}^{(1)}(a_p/a_{\rm in})},
\label{eq:eforced_ep_RapproxR}
\end{align}
where we have used the expressions of $A_p$ and $B_p(0)$ as given by Equations (4, PI) and (7, PI), respectively. The last approximation in Equation (\ref{eq:eforced_ep_RapproxR}) is valid to first order in $a_p/a$.

Finally, plugging Equation (\ref{eq:eforced_ep_RapproxR}) into Equation (\ref{eq:varpi_dot_app_1r}), and performing the integral of $\delta B_{d,p}$ (given by Equation (\ref{eq:Ajk-new})) over $a$, one can find after some straightforward algebra that
\begin{align}
  \dot{\varpi}_p &= A_{d,p} \times \left[ 1 + \frac{1}{2}\frac{\phi_2^{q=1}}{\phi_1} \frac{b_{3/2}^{(2)}(a_p/a_{\rm in})}{b_{3/2}^{(1)}(a_p/a_{\rm in})}\right] . 
\label{eq:varpidot_eq_3r}
\end{align}
In Equation (\ref{eq:varpidot_eq_3r}), the coefficients $\phi_1>0$ and $\phi_2<0$ are  determined by the power-law indices of  $\Sigma_d(a) \propto a^{-p}$ and $e_d(a) \propto a^{-q}$ characterizing the disk model, as well as ratios of the relevant semi-major axes, namely, $a_p/a\inn$ and $\delta = a\out/a\inn$; see Equations (A5, PI)--(A8, PI) and figure 13 of Paper I. By construction, we find that the coefficient $\phi_2$ appears as $\phi_2^{q=1}$ in Equation (\ref{eq:varpidot_eq_3r}), i.e., as evaluated at $q = 1$.
Equation (\ref{eq:wpdot_mod_theory}) follows directly from Equation (\ref{eq:varpidot_eq_3r}).

\bibliography{SRW23_DebrisGaps}{}
\bibliographystyle{aasjournal}


\end{document}